\documentclass[13pt]{article}
\usepackage{amssymb}
\usepackage{amsmath}
\usepackage{hyperref}
\usepackage{graphicx}
\usepackage{epstopdf}
\usepackage{setspace}
\usepackage{geometry}
\usepackage{xcolor}
\usepackage{caption}
\begin{document}
	\title{{Characteristics of Rayleigh Waves in Nonlocal Porous Orthotropic Thermoelastic Layer with Diffusion Under Three-Phase-Lag Model}}
	\author{\text{Abhishek Mallick  and Siddhartha Biswas\footnote{\\Corresponding author: Siddhartha Biswas\\Email: siddharthabsws957@gmail.com}} \\Department of
		Mathematics, \\University of North Bengal, Darjeeling, India }

	\date{}
	\maketitle
	
\begin{abstract}
\noindent
\onehalfspacing

This article delves into the intricate dynamics of Rayleigh wave propagation within a nonlocal orthotropic medium, where the presence of void and diffusion adds an intriguing layer to the analysis. Grounded in Eringen's nonlocal elasticity theory and embracing the three-phase-lag model of hyperbolic thermoelasticity, the study focuses on the interplay between the mass diffusion principles of Fick's and the Fourier's law under the framework of hyperbolic thermoelasticity. The investigation employs a methodological approach centered around normal mode analysis to navigate the complexities of the problem at hand. The derived frequency equation governing Rayleigh waves undergoes meticulous scrutiny through the exploration of specific cases.

\end{abstract}

\section*{Keywords}
Rayleigh wave; voids; nonlocal elasticity; TPL model; diffusion; orthotropic medium; normal mode analysis.

\section{Introduction}

\onehalfspacing
Generalized thermoelasticity provides a richer and more nuanced understanding of material behavior, allowing for more accurate predictions and designs in advanced engineering applications. Agarwal \cite{1} investigated the behavior of plane waves within the framework of generalized thermoelasticity, focusing on the unique wave propagation characteristics in this context. Sherief \cite{2} explored issues of uniqueness and stability within generalized thermoelasticity, that ensuring consistent and reliable solutions in advanced thermoelastic models. Hetnarski and Iganczak \cite{3} examined generalized thermoelasticity, investigating advanced theoretical frameworks and their implications for complex material behaviors. Youssef and Lehaibi \cite{4} explored a state-space methodology to address a one-dimensional problem in two-temperature generalized thermoelasticity. Othman and Kumar \cite{5} examined the reflection of magneto-thermoelastic waves, considering temperature-dependent properties within the framework of generalized thermoelasticity. Youssef \cite{6} investigated the theory of generalized thermoelasticity incorporating fractional order derivatives. Magdy et al. \cite{7} explored generalized thermoelasticity incorporating memory-dependent derivatives and the effects of two temperatures. Kartashov and Krylov \cite{8} investigated a generalized model of thermal shock within the context of dynamic thermoelasticity.\\

Nonlocal elasticity extends traditional elasticity theory by incorporating long-range interactions within a material, accounting for effects that span beyond local interactions. Eringen \cite{9} studied the theory of nonlocal thermoelasticity, focusing on its fundamental principles and implications. Polizzotto \cite{10} explored nonlocal elasticity and its associated variational principles, examining their theoretical foundations and applications. Pisano and Fuschi \cite{11} derived a closed-form solution for a nonlocal elastic bar subjected to tension. Pisano et al. \cite{12} developed 2D finite element solutions for nonlocal integral elasticity. Ruiz et al. \cite{13} applied Eringen's nonlocal elasticity theory to study the bending vibrations of rotating, nonuniform nanocantilevers. Salehipour et al. \cite{14} utilized a modified nonlocal elasticity theory to analyze functionally graded materials. Tuna and Kirca \cite{15} derived exact solutions for the bending of Euler–Bernoulli and Timoshenko beams using Eringen's nonlocal integral model. Eltaher et al. \cite{16} reviewed nonlocal elastic models for nanoscale beams, focusing on bending, buckling, vibrations, and wave propagation. Romano et al. \cite{17} reviewed iterative methods used to address problems in nonlocal elasticity. Patnaik et al. \cite{18} examined a displacement-based approach within the framework of nonlocal elasticity.\\

In various scientific and engineering contexts, a "void" refers to an empty space or absence of material within a solid structure. This concept is crucial in understanding and analyzing the behavior of materials and structures with internal cavities or porous regions. Voids can significantly affect the mechanical, thermal, and fluidic properties of materials, influencing their overall performance and stability. Iesan \cite{19} presented a theory addressing thermoelastic materials that include voids. Marin \cite{20} explored the impact of voids on the domain of influence within the context of thermoelastic bodies. Singh \cite{21} investigated wave propagation phenomena in generalized thermoelastic materials that contain voids. Quintanilla \cite{22} examined the impossibility of achieving localization in the context of linear thermoelasticity involving voids. Aouadi \cite{23} studied the uniqueness and existence theorems in thermoelasticity with voids, focusing on cases without energy dissipation. Abo-Dahab et al. \cite{24} analyzed the effects of voids and rotation on the behavior of plane waves in generalized thermoelasticity. Bucur et al. \cite{25} investigated Rayleigh surface waves within the framework of thermoelastic materials containing voids. Tomar et al. \cite{26} explored plane wave behavior in thermo-viscoelastic materials with voids, comparing various theories of thermoelasticity. Bachher and Sarkar \cite{27} investigated the nonlocal theory of thermoelastic materials with voids, incorporating fractional derivative heat transfer. Biswas \cite{28} examined the fundamental solution for steady oscillations in a thermoelastic medium containing voids.\\

Diffusion is a key factor in many scientific and engineering fields. It regulates the migration of atoms and molecules through the solid lattice, impacting processes like phase changes, alloy development, and the material's mechanical and thermal characteristics. Primarily, Nowacki \cite{29,30} contributed significantly to the advancement of diffusion theory within solid mechanics. Aouadi \cite{31} made significant contributions to the qualitative aspects of the coupled theory of thermoelastic diffusion, while Aoudi \cite{32} provided important qualitative results in the theory of thermoelastic diffusion mixtures. Kumar et al. \cite{33} explored several theorems related to generalized thermoelastic diffusion. Elhagary \cite{34} addressed the problem of generalized thermoelastic diffusion in an infinite medium containing a spherical cavity. Abbas \cite{35} utilized an eigenvalue approach to tackle the fractional order theory of thermoelastic diffusion in an infinite elastic medium with a spherical cavity. Zenkour and Kutbi \cite{36} applied multi-thermal relaxation techniques to the thermodiffusion problem within a thermoelastic half-space. Mallick and Biswas \cite{37} examined thermoelastic diffusion in a nonlocal orthotropic medium with porosity, exploring how these variables affect material behavior and diffusion processes. Debnath and  Singh \cite{38} examined the influence of thermal diffusivity on Rayleigh wave propagation in human tissue. Singh et al. \cite{39} investigated the effects of thermoelastic theories on refracted wave propagation in microstretch thermoelastic diffusion media.\\
	
 Rayleigh waves in layered media offer crucial insights into how surface waves interact with intricate material structures. These waves, which propagate along the surface of layered materials, are key to understanding the effects of varying material properties on wave behavior. Drake \cite{40} explored Love and Rayleigh waves in media with non-horizontal layering, investigating how these waves behave in materials where the layers are not aligned parallel to the surface. Singh and Tomar \cite{41} investigated Rayleigh–Lamb waves within a microstretch elastic plate that is coated with liquid layers, examining the interaction between the waves and the layered configuration. Gupta et al. \cite{42} examined the behavior of Rayleigh surface waves in anisotropic media situated above a prestressed orthotropic half-space, focusing on how the wave dynamics are influenced by the underlying stress and material properties. Li et al. \cite{43} investigated Rayleigh waves propagating on a half-space featuring a gradient piezoelectric layer and an imperfect interface, analyzing how these factors affect wave propagation. Biswas \cite{44} studied Rayleigh waves in a nonlocal thermoelastic layer positioned above a nonlocal thermoelastic half-space, examining the interaction between the waves and the layered thermoelastic medium. Chai et al. \cite{45} explored the scattering of Rayleigh waves caused by shallow cavities within layered half-spaces, focusing on how these cavities influence wave behavior in such configurations. Biswas \cite{46} examined Rayleigh waves in a porous, nonlocal orthotropic thermoelastic layer situated above a similar porous, nonlocal orthotropic thermoelastic half-space, investigating how these wave interactions are affected by the material's properties and structure. Mallick and Biswas \cite{47} presented thermoelastic diffusion within a nonlocal anisotropic medium with voids, analyzing how these factors influence material behavior and diffusion processes.
 Debnath et al. \cite{48} analyzed Rayleigh wave propagation in thermally affected skin tissues, considering variations due to age, gender, and morphology. Kumar et al. \cite{49} investigated Rayleigh-type wave propagation in thermo-poroelastic media incorporating dual-phase-lag heat conduction. Kumari et al. \cite{50} examined inhomogeneous wave propagation in a porothermoelastic medium considering dual-phase-lag heat conduction. Kumar et al. \cite{51} explored Rayleigh wave propagation in nonlocal generalized thermoelastic media.\\

 This article investigates Rayleigh waves in an orthotropic layer with voids, incorporating the effects of diffusion through nonlocal thermoelasticity. The primary aim is to develop an elasticity theory that accounts for void diffusion by introducing a novel nonlocal three-phase-lag (TPL) model. This model captures the nonlocal effects of mass transport and thermal conduction. To the best of the author's knowledge, this issue has not been addressed in the existing literature. The study derives its constitutive equations from Eringen's theory of nonlocal elasticity and employs normal mode analysis to solve the problem. It presents specific examples that align with previously reported findings in the literature. A key contribution of this paper is the introduction of graphical comparisons across different thermoelastic models, including the TPL, DPL, and LS systems. This comparative analysis, enhanced with visual illustrations, represents a novel aspect not explored before. The study observes distinct oscillatory behaviors in several physical parameters, including phase velocity, attenuation coefficient, specific loss, and penetration depth, as the wave number increases. The exploration of Rayleigh waves in a nonlocal orthotropic layer with porosity and diffusion has significant implications for applications in geothermal energy systems, implant design, and nanostructured materials. This research could lead to advancements in technology, improved efficiency, and solutions to complex challenges across various fields.\\
\textbf{Objective and novelty of the current study}: The main objective of this study is to analyze the behavior of Rayleigh waves in a nonlocal porous orthotropic thermoelastic layer with diffusion under the Three-Phase-Lag (TPL) model. Specifically, it aims to: Investigate the influence of nonlocality – examining how nonlocal effects impact wave propagation, attenuation, SL and penetration depth in a porous orthotropic medium. Understand the role of porosity – assessing how voids and porosity alter the mechanical and thermal responses of Rayleigh waves. Incorporate mass diffusion effects – studying how mass diffusion interacts with thermoelastic and poroelastic effects, influencing energy dissipation and wave characteristics. Compare with classical and other generalized models – evaluating the differences between the TPL model and other thermoelastic theories (e.g., LS, DPL) to determine its effectiveness in capturing wave behavior more accurately. Enhance predictive modeling – Providing insights for better material design and engineering applications where wave propagation in complex thermoelastic media is critical, such as geophysics, biomedical materials, and aerospace engineering. This study aims to offer a comprehensive understanding of how Rayleigh waves behave in advanced materials, leading to improved modeling accuracy and potential technological applications.

	\section{Fundamental Connections and Formulas}
\onehalfspacing
This subsection is dedicated to advancing the theoretical framework concerning nonlocal thermoelastic materials with voids and diffusion. The initial phase revolves around formulating material behavior laws and field formulations essential for comprehending the intricate behavior of such materials.\\

In the context of linear elasticity theory, the strain tensor $e_{ij}$ is precisely defined as the symmetric component derived from the displacement gradient.
\begin{equation}
 e_{ij}=\frac{1}{2}(u_{i,j}+u_{j,i}),
 \end{equation}
 where $u_{i}$ represent the components of the displacement vector.\\

 In the framework established by Iesan (reference \cite{19}), the solid material with voids demonstrating linear thermoelasticity is characterized by a key relationship wherein the mass density $\rho$ is given by the multiplication of the matrix density $\gamma$ and the void volume fraction $\nu$ ($0 < \nu \leq 1$). This fundamental expression underscores the importance of void volume fraction $\nu$ in capturing volumetric changes within the material due to void compaction or expansion.

In the context of studying thermoelastic phenomena inside porous orthotropic media exhibiting nonlocal behavior, $\nu$ assumes a crucial role, accounting for alterations in the volume of bulk substance due to either compression or expansion of voids. Introducing a novel kinematic variable $\phi$, dependent on both time $t$ and spatial coordinates $\mathbf{p} = (p_1, p_2, p_3)$ and its surrounding point denoted by $\mathbf{p}' = (p_1', p_2', p_3')$, facilitates the representation of volume changes relative to a reference configuration. If $\nu$ at the initial state is indicated by $\nu_R$, then $\phi = \nu(\mathbf{p}, t) - \nu_R$. This formulation provides a concise yet precise representation of the introduced kinematic variable in terms of void volume variations relative to the reference state.
 Consider a thermoelastic body with voids, occupying a region $B$ in $\mathbb{R}^3$ and bounded by the surface $S$. Let $V$ denote the volume of this body. The positions of individual points within $B$ are represented by $X_{i}$ in the undeformed state and $x_{i}$ in the deformed state. Thus, the displacement components of a particle are denoted by $u_{i}=x_{i}-X_{i}$. This succinctly expresses the relationship between undeformed and deformed states of points within the thermoelastic body, facilitating the characterization of displacement components.

Temperature variance compared to the baseline state, denoted as $T_0$, is defined as $T = \theta - T_0$, under the assumption that $|\frac{T}{T_0}| \ll 1$. Here, $\theta$ signifies temperature in absolute scale.

In starting state, the body is considered stress-free, implying the initial vanishing of the Cauchy stress tensor. Additionally, starting body is presumed to possess no inherent balanced force. This establishes foundational conditions of stress and force equilibrium for the system in its initial configuration.

At the designated point $\mathbf{p} = (p_1, p_2, p_3)$ within the context of baseline condition, the ensemble of essential parameter is denoted by:
    \begin{equation*}
	M=\textbf{\{}e_{ij}(\textbf{p}),\phi(\textbf{p}),\phi_{,i}(\textbf{p}),T(\textbf{p}),C(\textbf{p})\textbf{\}},
 \end{equation*}
 At the neighboring location $\mathbf{p'} = (p_{1}', p_{2}', p_{3}')$, the relevant collection of essential parameter is characterized by:
 \begin{equation}
	 \quad M^{'}=\textbf{\{}e_{ij}(\textbf{p}^{'}),\phi(\textbf{p}^{'}),\phi_{,i}(\textbf{p}^{'}),T(\textbf{p}^{'}),C(\textbf{p}^{'})\},
    \end{equation}
The formulation of the strain energy function $\Upsilon$, which characterizes the behavior of nonlocal thermoelastic materials incorporating diffusion effects, is elucidated by Eringen in the referenced publication \cite{9}.\\
\begin{equation}
  \begin{split}
2\Upsilon=c_{ijkl}e_{ij}(\textbf{p})e_{kl}(\textbf{p}^{'})+\xi \phi(\textbf{p})\phi(\textbf{p}^{'})+A_{ij} \phi_{,i}(\textbf{p})\phi_{,i}(\textbf{p}^{'})+B_{ij}[e_{ij}(\textbf{p})\phi(\textbf{p}^{'})+e_{ij}(\textbf{p}^{'})\phi(\textbf{p})]\\
+D_{ijk}[e_{ij}(\textbf{p})\phi_{,k}(\textbf{p}^{'})+e_{ij}(\textbf{p}^{'})\phi_{,k}(\textbf{p})]+d_{i}[\phi(\textbf{p})\phi_{,i}(\textbf{p}^{'})
+\phi(\textbf{p}^{'})\phi_{,i}(\textbf{p})]\\-\beta_{ij}[e_{ij}(\textbf{p})T(\textbf{p}^{'})+e_{ij}(\textbf{p}^{'})T(\textbf{p})]
-aT(\textbf{p})T(\textbf{p}^{'})-\nu_{1}[\phi(\textbf{p})T(\textbf{p}^{'})+\phi(\textbf{p}^{'})T(\textbf{p})]\\
-a_{i}[\phi_{,i}(\textbf{p})T(\textbf{p}^{'})+\phi_{,i}(\textbf{p}^{'})T(\textbf{p})]
-b_{ij}[e_{ij}(\textbf{p})C(\textbf{p}^{'})+e_{ij}(\textbf{p}^{'})C(\textbf{p})]+bC(\textbf{p})C(\textbf{p}^{'})\\
-\nu_{2}[\phi(\textbf{p})C(\textbf{p}^{'})+\phi(\textbf{p}^{'})C(\textbf{p})]-\mu_{i}[\phi_{,i}(\textbf{p})C(\textbf{p}^{'})+\phi_{,i}(\textbf{p}^{'})C(\textbf{p})]
-\nu_{3}[T(\textbf{p})C(\textbf{p}^{'})+T(\textbf{p}^{'})C(\textbf{p})],
\end{split}
\end{equation}

where $c_{ijkl},\xi,A_{ij},B_{ij},D_{ijk},d_{i},\beta_{ij},a,a_{i},b_{ij},b,\mu_{i}$,$\nu_{1}$,$\nu_{2}$ and $\nu_{3}$ are the material constants and predetermined function of the spatial coordinates $\textbf{p}$ and $\textbf{p}^{'}$.\\
Utilizing Eringen's theoretical structure \cite{9}, the core relations are deduced via the provided formulation:\\
\begin{equation}
\Gamma=\int_{V} \left[\left(\frac{\partial \Upsilon}{\partial{M}}\right)+\left(\frac{\partial \Upsilon}{\partial{M^{'}}}\right)^{S}\right]\,dV(\textbf{p}^{'}),
\end{equation}

Upper index $"S"$ signifies the uniformity of the quantity with respect to the replacement of $\textbf{p}$ with $\textbf{p}^{'}$. Structured set $\Gamma=\textbf{\{}\tau_{ij},-g,h_{i},-\rho S,P\textbf{\}}$ is correlated with $M$. As a result $\tau_{ij}$, $h_{i}$, $g$, $S$, and  $P$ are determined through the equations (3) and (4). These relations provide a systematic framework for deriving these quantities based on the prescribed conditions and symmetries.
\begin{equation}
\begin{split}
\tau_{ij}=\int_{V} [c_{ijkl}(\textbf{p},\textbf{p}^{'})e_{kl}(\textbf{p}^{'})+B_{ij}(\textbf{p},\textbf{p}^{'})\phi(\textbf{p}^{'})
+D_{ijk}(\textbf{p},\textbf{p}^{'})\phi_{,k}(\textbf{p}^{'})-\beta_{ij}(\textbf{p},\textbf{p}^{'})T(\textbf{p}^{'})-b_{ij}(\textbf{p},\textbf{p}^{'})C(\textbf{p}^{'})] \,dV(\textbf{p}^{'}),
\end{split}
\end{equation}

\begin{equation}
h_{i}=\int_{V} [D_{kli}(\textbf{p},\textbf{p}^{'})e_{kl}(\textbf{p}^{'})+d_{i}(\textbf{p},\textbf{p}^{'})\phi(\textbf{p}^{'})
+A_{ij}(\textbf{p},\textbf{p}^{'})\phi_{,i}(\textbf{p}^{'})-a_{i}(\textbf{p},\textbf{p}^{'})T(\textbf{p}^{'})
-\mu_{i}(\textbf{p},\textbf{p}^{'})C(\textbf{p}^{'})]\,dV(\textbf{p}^{'}),
\end{equation}

\begin{equation}
\begin{split}
g=-\int_{V} [\xi(\textbf{p},\textbf{p}^{'})\phi(\textbf{p}^{'})+B_{ij}(\textbf{p},\textbf{p}^{'})e_{ij}(\textbf{p}^{'})
+d_{i}(\textbf{p},\textbf{p}^{'})\phi_{,i}(\textbf{p}^{'})
-\nu_{1}(\textbf{p},\textbf{p}^{'})T(\textbf{p}^{'})-\nu_{2}(\textbf{p},\textbf{p}^{'})C(\textbf{p}^{'})]\,dV(\textbf{p}^{'}),
\end{split}
\end{equation}

\begin{equation}
\begin{split}
\rho S=\int_{V} [\beta_{ij}(\textbf{p},\textbf{p}^{'})e_{ij}(\textbf{p}^{'})+a(\textbf{p},\textbf{p}^{'})T(\textbf{p}^{'})
+\nu_{1}(\textbf{p},\textbf{p}^{'})\phi(\textbf{p}^{'})+a_{i}(\textbf{p},\textbf{p}^{'})\phi_{,i}(\textbf{p}^{'})
-\nu_{3}(\textbf{p},\textbf{p}^{'})C(\textbf{p}^{'})]\,dV(\textbf{p}^{'}),
\end{split}
\end{equation}
\begin{equation}
\begin{split}
P=\int_{V} [-b_{ij}(\textbf{p},\textbf{p}^{'})e_{ij}(\textbf{p}^{'})+b(\textbf{p},\textbf{p}^{'})T(\textbf{p}^{'})
-\nu_{2}(\textbf{p},\textbf{p}^{'})\phi(\textbf{p}^{'})-\mu_{i}(\textbf{p},\textbf{p}^{'}) \phi_{,i}(\textbf{p}^{'})-\nu_{3}(\textbf{p},\textbf{p}^{'})C(\textbf{p}^{'})]\,dV(\textbf{p}^{'}).
\end{split}
\end{equation}

For materials exhibiting centro-symmetry and orthotropy, the constitutive coefficients can be expressed as follows:\\
$D_{ijk}=d_{i}=a_{i}=\mu_{i}=0.$\\
The material coefficients $c_{ijkl},\beta_{ij},B_{ij}$ and $\xi$ are the function of $| \textbf{p}-\textbf{p}^{'}|$.\\
Therefore, the constitutive equations (5)-(9) are transformed accordingly
\begin{eqnarray}
\tau_{ij}=\int_{V} [c_{ijkl}(|\textbf{p}-\textbf{p}^{'}|)e_{kl}(\textbf{p}^{'})+B_{ij}(|\textbf{p}-\textbf{p}^{'}|)\phi(\textbf{p}^{'})
-\beta_{ij}(|\textbf{p}-\textbf{r}^{'}|)T(\textbf{r}^{'})-b_{ij}(|\textbf{p}-\textbf{p}^{'}|)C(\textbf{p}^{'})]\,dV(\textbf{p}^{'}),\\
h_{i}=\int_{V}
A_{ij}(|\textbf{p}-\textbf{p}^{'}|)\phi_{,i}(\textbf{p}^{'})\,dV(\textbf{p}^{'}),\\
g=-\int_{V} [\xi(|\textbf{p}-\textbf{p}^{'}|)\phi(\textbf{p}^{'})+B_{ij}(|\textbf{p}-\textbf{p}^{'}|)e_{ij}(\textbf{p}^{'})
-\nu_{1}(|\textbf{p}-\textbf{p}^{'}|)T(\textbf{p}^{'})-\nu_{2}(|\textbf{p}-\textbf{p}^{'}|)C(\textbf{p}^{'})]\,dV(\textbf{p}^{'}),\\
\rho S=\int_{V} [\beta_{ij}(|\textbf{p}-\textbf{p}^{'}|)e_{ij}(\textbf{p}^{'})+a(|\textbf{p}-\textbf{p}^{'}|)T(\textbf{p}^{'})+\nu_{1}(|\textbf{p}-\textbf{p}^{'}|)\phi(\textbf{p}^{'})
-\nu_{3}(|\textbf{p}-\textbf{p}^{'}|)C(\textbf{p}^{'})]\,dV(\textbf{p}^{'}),\\
P=\int_{V} [-b_{ij}(|\textbf{p}-\textbf{p}^{'}|)e_{ij}(\textbf{p}^{'})+b(|\textbf{p}-\textbf{p}^{'}|)C(\textbf{p}^{'})
-\nu_{2}(|\textbf{p}-\textbf{p}^{'}|)\phi(\textbf{p}^{'})-\nu_{3}(|\textbf{p}-\textbf{p}^{'}|)C(\textbf{p}^{'})]\,dV(\textbf{p}^{'}).
\end{eqnarray}

All constitutive coefficients depend on the term $(|\textbf{p}-\textbf{p}'|)$, which represents the distance between two points. These coefficients decrease with increasing distance, a behavior that can be referable to the small area of the cohesive patch in most materials. Within this zone, molecular attractions weaken quickly as the separation from the reference point grows. Consequently, it is hypothesized that all constitutive parameters weaken with increasing length\\
\begin{eqnarray*}
\lim_{(|\textbf{p}-\textbf{p}^{'}|)\rightarrow \infty} a(|\textbf{p}-\textbf{p}^{'}|)\rightarrow 0.
\end{eqnarray*}
 Additionally, we posit a constant attenuation level across all material parameters, in each case attaining its maximum value at a designated location where $\textbf{p}=\textbf{p}^{'}$.\\

 Accordingly, we can conclude the consequent bonds between nonlocal and local coefficient:

\begin{equation}
\begin{split}
 \frac{c_{ij}(|\textbf{p}-\textbf{p}^{'}|)}{c^{0}_{ij}}=\frac{A_{ij}(|\textbf{p}-\textbf{p}^{'}|)}{A^{0}_{ij}}=
 \frac{B_{ij}(|\textbf{p}-\textbf{p}^{'}|)}{B^{0}_{ij}}=\frac{\xi(|\textbf{p}-\textbf{p}^{'}|)}{\xi^{0}}=
 \frac{\beta_{ij}(|\textbf{p}-\textbf{p}^{'}|)}{\beta^{0}_{ij}}=\frac{b_{ij}(|\textbf{p}-\textbf{p}^{'}|)}{b^{0}_{ij}}\\
 =\frac{a(|\textbf{p}-\textbf{p}^{'}|)}{a^{0}}=\frac{b(|\textbf{p}-\textbf{r}^{'}|)}{b^{0}}
 =\frac{\nu_{1}(|\textbf{p}-\textbf{p}^{'}|)}{\nu_{1}^{0}}=\frac{\nu_{2}(|\textbf{p}-\textbf{p}^{'}|)}{\nu_{2}^{0}}
 =\frac{\nu_{3}(|\textbf{p}-\textbf{p}^{'}|)}{\nu_{3}^{0}}=G(|\textbf{p}-\textbf{p}^{'}|).
\end{split}
\end{equation}
Furthermore, the nonlocal kernel function $G(|\textbf{p}-\textbf{p}^{'}|)$ exhibits the following properties:

 (1) $\int_{V} G(|\textbf{p}-\textbf{p}^{'}|)\,dV(\textbf{p}^{'})=1$\\

 (2) The function $G$ reaches its peak when $(|\textbf{p}-\textbf{p}^{'}|)=0$ and typically decreases as the distance $(|\textbf{p}-\textbf{p}^{'}|)$ between points expands.\\

 (3) According to Eringen's work \cite{9}, we observe
  \begin{eqnarray}
    (1-\varepsilon^2\nabla^2) G(|\textbf{p}-\textbf{p}^{'}|)=\delta(|\textbf{p}-\textbf{p}^{'}|)
  \end{eqnarray}
  where $\nabla^{2}=\frac{\partial^{2}}{\partial x^{2}}+\frac{\partial^{2}}{\partial y^{2}}+\frac{\partial^{2}}{\partial z^{2}}$ and $\varepsilon=e_{0}a_{cl}$ represents the elastic nonlocal parameter,with,
  $a_{cl}$ being the internal characteristic length and $e_{0}$ is a material constant.

Implement $(1-\varepsilon^2\nabla^2)$ in the equations (11)-(15) and using equation (16), the material behavior equations is developed for a thermoelastic material that is orthotropic, nonlocal, and includes the effects of voids and diffusion. In this derivation, the subscript "0" is omitted from the constitutive coefficients for simplicity. This approach allows for the formulation of the material's behavior by incorporating nonlocal effects, leading to a comprehensive set of constitutive relations\\

\begin{eqnarray}
(1-\varepsilon^2\nabla^2)\tau_{ij}=\tau^{L}_{ij}= c_{ijkl}e_{kl}(\textbf{p})+B_{ij}\phi(\textbf{p})
-\beta_{ij}T(\textbf{p})-b_{ij}C(\textbf{p}),\\
(1-\varepsilon^2\nabla^2)h_{i}=h^{L}_{i}=A_{ij}\phi_{,i}(\textbf{p}),\\
(1-\varepsilon^2\nabla^2)g=g^{L}=- \xi\phi(\textbf{p})+B_{ij}e_{ij}(\textbf{p})
-\nu_{1}T(\textbf{p})-\nu_{2}C(\textbf{p}),\\
(1-\varepsilon^2\nabla^2)(\rho S)=(\rho S)^{L}=\beta_{ij}e_{ij}(\textbf{p})+aT(\textbf{p})+\nu_{1}\phi(\textbf{p})-\nu_{3}C(\textbf{p}),\\
(1-\varepsilon^2\nabla^2)P=P^{L}= -b_{ij}e_{ij}(\textbf{p})-\nu_{2}\phi(\textbf{p})
-\nu_{3}T(\textbf{p})+bC(\textbf{p}),
\end{eqnarray}

wherein the formula
\begin{equation}
\int f(x)\delta(x-a)\ dx=f(a)
\end{equation}
has been employed. Here the quantities $\tau^{L}_{ij},h^{L}_{i},g^{L},(\rho S)^{L}$ and $P^{L}$ locally formulated model is proposed for a solid exhibiting thermoelastic behavior, incorporating the inclusion of diffusion and voids effects.\\
We advance a variant of Eringen-type Fourier's law to delineate the nonlocally expansion under TPL model as delineated below:\\
\begin{equation}
(1-\varepsilon^2\nabla^2)(1+\tau_{q}\frac{\partial}{\partial{t}}+\frac{\tau^2_{q}}{2}\frac{\partial^2}{\partial{t^2}})q_{i}
=(1+\tau_{q}\frac{\partial}{\partial{t}}+\frac{\tau^2_{q}}{2}\frac{\partial^2}{\partial{t^2}})q^{L}_{i}
=-K_{ij}(1+\tau_{T}\frac{\partial}{\partial{t}})T_{,j}-K^{\ast}_{ij}(1+\tau_{\nu}\frac{\partial}{\partial{t}})\nu_{,j}
\end{equation}
where $\frac{\partial \nu_{,ij}}{\partial{t}}=T_{,ij}$\\

The energy equation articulated as:

\begin{equation}
-\rho T_{0}\dot{S}=q_{i,i},
\end{equation}
Now operating $(1-\varepsilon^2\nabla^2)$ on the equation (24) and using equations (20) and (23) we obtain \\

\begin{equation}
K_{ij}(1+\tau_{T}\frac{\partial}{\partial{t}})\dot{T_{,ij}}+K^{\ast}_{ij}(1+\tau_{v}\frac{\partial}{\partial{t}})T_{,ij}
=(1+\tau_{q}\frac{\partial}{\partial{t}}+\frac{\tau^2_{q}}{2}\frac{\partial^2}{\partial{t^2}})(\rho C_{v}\ddot{T}+T_{0}\beta_{ij}\ddot{e_{ij}}+aT_{0}\ddot{C}
+\nu_{1}T_{0}\ddot{\phi}),
\end{equation}

 where $K_{ij}$ are the thermal conductivity tensor, $K^{\ast}_{ij}$ are the material constant characteristic
of the theory, $\tau_{q}$ is the phase lag due to heat flux vector, $\tau_{T}$ is the phase-lag of the temperature gradient and $\tau_{v}$ is the thermal displacement gradient.\\

 Fick's principle of diffusion is

 \begin{equation}
 (1-\varepsilon^2\nabla^2)(1+\tau_{1}\frac{\partial}{\partial{t}}+\frac{\tau^2_{1}}{2}\frac{\partial^2}{\partial{t^2}})\eta_{i}
 =(1+\tau_{1}\frac{\partial}{\partial{t}}+\frac{\tau^2_{1}}{2}\frac{\partial^2}{\partial{t^2}})\eta^{L}_{i}
 =-(1+\tau_{2}\frac{\partial}{\partial{t}})d_{ij}P^{L}_{,j}-(1+\tau_{3}\frac{\partial}{\partial{t}})d^{\ast}_{ij}P^{{\ast}L}_{,j},
 \end{equation}
 where $\dot{P^{{\ast}}}=P$

 The equation of conservation of mass
 \begin{equation}
\eta_{i,i}=-\dot{C},
\end{equation}

As a consequence of equation (26), we derive
\begin{equation}
(1-\varepsilon^2\nabla^2)(1+\tau_{1}\frac{\partial}{\partial{t}}++\frac{\tau^2_{1}}{2}\frac{\partial^2}{\partial{t^2}})\eta_{i,i}
=(1+\tau_{1}\frac{\partial}{\partial{t}}+\frac{\tau^2_{1}}{2}\frac{\partial^2}{\partial{t^2}})\eta^{L}_{i,i}
 =-(1+\tau_{2}\frac{\partial}{\partial{t}})d_{ij}P^{L}_{,ij}-(1+\tau_{3}\frac{\partial}{\partial{t}})d^{\ast}_{ij}P^{{\ast}L}_{,ij},
\end{equation}

Using equation (27) and (28), we get

\begin{equation}
(1+\tau_{2}\frac{\partial}{\partial{t}})d_{ij}\dot{P^{L}_{,ij}}+(1+\tau_{3}\frac{\partial}{\partial{t}})d^{\ast}_{ij}P^{L}_{,ij}
=(1-\varepsilon^2\nabla^2)(1+\tau_{1}\frac{\partial}{\partial{t}}+\frac{\tau^2_{1}}{2}\frac{\partial^2}{\partial{t^2}})\ddot{C},
\end{equation}

Equation describing stress motion without the influence of body forces:

 \begin{equation}
\tau^{L}_{ij,j}=(1-\varepsilon^2\nabla^2)\rho \ddot{u_{i}},
\end{equation}

Formulations of equations illustrating equilibrated forces, without extrinsic equilibrated force, are presented as:
\begin{equation}
h^{L}_{i,i} + g^{L} =\rho \chi \ddot{\phi}.
\end{equation}

\begin{table}[htbp]
    \centering
    \caption{Description of Symbols}
    \begin{tabular}{|c|c|c|c|}
    \hline
    Symbol & Description & Symbol & Description \\
    \hline
    $q_{i}$ & Heat flux components & $u_{i}$ & Displacement components \\
    $c_{ijkl}$ & Elastic constants & $C_{v}$ & Specific heat at constant strain \\
    $\tau_{ij}$ & Stress tensor components & $\tau_{1}$, $\tau_{2}$, $\tau_{3}$ & Diffusion relaxation time \\
    $b_{ij}$ & Diffusion moduli tensor & $\eta$ & Diffusion mass flow \\
    $S$ & Entropy per unit mass & $b$ & Diffusive effect measure \\
    $\chi$ & Equilibrated inertia & $d^{\ast}_{ij}$ & Diffusion rate \\
    $A_{ij}$, & Voids parameters & $B_{ij}$ & Voids parameters\\
    \hline
    \end{tabular}
    \label{tab:symbols}
\end{table}

	\section{Establishing the problem framework}
 We consider a medium composed of two distinct regions: an orthotropic, homogeneous, nonlocal layer of constant thickness $H$, which lies atop an orthotropic, homogeneous, nonlocal half-space. Both the layer and the half-space exhibit orthotropic material properties, meaning their mechanical and thermal behaviors vary along different principal directions, and are characterized by nonlocal effects, where stress and thermal responses at a point depend on the surrounding material behavior. The layer has a finite thickness $H$, while the half-space extends infinitely beneath it.\\
 Assume $\vec{u_{1}}=(u_{1},0,w_{1})$ represents movement vector in the thermoelastic layer and $\vec{u_{2}}=(u_{2},0,w_{2})$ represents movement vector in the half-space..\\
 Considered variables $u_{1},v_{1},w_{1},\phi_{1},T_{1},$ ,$C_{1}$ and  $u_{2},v_{2},w_{2},\phi_{2},T_{2},$ and $C_{2}$ outlined in the following manner:
	\begin{equation}
\begin{split}
	(u_{1},u_{2})=(u_{1},u_{2})(x,z,t), (v_{1},v_{2})=(0,0), (w_{1},w_{2})=(w_{1},w_{2})(x,z,t), (\phi_{1},\phi_{2})=(\phi_{1},\phi_{2})(x,z,t), \\  (T_{1},T_{2})=(T_{1},T_{2})(x,z,t) , (C_{1},C_{2})=(C_{1},C_{2})(x,z,t).
\end{split}
	\end{equation}
	The interrelations between $(\tau_{ij})_{1},u_{1},w_{1},\phi_{1},T_{1},$ and $C_{1}$ are articulated in the following manner:
	\begin{equation}
		(1-\varepsilon^2\nabla^2)(\tau_{xx})_{1}=(\tau^L_{xx})_{1}=c_{11}u_{1,x}+c_{13}w_{1,z}+B_{1}\phi_{1}-\beta_{1}T_{1}-b_{1}C_{1},
	\end{equation}
	\begin{equation}
(1-\varepsilon^2\nabla^2)(\tau_{xz})_{1}=(\tau^L_{xz})_{1}=c_{55}(u_{1,z}+w_{1,x}),
\end {equation}
\begin{equation}
	(1-\varepsilon^2\nabla^2)(\tau_{zz})_{1}=(\tau^{L}_{zz})_{1}=c_{11}u_{1,x}+c_{33}w_{1,z}+B_{3}\phi_{1}-\beta_{3}T_{1}-b_{3}C_{1}.
 \end{equation}
	The dynamic equations dictating thermoelastic behavior within the realm of a nonlocal thermoelastic medium are presented as follows:
 \begin{equation}
 (1-\varepsilon^2\nabla^2)((\tau_{xx,x})_{1}+(\tau_{xz,z})_{1})=(1-\varepsilon^2\nabla^2)\rho_{1}\ddot{u_{1}},
 \end{equation}
 \begin{equation}
 (1-\varepsilon^2\nabla^2)((\tau_{xz,x})_{1}+(\tau_{zz,z})_{1})=(1-\varepsilon^2\nabla^2)\rho_{1}\ddot{w_{1}}.
 \end{equation}
 The fundamental relations under the TPL model can be concisely represented as:
 \begin{equation}
 c_{11}u_{1,xx}+c_{55}u_{1,zz}+(c_{13}+c_{55})w_{1,xz}+B_{1} \phi_{1,x}-\beta_{1}T_{1,x}-b_{1}C_{1,x}=(1-\varepsilon^2\nabla^2)\rho\ddot{u_{1}},
 \end{equation}
 \begin{equation}
 (c_{13}+c_{55})u_{1,xz}+c_{55}w_{1,xx}+c_{33}w_{1,zz}+B_{3} \phi_{1,z}-\beta_{3}T_{1,z}-b_{3}C_{1,z}=(1-\varepsilon^2\nabla^2)\rho\ddot{w_{1}},
 \end{equation}
 \begin{equation}
 (A_{1}\phi_{1,xx}+A_{3}\phi_{1,zz})-(B_{1}u_{1,x}+B_{3}w_{1,z})-\xi \phi_{1}+\nu_{1}T_{1}-\nu_{2} C_{1}=(1-\varepsilon^2\nabla^2)\rho \chi \ddot{\phi_{1}},
 \end{equation}
 \begin{equation}
 \begin{split}
 (1+\tau_{q}\frac{\partial}{\partial{t}}+\frac{\tau^2_{q}}{2}\frac{\partial^2}{\partial{t^2}})[\rho C_{v}\ddot{T}+(\beta_{1}\ddot u_{1,x}+\beta_{3}\ddot w_{1,z})T_{0}+\nu_{1}T_{0}\ddot{\phi_{1}}+aT_{0} \ddot{C_{1}}]=(1+\tau_{T} \frac{\partial}{\partial{t}})(K_{1}\dot{T}_{1,xx}+K_{3}\dot{T}_{1,zz})\\+(1+\tau_{T} \frac{\partial}{\partial{v}})(K^{\ast}_{1}T_{1,xx}+K^{\ast}_{3}T_{1,zz}),
 \end{split}
 \end{equation}
 \begin{equation}
 \begin{split}
 -d_{1}(1+\tau_{2}\frac{\partial}{\partial{t}})(b_{1}\dot{u}_{1,xxx}+b_{3}\dot{w}_{1,xxz}) -d_{3}(1+\tau_{2}\frac{\partial}{\partial{t}})(b_{1}\dot{u}_{1,xzz}+b_{3}\dot{w}_{1,zzz})
 -d^{\ast}_{1}(1+\tau_{3}\frac{\partial}{\partial{t}})(b_{1}u_{1,xxx}+b_{3}w_{1,xxz})\\ -d^{\ast}_{3}(1+\tau_{3}\frac{\partial}{\partial{t}})(b_{1}u_{1,xzz}+b_{3}w_{1,zzz})
 -\nu_{3}d_{1}(1+\tau_{2}\frac{\partial}{\partial{t}})\dot{\phi}_{1,xx}-\nu_{3}d_{3}(1+\tau_{2}\frac{\partial}{\partial{t}})\dot{\phi}_{1,zz}
 -\nu_{3}d^{\ast}_{1}(1+\tau_{3}\frac{\partial}{\partial{t}})\phi_{1,xx}\\-\nu_{3}d^{\ast}_{3}(1+\tau_{3}\frac{\partial}{\partial{t}})\phi_{1,zz}
 -\nu_{3}d_{1}(1+\tau_{2}\frac{\partial}{\partial{t}})\dot{T}_{1,xx}-\nu_{3}d_{3}(1+\tau_{2}\frac{\partial}{\partial{t}})\dot{T}_{1,zz}
 -\nu_{3}d^{\ast}_{1}(1+\tau_{3}\frac{\partial}{\partial{t}})T_{1,xx}-\nu_{3}d^{\ast}_{3}(1+\tau_{3}\frac{\partial}{\partial{t}})T_{1,zz}\\
 +bd_{1}(1+\tau_{2}\frac{\partial}{\partial{t}})\dot{C}_{1,xx}
 +bd_{3}(1+\tau_{2}\frac{\partial}{\partial{t}})\dot{C}_{1,zz}
 +bd^{\ast}_{1}(1+\tau_{3}\frac{\partial}{\partial{t}})C_{1,xx}
 +bd^{\ast}_{3}(1+\tau_{3}\frac{\partial}{\partial{t}})C_{1,zz}\\=(1-\varepsilon^2\nabla^2)(1+\tau_{1}\frac{\partial}{\partial{t}}+\frac{\tau^2_{1}}{2}\frac{\partial^2}{\partial{t^2}})\ddot{C_{1}},
 \end{split}
 \end{equation}
where $K_{1}$ and  $K_{3}$ are the thermal conductivity, $K^{\ast}_{1}$ and  $K^{\ast}_{3}$ are the material constant characteristic
of the theory, $\rho_{1}$ is the mass density and $(\tau_{xx})_{1},(\tau_{xz})_{1},(\tau_{zz})_{1}$ are the stress components in the case of layer and $\beta_{1},\beta_{3}$ are the thermal moduli.\\
 \\
The elastic constants in the half-space are denoted by $c'_{ij}$. Additionally, $\rho_{2}$ represents the mass density, and $T_{2}$  is the temperature in the half-space measured relative to a reference temperature.\\
	The interrelations between $(\tau_{ij})_{2},u_{2},w_{2},\phi_{2},T_{2},$ and $C_{2}$ are articulated in the following manner:
	\begin{equation}
		(1-\varepsilon^2\nabla^2)(\tau_{xx})_{2}=\tau^L_{xx})_{2}=c'_{11}u_{2,x}+c'_{13}w_{2,z}+B'_{1}\phi_{2}-\beta'_{1}T_{2}-b_{1}C_{2},
	\end{equation}
	\begin{equation}
(1-\varepsilon^2\nabla^2)(\tau_{xz})_{2}=(\tau^L_{xz})_{2}=c'_{55}(u_{2,z}+w_{2,x}),
\end {equation}
\begin{equation}
	(1-\varepsilon^2\nabla^2)(\tau_{zz})_{2}=(\tau^{L}_{zz})_{2}=c'_{13}u_{2,x}+c'_{33}w_{2,z}+B_{3}\phi_{2}-\beta'_{3}T_{2}-b'_{3}C_{2}.
 \end{equation}
	The dynamic equations dictating thermoelastic behavior within the realm of a nonlocal thermoelastic medium are presented as follows:
 \begin{equation}
 (1-\varepsilon^2\nabla^2)((\tau_{xx,x})_{2}+(\tau_{xz,z})_{2})=(1-\varepsilon^2\nabla^2)\rho_{2}\ddot{u_{2}},
 \end{equation}
 \begin{equation}
 (1-\varepsilon^2\nabla^2)((\tau_{xz,x})_{2}+(\tau_{zz,z})_{2})=(1-\varepsilon^2\nabla^2)\rho_{2}\ddot{w_{2}}.
 \end{equation}
 The fundamental relations under the TPL model can be concisely represented as:
 \begin{equation}
 c'_{11}u_{2,xx}+c'_{55}u_{2,zz}+(c'_{13}+c'_{55})w_{2,xz}+B'_{1} \phi_{2,x}-\beta'_{1}T_{2,x}-b'_{1}C_{2,x}=(1-\varepsilon^2\nabla^2)\rho_{2}\ddot{u_{2}},
 \end{equation}
 \begin{equation}
 (c'_{13}+c'_{55})u_{2,xz}+c'_{55}w_{2,xx}+c'_{33}w_{2,zz}+B'_{3} \phi_{2,z}-\beta'_{3}T_{1,z}-b'_{3}C_{1,z}=(1-\varepsilon^2\nabla^2)\rho_{2}\ddot{w_{1}},
 \end{equation}
 \begin{equation}
 (A'_{1}\phi_{2,xx}+A'_{3}\phi_{2,zz})-(B'_{1}u_{2,x}+B'_{3}w_{2,z})-\xi' \phi_{2}+\nu'_{1}T_{2}-\nu'{2} C_{2}=(1-\varepsilon^2\nabla^2)\rho_{2} \chi' \ddot{\phi_{2}},
 \end{equation}
 \begin{equation}
 \begin{split}
 (1+\tau_{q}\frac{\partial}{\partial{t}}+\frac{\tau^2_{q}}{2}\frac{\partial^2}{\partial{t^2}})[\rho_{2} C'_{v}\ddot{T}+(\beta_{1}\ddot u_{2,x}+\beta_{3}\ddot w_{2,z})T_{0}+\nu_{1}T_{0}\ddot{\phi_{2}}+a'T_{0} \ddot{C_{2}}]=(1+\tau_{T} \frac{\partial}{\partial{t}})(K'_{1}\dot{T}_{2,xx}+K'_{3}\dot{T}_{2,zz})\\+(1+\tau_{T} \frac{\partial}{\partial{v}})(K'^{\ast}_{1}T_{2,xx}+K'^{\ast}_{3}T_{2,zz}),
 \end{split}
 \end{equation}
 \begin{equation}
 \begin{split}
 -d'_{1}(1+\tau_{2}\frac{\partial}{\partial{t}})(b'_{1}\dot{u}_{2,xxx}+b'_{3}\dot{w}_{2,xxz}) -d'_{3}(1+\tau_{2}\frac{\partial}{\partial{t}})(b'_{1}\dot{u}_{2,xzz}+b'_{3}\dot{w}_{2,zzz})
 -d'^{\ast}_{1}(1+\tau_{3}\frac{\partial}{\partial{t}})(b'_{1}u_{2,xxx}+b_{3}w_{2,xxz})\\ -d'^{\ast}_{3}(1+\tau_{3}\frac{\partial}{\partial{t}})(b'_{1}u_{2,xzz}+b_{3}w_{2,zzz})
 -\nu'_{3}d'_{1}(1+\tau_{2}\frac{\partial}{\partial{t}})\dot{\phi}_{2,xx}-\nu'_{3}d'_{3}(1+\tau_{2}\frac{\partial}{\partial{t}})\dot{\phi}_{2,zz}
 -\nu'_{3}d'^{\ast}_{1}(1+\tau_{3}\frac{\partial}{\partial{t}})\phi_{2,xx}\\-\nu'_{3}d'^{\ast}_{3}(1+\tau_{3}\frac{\partial}{\partial{t}})\phi_{2,zz}
 -\nu'_{3}d'_{1}(1+\tau_{2}\frac{\partial}{\partial{t}})\dot{T}_{2,xx}-\nu'_{3}d'_{3}(1+\tau_{2}\frac{\partial}{\partial{t}})\dot{T}_{2,zz}
 -\nu'_{3}d'^{\ast}_{1}(1+\tau_{3}\frac{\partial}{\partial{t}})T_{2,xx}-\nu'_{3}d'^{\ast}_{3}(1+\tau_{3}\frac{\partial}{\partial{t}})T_{2,zz}\\
 +b'd'_{1}(1+\tau_{2}\frac{\partial}{\partial{t}})\dot{C}_{2,xx}
 +b'd'_{3}(1+\tau_{2}\frac{\partial}{\partial{t}})\dot{C}_{2,zz}
 +b'd'^{\ast}_{1}(1+\tau_{3}\frac{\partial}{\partial{t}})C_{2,xx}
 +b'd'^{\ast}_{3}(1+\tau_{3}\frac{\partial}{\partial{t}})C_{2,zz}\\=(1-\varepsilon^2\nabla^2)(1+\tau_{1}\frac{\partial}{\partial{t}}+\frac{\tau^2_{1}}{2}\frac{\partial^2}{\partial{t^2}})\ddot{C_{2}}.
 \end{split}
 \end{equation}
where $K'_{1}$ and  $K'_{3}$ are the thermal conductivity, $K'^{\ast}_{1}$ and  $K'^{\ast}_{3}$ are the material constant characteristic of the theory, $\rho_{2}$ is the mass density and $(\tau_{xx})_{2},(\tau_{xz})_{2},(\tau_{zz})_{2}$ are the stress components in the case of half-space and $\beta'_{1},\beta'_{3}$ are the thermal moduli.
 \section{Boundary conditions}
On the surface where thermal stress is absent, indicated by $z=0$, the defined mechanical and thermal parameters entail a set of specific requirements. These conditions dictate the behavior of the system at this boundary.
	\begin{description}
\item[(a)] The layer surface at $z=0$,
\end{description}
\begin{eqnarray}
  (\tau^L_{zz})_{1}=0.\\
  (\tau^L_{xz})_{1}=0.\\
  (\phi^L_{z})_{1}=0.\\
T_{1}=0\\
P^{L}_{,z}=0.
\end{eqnarray}
\begin{description}
\item[(b)] Interface between the layer and half-space $z=H$:
\end{description}
  \begin{eqnarray}
 u_{1}=u_{2}.\\
 w_{1}=w_{2}.\\
 (\tau_{xz})^L_{1}=(\tau_{xz})^L_{2}.\\
 (\tau_{zz})^L_{1}=(\tau_{zz})^L_{2}.\\
 A_{3}\frac{\partial \phi_{1}}{\partial{z}}=A'_{3}\frac{\partial \phi_{2}}{\partial{z}}.\\
 \phi_{1}=\phi_{2}.\\
 (\eta^L_{z})_{1}=(\eta^L_{z})_{2}.\\
 C_{1}=C_{2}.
 \end{eqnarray}
\begin{description}
\item[(c)] Infinite depth $z\rightarrow\infty$:\\
We require the displacements and temperature to diminish to zero at large depths,\\
 $u_{2}\rightarrow0$,$w_{2}\rightarrow0$,$\phi_{2}\rightarrow0$,$\theta_{2}\rightarrow0$,$C_{2}\rightarrow0$.
\end{description}
From equation (27),we get
\begin{eqnarray*}
&&(1+\tau_{1}\frac{\partial}{\partial{t}}+\frac{\tau^2_{1}}{2}\frac{\partial^2}{\partial{t^2}})\eta^{L}_{z}
 =-(1+\tau_{2}\frac{\partial}{\partial{t}})d_{3}P^{L}_{,z}-(1+\tau_{3}\frac{\partial}{\partial{t}})d^{\ast}_{3}P^{{\ast}L}_{,z},\\
&\Rightarrow& \eta^{L}_{z}=-[(1+\tau_{2}\frac{\partial}{\partial{t}})d_{3}+\frac{1}{D'}(1+\tau_{3}\frac{\partial}{\partial{t}})d^{\ast}_{3}]
(1-\tau_{1}\frac{\partial}{\partial{t}}+\frac{\tau^2_{1}}{2}\frac{\partial^2}{\partial{t^2}})P^{L}_{,z},(\text{neglecting higher order term of}\quad \tau_{1})\\
&\Rightarrow& \eta^{L}_{z}=\overline{\delta} P^{L}_{,z}
\end{eqnarray*}
where $\overline{\delta}=-[(1-ikc\tau_{2})d_{3}-\frac{1}{ikc}(1-ikc\tau_{3})d^{\ast}_{3}]
(1+ikc\tau_{1}-\frac{k^2c^2\tau^2_{1}}{2}).$\\
\section{Solution methodology for a nonlocal thermoelastic layer}
We choose
  \begin{equation}
  (u_{1},w_{1},\phi_{1},T_{1},C_{1})(x_{1},z_{1},t_{1})=(\overline{u_{1}},\overline{w_{1}},\overline{\phi_{1}},\overline{T_{1}},\overline{C_{1}})(z)\exp[ik(x-ct)].
  \end{equation}
   In this context, $\overline{u_{1}}$, $\overline{w_{1}}$, $\overline{\phi_{1}}$, $\overline{T_{1}}$, and $\overline{C_{1}}$ symbolize the magnitudes of the physical parameters, with $c$ denoting the phase velocity and $k$ representing the wave number along the $x$-axis.
\\Applying equation (66) to equations (38)-(42), we obtain:
  \begin{eqnarray}
 a_{11}\overline{u_{1}}+a_{12}D^{2}\overline{u_{1}}+a_{13}D\overline{w_{1}}+a_{14}\overline{\phi_{1}}+a_{15}\overline{T_{1}}+a_{16}\overline{C_{1}}=0,\\
 a_{21}D\overline{u_{1}}+a_{22}\overline{w_{1}}+a_{23}D^2\overline{w_{1}}+a_{24}D\overline{\phi_{1}}+a_{25}D\overline{T_{1}}+a_{26}D\overline{C_{1}}=0,\\
 a_{31}\overline{u_{1}}+a_{32}D\overline{w_{1}}+a_{33}\overline{\phi_{1}}+a_{34}D^2\overline{\phi_{1}}+a_{35}\overline{T_{1}}+a_{36}\overline{C_{1}}=0,\\
 a_{41}\overline{u_{1}}+a_{42}D\overline{w_{1}}+a_{43}\overline{\phi_{1}}+a_{44}\overline{T_{1}}+a_{45}D^2\overline{T_{1}}+a_{46}\overline{C_{1}}=0,\\
 a_{51}\overline{u_{1}}+a_{52}D^2\overline{u_{1}}+a_{53}D\overline{w_{1}}+a_{54}D^3\overline{w_{1}}+a_{55}\overline{\phi_{1}}+a_{56}D^2\overline{\phi_{1}}
+a_{57}\overline{T_{1}}+a_{58}D^2\overline{T_{1}}
+a_{59}\overline{C_{1}}+a_{5(10)}D^2\overline{C_{1}}=0.
 \end{eqnarray}

where $a_{ij}$ are given in the Appendix A.\\
Removing $\overline{u},\overline{w},\overline{\phi},\overline{T},\overline{C}$ from equations (67)-(71), we acquire
\begin{eqnarray*}
   (D^{10}+R_{1}D^{8}+R_{2}D^{6}+R_{3}D^{4}+R_{4}D^{2}+R_{5})(\overline{u},\overline{w},\overline{\phi},\overline{T},\overline{C})(z)=0.
  \end{eqnarray*}
   This equation further expressed as:
  \begin{eqnarray*}
  ( D^{2}+\lambda^2_{1})( D^{2}+\lambda^2_{2})( D^{3}+
  \lambda^2_{3})( D^{2}+\lambda^2_{4})( D^{2}+\lambda^2_{5})
  (\overline{u},\overline{w},\overline{\phi},\overline{T},\overline{C})(z)=0.
\end{eqnarray*}
The defining formula is expressed as follows
  \begin{eqnarray*}
  \Psi^{10}+R_{1}\Psi^{8}+R_{2}\Psi^{6}+R_{3}\Psi^{4}+R_{4}\Psi^{2}+R_{5}=0.
  \end{eqnarray*}
  where $R_{1},R_{2},R_{3},R_{4},R_{5}$ are given in the Appendix A.\\
  Consequently, we derive the following expressions:
\begin{eqnarray*}
\begin{split}
\overline{u_{1}}(z)=\sum_{n=1}^{5}A^{(1)}_{n}cos\lambda_{n}z+\sum_{n=1}^{5}B^{(1)}_{n}sin\lambda_{n}z,\\
\overline{w_{1}}(z)=\sum_{n=1}^{5}C^{(1)}_{n}cos\lambda_{n}z+\sum_{n=1}^{5}D^{(1)}_{n}sin\lambda_{n}z,\\
\overline{\phi_{1}}(z)=\sum_{n=1}^{5}E^{(1)}_{n}cos\lambda_{n}z+\sum_{n=1}^{5}F^{(1)}_{n}sin\lambda_{n}z,\\
\overline{T_{1}}(z)=\sum_{n=1}^{5}G^{(1)}_{n}cos\lambda_{n}z+\sum_{n=1}^{5}H^{(1)}_{n}sin\lambda_{n}z,\\
\overline{C_{1}}(z)=\sum_{n=1}^{5}I^{(1)}_{n}cos\lambda_{n}z+\sum_{n=1}^{5}J^{(1)}_{n}sin\lambda_{n}z,
\end{split}
\end{eqnarray*}
As a result, we obtain:
\begin{equation}
\left.
\begin{split}
 u_{1}(x,z,t)=\left[\sum_{n=1}^{5}A^{(1)}_{n}cos\lambda_{n}z+\sum_{n=1}^{5}B^{(1)}_{n}sin\lambda_{n}z]\right]\ exp[ik(x-ct)],\\
w_{1}(x,z,t)=\left[\sum_{n=1}^{5}C^{(1)}_{n}cos\lambda_{n}z+\sum_{n=1}^{5}D^{(1)}_{n}sin\lambda_{n}z \right] \exp[ik(x-ct)],\\
\phi_{1}(x,z,t)=\left[\sum_{n=1}^{5}E^{(1)}_{n}cos\lambda_{n}z+\sum_{n=1}^{5}F^{(1)}_{n}sin\lambda_{n}z\right] \exp[ik(x-ct)],\\
T_{1}(x,z,t)=\left[\sum_{n=1}^{5}G^{(1)}_{n}cos\lambda_{n}z+\sum_{n=1}^{5}H^{(1)}_{n}sin\lambda_{n}z \right]\ exp[ik(x-ct)],\\
C_{1}(x,z,t)=  \left[\sum_{n=1}^{5}I^{(1)}_{n}cos\lambda_{n}z+\sum_{n=1}^{5}J^{(1)}_{n}sin\lambda_{n}z \right]\ exp[ik(x-ct),\\
\end{split}
\right\}
\end{equation}
where we take \\$C^{(1)}_{n}=c^{(1)}_{n}A^{(1)}_{n}$,$D^{(1)}_{n}=d^{(1)}_{n}B^{(1)}_{n}$,$E^{(1)}_{n}=e^{(1)}_{n}A^{(1)}_{n}$,$F^{(1)}_{n}=j^{(1)}_{n}B^{(1)}_{n}$,
$G^{(1)}_{n}=g^{(1)}_{n}A^{(1)}_{n}$,
$H^{(1)}_{n}=h^{(1)}_{n}B^{(1)}_{n}$,$I^{(1)}_{n}=i^{(1)}_{n}A^{(1)}_{n}$,$J^{(1)}_{n}=j^{(1)}_{n}B^{(1)}_{n}$.\\
Since the stress are obtained as follows:\\

\begin{equation*}
\begin{split}
(\tau_{xx})^{L}_{1} &= c_{11}u_{1,x} + c_{13}w_{1,z} +B_{1}\phi_{1}- \beta_{1}T_{1} - b_{1}C_{1}, \\
&= \left[\sum_{n=1}^{4} \left( ikc_{11}A^{(1)}_{n} + c_{13} \lambda_{n}d^{(1)}_{n}B^{(1)}_{n} +B_{1}e^{(1)}_{n}A^{(1)}_{n}- \beta_{1}g^{(1)}_{n}A^{(1)}_{n} - b_{1}i^{(1)}_{n}A^{(1)}_{n} \right) \cos(\lambda_{n}z) \right. \\
&\quad + \left. \sum_{n=1}^{4} \left( ikc_{11}B^{(1)}_{n} -\lambda_{n}c_{13}c^{(1)}_{n}A^{(1)}_{n}+B_{1}f^{(1)}_{n}B^{(1)}_{n} - \beta_{1}h^{(1)}_{n}B^{(1)}_{n} - b_{1}j^{(1)}_{n}B^{(1)}_{n} \right) \sin(\lambda_{n}z) \right] \exp[ik(x - ct)],
\end{split}
\end{equation*}
\begin{equation*}
\begin{split}
(\tau_{xz})^{L}_{1} &= c_{55}(u_{1,z} + w_{1,x}),\\
&= \left[\sum_{n=1}^{4} c_{55}\left( \lambda_{n}B^{(1)}_{n} + ikc^{(1)}_{n}A^{(1)}_{n} \right) \cos(\lambda_{n}z) \right.+
\left. \sum_{n=1}^{4} c_{55}\left( -\lambda_{n}A^{(1)}_{n} + ikb^{(1)}_{n}B^{(1)}_{n} \right) \sin(\lambda_{n}z) \right] \exp[ik(x - ct)],
\end{split}
\end{equation*}

\begin{equation*}
\begin{split}
(\tau_{zz})^{L}_{1} &= c_{11}u_{1,x} + c_{33}w_{1,z} +B_{3}\phi_{1}- \beta_{3}T_{1} - b_{3}C_{1}, \\
&= \left[\sum_{n=1}^{4} \left( ikc_{11}A^{(1)}_{n} + c_{33} \lambda_{n}d^{(1)}_{n}B^{(1)}_{n} +B_{3}e^{(1)}_{n}A^{(1)}_{n}- \beta_{3}g^{(1)}_{n}A^{(1)}_{n} - b_{3}i^{(1)}_{n}A^{(1)}_{n} \right) \cos(\lambda_{n}z) \right. \\
&\quad + \left. \sum_{n=1}^{4} \left( ikc_{11}B^{(1)}_{n} -\lambda_{n}c_{33}c^{(1)}_{n}A^{(1)}_{n}+B_{3}f^{(1)}_{n}B^{(1)}_{n} - \beta_{3}h^{(1)}_{n}B^{(1)}_{n} - b_{3}j^{(1)}_{n}B^{(1)}_{n} \right) \sin(\lambda_{n}z) \right] \exp[ik(x - ct)].
\end{split}
\end{equation*}

\section{Solution methodology for a nonlocal thermoelastic half space}
We utilize the normal mode approach to obtain solutions for the problem. \\
We choose
  \begin{equation}
  (u_{2},w_{2},\phi_{2},T_{2},C_{2})(x,z,t)=(\overline{u_{2}},\overline{w_{2}},\overline{\phi_{2}},\overline{T_{2}},\overline{C_{2}})(z)\exp[ik(x-ct)].
  \end{equation}
        In this context, $\overline{u_{2}}$, $\overline{w_{2}}$, $\overline{\phi_{2}}$, $\overline{T_{2}}$, and $\overline{C_{2}}$ symbolize the magnitudes of the physical parameters, with $c$ denoting the phase velocity and $k$ representing the wave number along the $x$-axis.
\\Applying equation (73) to equations (48)-(52), we obtain:
\begin{eqnarray}
 a'_{11}\overline{u_{2}}+a'_{12}D^{2}\overline{u_{2}}+a'_{13}D\overline{w_{2}}+a'_{14}\overline{\phi_{2}}+a'_{15}\overline{T_{2}}+a'_{16}\overline{C_{2}}=0,\\
a'_{21}D\overline{u_{2}}+a'_{22}\overline{w_{2}}+a'_{23}D^2\overline{w_{2}}+a'_{24}D\overline{\phi_{2}}+a'_{25}D\overline{T_{2}}+a'_{26}D\overline{C_{2}}=0,\\
a'_{31}\overline{u_{2}}+a'_{32}D\overline{w_{2}}+a'_{33}\overline{\phi_{2}}+a'_{34}D^2\overline{\phi_{2}}+a'_{35}\overline{T_{2}}+a'_{36}\overline{C_{2}}=0,\\
a'_{41}\overline{u_{2}}+a'_{42}D\overline{w_{2}}+a'_{43}\overline{\phi_{2}}+a'_{44}\overline{T_{2}}+a'_{45}D^2\overline{T_{2}}+a'_{46}\overline{C_{2}}=0,\\
a'_{51}\overline{u_{2}}+a'_{52}D^2\overline{u_{2}}+a'_{53}D\overline{w_{2}}+a'_{54}D^3\overline{w_{2}}+a'_{55}\overline{\phi_{2}}+a'_{56}D^2\overline{\phi_{2}}
+a'_{57}\overline{T_{2}}+a'_{58}D^2\overline{T_{2}}
+a'_{59}\overline{C_{2}}+a'_{5(10)}D^2\overline{C_{2}}=0.
\end{eqnarray}
where $a'_{ij}$ are given in the Appendix B.\\
Removing $\overline{u_{2}},\overline{w_{2}},\overline{\phi_{2}},\overline{T_{2}},\overline{C_{2}}$ from equations (74)-(78), we acquire
\begin{eqnarray*}
   (D^{10}+R'_{1}D^{8}+R'_{2}D^{6}+R'_{3}D^{4}+R'_{4}D^{2}+R'_{5})(\overline{u_{2}},\overline{w_{2}},\overline{\phi_{2}},\overline{T_{2}},\overline{C_{2}})(z)=0.
  \end{eqnarray*}
    This equation further expressed as:
  \begin{eqnarray*}
  ( D^{2}-\varrho^2_{1})( D^{2}-\varrho^2_{2})( D^{3}-
  \varrho^2_{3})( D^{2}-\varrho^2_{4})( D^{2}-\varrho^2_{5})
  (\overline{u_{2}},\overline{w_{2}},\overline{\phi_{2}},\overline{T_{2}},\overline{C_{2}})(z)=0.
\end{eqnarray*}
  The defining formula is expressed as follows
  \begin{eqnarray*}
  \Psi^{10}+R'_{1}\Psi^{8}+R'_{2}\Psi^{6}+R'_{3}\Psi^{4}+R'_{4}\Psi^{2}+R'_{5}=0,
  \end{eqnarray*}
  where $R'_{1},R'_{2},R'_{3},R'_{4},R'_{5}$ are given in the Appendix B.

We will strictly utilize the form of $\varrho_{n}$ that conforms to the prescribed conventional criterion:\\
As $z$ approaches $\infty$, $\overline{u_{2}},\overline{w_{2}},\overline{\phi_{2}},\overline{T_{2}},\overline{C_{2}}$ tend towards zero.\\
Consequently, we derive the following expressions:
\begin{eqnarray}
\left.
\begin{split}
\overline{u_{2}}(z)=\sum_{n=1}^{5}S^{(1)}_{n}\exp{(-\varrho_{n}z)},\\
\overline{w_{2}}(z)=\sum_{n=1}^{5}S^{(2)}_{n}\exp{(-\varrho_{n}z)},\\
\overline{\phi_{2}}(z)=\sum_{n=1}^{5}S^{(3)}_{n}\exp{(-\varrho_{n}z)},\\
\overline{T_{2}}(z)=\sum_{n=1}^{5}S^{(4)}_{n}\exp{(-\varrho_{n}z)},\\
\overline{C_{2}}(z)=\sum_{n=1}^{5}S^{(5)}_{n}\exp{(-\varrho_{n}z)}.
\end{split}
\right\}
\end{eqnarray}

where $S^{(1)}_{n}, S^{(2)}_{n}, S^{(3)}_{n}, S^{(4)}_{n}, S^{(5)}_{n}$  (where $n$ varies from 1 to 5) are some constants and radicals of the optimistic real components of $\varrho^2_{n}$ are denoted as $\varrho_{n}$, where $n$ ranges from 1 to 5.\\
By applying equations (79) to equations (74)-(78), we obtain:\\
$S^{(2)}_{n}=f_{n}S^{(1)}_{n}$,$S^{(3)}_{n}=j_{n}S^{(1)}_{n}$,$S^{(4)}_{n}=k_{n}S^{(1)}_{n}$, $S^{(5)}_{n}=l_{n}S^{(1)}_{n}$.\\
Now replacing $z$ by $(z-H)$, we get
\begin{equation}
\left.
\begin{split}
u_{2}(x,z,t)=\sum_{n=1}^{5}S^{(1)}_{n}\exp{(-\varrho_{n}(z-H))}\exp[ik(x-ct)],\\
w_{2}(x,z,t)=\sum_{n=1}^{5}f_{n}S^{(1)}_{n}\exp{(-\varrho_{n}(z-H))}\exp[ik(x-ct)],\\
\phi_{2}(x,z,t)=\sum_{n=1}^{5}j_{n}S^{(1)}_{n}\exp{(-\varrho_{n}(z-H))}\exp[ik(x-ct)],\\
T_{2}(x,z,t)=\sum_{n=1}^{5}k_{n}S^{(1)}_{n}\exp{(-\varrho_{n}(z-H))}\exp[ik(x-ct)],\\
C_{2}(x,z,t)=\sum_{n=1}^{5}l_{n}S^{(1)}_{n}\exp{(-\varrho_{n}(z-H))}\exp[ik(x-ct)],
\end{split}
\right\}
\end{equation}

\begin{eqnarray*}
\begin{split}
        \tau^{L}_{xx}=\sum_{n=1}^{5}(ikc_{11}-c'_{13}\varrho_{n}f_{n}+B'_{1}j_{n}-\beta'_{1}k_{n}-b'_{1}l_{n})S^{(1)}_{n}\exp[-\varrho_{n}z+ik(x-ct)],\\
        \tau^{L}_{xz}=\sum_{n=1}^{5}c'_{55}(ikf_{n}-\varrho_{n})S^{(1)}_{n}\exp[-\varrho_{n}z+ik(x-ct)],\\
        \tau^{L}_{zz}=\sum_{n=1}^{5}(ikc'_{13}-c'_{33}\varrho_{n}f_{n}+B'_{3}j_{n}-\beta'_{3}k_{n}-b'_{3}l_{n})S^{(1)}_{n}\exp[-\varrho_{n}z+ik(x-ct)].
        \end{split}
        \end{eqnarray*}

where $f_{n},j_{n},k_{n},l_{n}$ are given in the Appendix B.

  \section{ Evaluation of the frequency equation}
   Using boundary conditions (53)-(57),we get\\
  \begin{equation}
  \sum_{n=1}^{5}(\lambda_{n}B^{(1)}_{n}+ikc^{(1)}_{n}A^{(1)}_{n})=0,
  \end{equation}
  \begin{equation}
  \sum_{n=1}^{5}[(ikc_{11}+B_{3}e^{(1)}_{n}-\beta_{3}g^{(1)}_{n}-b_{3}i^{(1)}_{n})A^{(1)}_{n}+c_{33}\lambda_{n}d^{(1)}_{n}B^{(1)}_{n}]=0,
  \end{equation}
  \begin{equation}
  \sum_{n=1}^{5}\lambda_{n}e^{(1)}_{n}B^{(1)}_{n}=0,
  \end{equation}
  \begin{equation}
  \sum_{n=1}^{5}g^{(1)}_{n}A^{(1)}_{n}=0,
  \end{equation}
  \begin{equation}
  \sum_{n=1}^{5}[(ikb_{1}-b_{3}\lambda_{n}e^{(1)}_{n}-\nu_{2}\lambda_{n}g^{(1)}_{n}-\nu_{3}\lambda_{n}i^{(1)}_{n})A^{(1)}_{n}-b_{3}\lambda^2_{n}d^{(1)}_{n}B^{(1)}_{n}]=0,
  \end{equation}
  From equation (81) we get
   \begin{equation}
  B^{(1)}_{n}=\daleth_{n}A^{(1)}_{n},
  \end{equation}
  Where $\daleth_{n}=-\frac{ikc^{(1)}_{n}}{\lambda_{n}}$\\
 Using eqn (84) we get\\
  \begin{equation}
  A^{(1)}_{5}=-\frac{g^{(1)}_{1}}{g^{(1)}_{5}}A^{(1)}_{1}-\frac{g^{(1)}_{2}}{g^{(1)}_{5}}A^{(1)}_{2}-\frac{g^{(1)}_{3}}{g^{(1)}_{5}}A^{(1)}_{3}-\frac{g^{(1)}_{4}}{g^{(1)}_{5}}A^{(1)}_{4},
  \end{equation}
  By incorporating equation (86) in equation (72) and (67) through (71), we derive the following set of outcomes:\\
  \begin{equation*}
  e^{(1)}_{n}=\frac{(X_{4}X_{18}-X_{2}X_{20})(X_{4}X_{9}-X_{1}X_{12})-(X_{4}X_{10}-X_{2}X_{12})(X_{4}X_{17}-X_{1}X_{20})}
  {(X_{4}X_{10}-X_{2}X_{12})(X_{4}X_{17}-X_{1}X_{20})-(X_{4}X_{18}-X_{2}X_{20})(X_{2}X_{20}-X_{3}X_{12})},
  \end{equation*}
  \begin{equation*}
   f^{(1)}_{n}=\frac{(X_{8}X_{22}-X_{6}X_{24})(X_{8}X_{13}-X_{5}X_{16})-(X_{8}X_{14}-X_{6}X_{16})(X_{8}X_{21}-X_{5}X_{24})}
  {(X_{8}X_{14}-X_{6}X_{16})(X_{8}X_{23}-X_{7}X_{24})-(X_{8}X_{22}-X_{6}X_{24})(X_{8}X_{15}-X_{7}X_{16})},
  \end{equation*}
  \begin{equation*}
  c^{(1)}_{n}=r^1_{n}\delta_{n},
  d^{(1)}_{n}=\frac{r^2_{n}}{\delta_{n}},
  \end{equation*}
  \begin{equation*}
  r^1_{n}=-\frac{(X_{8}X_{13}-X_{5}X_{16})+(X_{8}X_{15}-X_{7}X_{16})f^{(1)}_{n}}{(X_{8}X_{14}-X_{6}X_{16})},
   r^2_{n}=-\frac{(X_{4}X_{9}-X_{1}X_{12})+(X_{4}X_{11}-X_{3}X_{12})e^{(1)}_{n}}{(X_{8}X_{14}-X_{6}X_{16})},
   \end{equation*}
   \begin{equation*}
  g^{(1)}_{n}=-\frac{X_{1}+X_{2}r^2_{n}+X_{3}e^{(1)}_{n}}{X_{4}},
  h^{(1)}_{n}=-\frac{X_{5}+X_{6}r^2_{n}+X_{7}f^{(1)}_{n}}{X_{8}},
  \end{equation*}
  \begin{equation*}
  i^{(1)}_{n}=-\frac{(n_{1}-n_{2}\lambda^2_{n})+n_{3}\lambda_{n}r^1_{n}+n_{4}e^{(1)}_{n}+n_{5}g^{(1)}_{n}}{n_{6}},
  j^{(1)}_{n}=-\frac{(n_{1}-n_{2}\lambda^2_{n})+n_{3}\lambda_{n}r^2_{n}+n_{4}f^{(1)}_{n}+n_{5}h^{(1)}_{n}}{n_{6}},
  \end{equation*}
  where $X_{i}$ are in Appendix C.\\

  By applying the boundary conditions (57)-(65) and utilizing equations (86) and (87), we derive a set of seven homogeneous equations that involve seven unknowns, denoted as $A^{(1)}_{i}$(for i=1,2,3,4) and $S_{i}$(for i=1,2,3,4,5). This system of equations is essential for solving the problem under investigation and is constructed as follows:

\begin{equation}
\beth_{11}A^{(1)}_{1}+\beth_{12}A^{(1)}_{2}+\beth_{13}A^{(1)}_{3}+\beth_{14}A^{(1)}_{4}-S_{1}-S_{2}-S_{3}-S_{4}-S_{5}=0,
\end{equation}
\begin{equation}
\beth_{21}A^{(1)}_{1}+\beth_{22}A^{(1)}_{2}+\beth_{23}A^{(1)}_{3}+\beth_{24}A^{(1)}_{4}-f_{1}S_{1}-f_{2}S_{2}-f_{3}S_{3}-f_{4}S_{4}-f_{5}S_{5}=0,
\end{equation}
\begin{equation}
\beth_{31}A^{(1)}_{1}+\beth_{32}A^{(1)}_{2}+\beth_{33}A^{(1)}_{3}+\beth_{34}A^{(1)}_{4}-\beth_{35}S_{1}-\beth_{36}S_{2}-\beth_{37}S_{3}-\beth_{38}S_{4}-\beth_{39}S_{5}=0,
\end{equation}
\begin{equation}
\beth_{41}A^{(1)}_{1}+\beth_{42}A^{(1)}_{2}+\beth_{43}A^{(1)}_{3}+\beth_{44}A^{(1)}_{4}-\beth_{45}S_{1}-\beth_{46}S_{2}-\beth_{47}S_{3}-\beth_{48}S_{4}-\beth_{49}S_{5}=0,
\end{equation}
\begin{equation}
\beth_{51}A^{(1)}_{1}+\beth_{52}A^{(1)}_{2}+\beth_{53}A^{(1)}_{3}+\beth_{54}A^{(1)}_{4}-\beth_{55}S_{1}-\beth_{56}S_{2}-\beth_{57}S_{3}-\beth_{58}S_{4}-\beth_{59}S_{5}=0,
\end{equation}
\begin{equation}
\beth_{61}A^{(1)}_{1}+\beth_{62}A^{(1)}_{2}+\beth_{63}A^{(1)}_{3}+\beth_{64}A^{(1)}_{4}-j_{1}S_{1}-j_{2}S_{2}-j_{3}S_{3}-j_{4}S_{4}-j_{5}S_{5}=0,
\end{equation}
\begin{equation}
\beth_{71}A^{(1)}_{1}+\beth_{72}A^{(1)}_{2}+\beth_{73}A^{(1)}_{3}+\beth_{74}A^{(1)}_{4}-k_{1}S_{1}-k_{2}S_{2}-k_{3}S_{3}-k_{4}S_{4}-k_{5}S_{5}=0,
\end{equation}
\begin{equation}
\beth_{81}A^{(1)}_{1}+\beth_{82}A^{(1)}_{2}+\beth_{83}A^{(1)}_{3}+\beth_{84}A^{(1)}_{4}-\beth_{85}S_{1}-\beth_{86}S_{2}-\beth_{87}S_{3}-\beth_{88}S_{4}-\beth_{89}S_{5}=0,
\end{equation}
\begin{equation}
\beth_{91}A^{(1)}_{1}+\beth_{92}A^{(1)}_{2}+\beth_{93}A^{(1)}_{3}+\beth_{94}A^{(1)}_{4}-l_{1}S_{1}-l_{2}S_{2}-l_{3}S_{3}-l_{4}S_{4}-l_{5}S_{5}=0.
\end{equation}

The system of equations has nontrivial solution if
\begin{equation}
        \left|\begin{array}{ccccccccc}
        \beth_{11} & \beth_{12} & \beth_{13} & \beth_{23} & -1 & -1 & -1 & -1 & -1\\
\beth_{21} & \beth_{22} & \beth_{23} & \beth_{24} & -f_{1} & -f_{2} & -f_{3} & -f_{4} & -f_{5}\\
\beth_{31} & \beth_{32} & \beth_{33} & \beth_{34} & -\beth_{35} & -\beth_{36} & -\beth_{37} & -\beth_{38} & \beth_{39}\\
\beth_{41} & \beth_{42} & \beth_{43} & \beth_{44} & -\beth_{45} & -\beth_{46} & -\beth_{47} & -\beth_{48} & \beth_{49}\\
\beth_{51} & \beth_{52} & \beth_{53} & \beth_{54} & -\beth_{55} & -\beth_{56} & -\beth_{57} & -\beth_{58} & \beth_{59}\\
\beth_{61} & \beth_{62} & \beth_{63} & \beth_{64} & -j_{1} & -j_{2} & -j_{3} & -j_{4} & -j_{5}\\
\beth_{71} & \beth_{72} & \beth_{73} & \beth_{74} & -k_{1} & -k_{2} & -k_{3} & -k_{4} & -k_{5}\\
\beth_{81} & \beth_{82} & \beth_{83} & \beth_{84} & -\beth_{85} & -\beth_{86} & -\beth_{87} & -\beth_{88} & \beth_{89}\\
\beth_{91} & \beth_{92} & \beth_{93} & \beth_{94} & -l_{1} & -l_{2} & -l_{3} & -l_{4} & -l_{5}
          \end{array}\right|=0.
  \end{equation}
  Expanding the determinant, we get
  \begin{eqnarray}
\beth_{11}M_{1}-\beth_{12}M_{2}+\beth_{13}M_{3}-\beth_{14} M_{4}-M_{5}+M_{6}-M_{7}+M_{8}-M_{9}=0.
\end{eqnarray}
where $M_{i}$ where (i=1,...,9) are in Appendix D.
  \section{ Study of the frequency equation}
     By investigating specific values of parameters, we can obtain unique results as outlined below:\\
     \\
     (a) Frequency equation of Rayleigh wave for local thermoelastic model:\\
      Setting $\varepsilon=0$ in the frequency equation (98) leads to the reduction of the case to local thermoelasticity.\\
      \\
     (b)  Rayleigh waves in the context of nonlocal thermoelasticity inclusion of diffusion and voids , within the framework of the CT model:\\
     The problem conforms to CT models when the parameters $\tau_{T}$ and $\tau_{q}$ are both set to zero, along with $\tau_{1}$ and $\tau_{2}$. This condition indicates a scenario where thermal and mechanical couplings are absent, aligning with the principles of classical coupled thermoelasticity.

          \section{Evaluation and confirmation of the current study}
    This section undertakes an examination of various scenarios by evaluating specific parameter values, followed by a comparative analysis of the results with those found in existing literature.\\
     \\
     \textbf{(a) Rayleigh waves in the context of nonlocal thermoelasticity without voids and diffusion, within the framework of the Lord-Shulman (LS) model:} \\
          When $\tau_{\theta}=0$ and $\tau_{2}=0$, along with $\xi=\nu_{2}=A_{1}=A_{3}=B_{1}=B_{3}=0$, $b_{1}=b_{3}=d_{1}=d_{3}=d^{\ast}_{1}=d^{\ast}_{3}=C=0$, the problem undergoes a transformation into the nonlocal theory resembling the LS model. This adjustment in parameters leads to a configuration akin to the findings outlined by Biswas \cite{52}, albeit following certain modifications.\\ \\
      \textbf{(b) Rayleigh waves in Local thermoelasticity with voids, excluding diffusion, in an isotropic medium:}\\
       If we choose $\varepsilon=0$, $b_{1}=b_{3}=d_{1}=d_{3}=d^{\ast}_{1}=d^{\ast}_{3}=C=0$, then the situation becomes to a locally heat-conductive medium with prosity and without diffusion, consistent with Isean \cite{19} after certain modifications.\\
        \\
     \textbf{(c) Rayleigh waves under the context of nonlocally heat-conductive media inclusion of void and exclusion of diffusion in an isotropic medium:}\\
     If we choose $c_{11}=c_{33}=\lambda+2\mu$, $c_{13}=\lambda$, $c_{55}=\mu$, $K_{1}=K_{3}=K$, $\beta_{1}=\beta_{3}=\beta$, $b_{1}=b_{3}=\overline{b}$, $d_{1}=d_{3}=d$, $d^{\ast}_{1}=d^{\ast}_{3}=d^{\ast}$ and set $\overline{b}=0,d=0,d^{\ast}=0,C=0$, then the situation becomes to an isotropic medium without diffusion, in agreement in conjunction with Biswas \cite{44} after particular changes.\\
        \\
     \textbf{(d) Rayleigh wave in in the context of local heat-conductive medium without voids, incorporating diffusion, in an orthotropic medium:}\\
      Taking $\varepsilon=0$ and $\xi=\nu_{2}=A_{1}=A_{3}=B_{1}=B_{3}=0$ then the situation becomes one of local thermoelasticity inclusion of  diffusion and validates Yadav's \cite{55} argument.\\
      \\
          \section{Technical applications}
          \textbf{Geophysical Exploration:}\\ Understanding Rayleigh wave propagation in porous, orthotropic materials is crucial for seismic exploration, especially for identifying subsurface characteristics such as fluid-filled rock layers.\\
          \\
\textbf{Non-Destructive Testing (NDT):}\\ In engineering structures made of advanced composite materials (like porous ceramics or fiber-reinforced composites), analyzing Rayleigh waves can help detect defects, delaminations, or voids.\\
\\
\textbf{Material Design:}\\ The influence of porosity, nonlocality, and thermal effects helps in designing materials with tailored properties, particularly in aerospace and civil engineering where high-performance materials are used.\\
\\
\textbf{Environmental Sensing:}\\ Rayleigh waves in porous media can be used to monitor environmental conditions, like the detection of pollutants or underground fluid movement.
\section{Diverse features of Rayleigh waves}
\textbf{Propagation speed and attenuation coefficient:}\\
   In general, the phase velocity $(c)$ is complex quantity. \\We take
   \begin{eqnarray*}
   c^{-1}=V^{-1}+i\omega^{-1}Q.
   \end{eqnarray*}
    In this context, V represents the speed of propagation, Q denotes the attenuation coefficient, and $\omega$ stands for the angular frequency.\\
\\
   \textbf{Specific loss:}\\
   The specific loss (SL) quantifies the internal friction within a material by comparing the energy dissipated during a cycle, denoted as $\Delta W$, to the elastic energy stored in the specimen, represented as W, when the strain is at its peak. According to Puri and Cowin \cite{53}, specific loss offers a direct method for defining internal friction. Kolsky \cite{54} demonstrated that for a sinusoidal surface wave of small amplitude, the specific loss $\frac{\Delta W}{W}$ can be expressed mathematically as:
    \begin{eqnarray*}
   SL=\frac{\Delta W}{W}=4\pi \bigg |\frac{VQ}{\omega} \bigg|.
    \end{eqnarray*}
    \\
    \textbf{Penetration depth:}\\
    The penetration depth is defined by
     \begin{eqnarray*}
    \delta=\frac{1}{|Q|}
    \end{eqnarray*}

\section{Special cases}

     In the absence of layer, voids and diffusion i.e., by taking $H=0$, $\xi'=\nu'_{2}=A'_{1}=A'_{3}=B'_{1}=B'_{3}=0$, $b'_{1}=b'_{3}=d'_{1}=d'_{3}=d'^{\ast}_{1}=d'^{\ast}_{3}=C_{2}=0$, we consider some special cases.\\

In the context of transversely isotropic materials, we adopt the following relationships:\\ $c'_{11}=c'_{33},2c'_{55}=c'_{11}-c'_{13}$ and $\beta'_{1}=\beta'_{3}$.\\
    Subsequently, we consider the simplification $\beta'_{1}=\beta'_{3}=\beta$ and proceed to analyze the ensuing cases:
   .\\
    \textbf{Case 1:} Frequency equation of classical coupled thermoelasticity in transversely isotropic half-space is obtained by taking $\tau_{q}=0$\\
    which is
    \begin{equation}
    \begin{split}
    \left[2-\frac{2\rho_{2} c^2}{(c'_{11}-c'_{13})}\right]^2\left(\gamma^2_{1}+\gamma_{1}\gamma_{2}+\gamma^2_{2}-1+\frac{\rho c^2}{c'_{11}}\right)-4\gamma_{1}\gamma_{2}\gamma_{3}(\gamma_{1}+\gamma_{2})\\
    -\frac{m}{k} \left[\left(2-\frac{2\rho_{2} c^2}{(c'_{11}-c'_{13})}\right)^2 (\gamma_{1}+\gamma_{2})-4\gamma_{3}\left(\gamma_{1}\gamma_{2}+1-\frac{\rho_{2} k^2 c^2}{c'_{11}}\right)\right]=0
    \end{split}
    \end{equation}where $m\longrightarrow 0$ equates to a thermally insulated surface and
$m\longrightarrow \infty $equates to an isothermal surface.\\
     $\gamma^2_{1}=1-\frac{\xi^2_{1}}{k^2}, \gamma^2_{2}=1-\frac{\xi^2_{2}}{k^2}, \gamma^2_{3}=1-\frac{\zeta^2}{k^2}, \zeta^2=\frac{2\rho k^2c^2}{c'_{11}-c'_{13}},$ and $\xi^2_{1}$ and $\xi^2_{2}$ are the roots of biquadratic equation\\
     \begin{equation}
    \begin{split}
     \xi^4-\left[(k^2-\frac{K_{1}k^2}{K_{3}})+\frac{\rho_{2} k^2c^2}{c_{11}}+(1+\kappa)\frac{ikc\rho_{2} C'_{v}}{K_{3}}\right]\xi^2\\+ \left(\frac{\rho_{2} k^2c^2}{c'_{11}}-\frac{K'_{1}}{K'_{3}}\frac{\rho_{2} k^2c^2}{c'_{11}}\right)k^2+\frac{ik^3c^3\rho_{2} C'_{v}}{c_{11}K}=0
     \end{split}
    \end{equation}
    in which $\kappa=\frac{\beta'^2T_{0}}{\rho^2_{2}c^2C'_{v}}.$\\
    Derived frequency Equation (99)  is in agreement with the outcome achieved by Biswas \cite{52}.
    Now if we assign $c'_{13}=\lambda, c'_{11}=c'_{33}=\lambda+2\mu, K'_{1}=K'_{3}=K,$ then we obtain the frequency equation of Rayleigh waves for an isotropic half-space in the case of classical coupled thermoelasticity as follows:\\
     \begin{equation}
    \begin{split}
    (2-\frac{c^2}{c^2_{2}})^2\left(\gamma^2_{1}+\gamma_{1}\gamma_{2}+\gamma^2_{2}-1+\frac{c^2}{c^2_{1}}\right)-4\gamma_{1}\gamma_{2}\gamma_{3}(\gamma_{1}+\gamma_{2})\\
    -\frac{m}{k}\left[(2-\frac{c^2}{c^2_{2}})^2(\gamma_{1}+\gamma_{2})-4\gamma_{3}(\gamma_{1}\gamma_{2}+1-\frac{k^2c^2}{c^2_{1}})\right]=0
     \end{split}
    \end{equation}
    Equation (101) is similar to the finding reached in Nowinski \cite{57},\\
    where $\gamma^2_{1}=1-\frac{\xi^2_{1}}{k^2}, \gamma^2_{2}=1-\frac{\xi^2_{2}}{k^2}, \gamma^2_{3}=1-\frac{\zeta^2}{k^2}, \zeta^2=\frac{k^2c^2}{c^2}$ and $\xi^2_{1}$ and $\xi^2_{2}$ are the roots of biquadratic equation\\
    \begin{equation}
    \xi^4-\left[\frac{\rho k^2c^2}{c^2_{1}}+(1+\kappa)\frac{ikc\rho C_{v}}{K}\right]\xi^2+\frac{ik^3c^3\rho C'_{v}}{Kc^2_{1}}=0
    \end{equation}
    in which $\kappa=\frac{\beta^2T_{0}}{\rho^2 c^2_{1}C'_{v}}$ and $c^2_{1}=\frac{\lambda +2\mu}{\rho},c^2_{2}=\frac{\mu}{\rho}.$\\

     \textbf{Case 2:} Excluding thermal parameters, i.e., in the scenario where strain field and temperature are not coupled, the frequency equation of the orthotropic elastic half-space can be formulated as:\\

     \begin{equation}
     \begin{split}
     2(c'_{33}-c'_{13})\left[\left(\frac{\rho c^2-c'_{11}}{c'_{13}+2c'_{55}}\right)\left(\frac{\rho c^2-c'_{55}}{c'_{33}-c'_{55}-c'_{13}}\right)\right]^\frac{1}{2}=\left(\frac{\rho c^2-c'_{55}}{c'_{33}-c'_{55}-c'_{13}}-1\right)\\
     \left(\frac{c'_{33}(\rho_{2} c^2-c'_{11})}{c'_{13}+2c'_{55}}+c'_{13}\right)
     \end{split}
     \end{equation}
     This corresponds to the result of Abd-Alla et al. \cite{56}.\\
      Now putting $c'_{11}=c'_{33}=\lambda+2\mu, c'_{13}=\lambda, c'_{55}=\mu$ in the equation (103), \\we get the frequency equation  for the isotropic elastic half-space as follows:
     \begin{equation}
     \left(2-\frac{c^2}{c^2_{2}}\right)^2=4\left(1-\frac{c^2}{c^2_{1}}\right)^\frac{1}{2}\left(1-\frac{c^2}{c^2_{2}}\right)^\frac{1}{2}
     \end{equation}
     where $c^2_{1}=\frac{\lambda+2\mu}{\rho},c^2_{2}=\frac{\mu}{\rho}.$\\

    \section{Concluding remarks}
     The paper presents an innovative model of Rayleigh waves in an orthotropic medium, utilizing the three-phase-lag model to explore the effects of voids and diffusion. The problem is analyzed using the normal mode technique, a method chosen for its effectiveness in examining the system's behavior comprehensively. Graphical representations are utilized to visually demonstrate how the nonlocal parameter $\varepsilon$, the presence of voids, and the diffusion parameter influence different aspects of Rayleigh waves concerning the wave number. These illustrations provide valuable insights into how variations in these parameters affect the propagation characteristics, attenuation, and penetration depth of Rayleigh waves, offering a thorough understanding of their behavior in the given context. Through extensive theoretical and numerical analyses, several key observations have been made:\\

\textbf{(a)} The addition of $\varepsilon$ generally enhances propagation speed (V) in the TPL and DPL models, suggesting that nonlocal effects positively influence wave propagation dynamics. Initially, the TPL model with $\varepsilon=0$ shows the highest V values, but nonlocal effects lead to increased V in the presence of $\varepsilon$.\\
\textbf{(b)} The parameter Q exhibits wave-like behavior in the TPL and DPL models with $\varepsilon$, while it remains stable in the LS model. The nonlocal parameter $\varepsilon$ significantly affects Q, leading to oscillations in the TPL and DPL models and stability in the LS model.  SL shows varying behavior depending on the model and the value of $\varepsilon$.\\
\textbf{(c)} The parameter $\varepsilon$ enhances SL values in the DPL and LS models but reduces them in the TPL model, indicating its significant influence on specific loss. \\
\textbf{(d)} The parameter $\delta$ increases steadily within a certain range before exhibiting wave-like behavior. The nonlocal parameter $\varepsilon$ leads to higher initial $\delta$ values in the LS model, highlighting its impact on penetration depth.\\
  \textbf{(e)} Overall, the study underscores the critical impact of the nonlocal parameter $\varepsilon$ on the thermoelastic response of different models. The nonlocal effects contribute to an increase in wave speed, introduce oscillatory behavior in attenuation coefficient, and significantly alter specific loss  patterns. These findings provide valuable insights into the behavior of nonlocal thermoelastic systems and emphasize the necessity of considering nonlocal parameters when analyzing heat conduction and wave propagation in such materials. The detailed graphical analysis further solidifies these conclusions, offering a visual representation of the variations in key parameters across different models and validating the theoretical framework employed in this study.\\
\textbf{(f)}The incorporation of voids into the TPL, DPL, and LS models significantly impacts thermoelastic wave behavior, influencing propagation speed. The presence of voids initially enhances the propagation speed V in the DPL model, demonstrating a more dynamic response compared to other configurations. However, as the wave number $k$ increases, V decreases consistently across all models, with voids affecting the rate of this decline differently. Despite the initial advantage, the general trend remains a reduction in V with increasing $k$, irrespective of void inclusion.\\
\textbf{(g)} Attenuation Coefficient (Q): Voids have a pronounced effect on the attenuation coefficient Q, especially in the DPL model where Q fluctuates more significantly with $k$. Initially, voids increase Q values, but as $k$ grows, Q exhibits wave-like behavior. This variation highlights the complex interaction between voids and energy attenuation, particularly in the DPL model.\\
\textbf{(h)}  Specific Loss (SL): The specific loss SL remains relatively stable across models and wave numbers until $k$ surpasses 7900, after which it experiences a sharp decline. Voids influence SL differently across models, enhancing it in the DPL and LS models while reducing it in the TPL model. This variability underscores the role of voids in modifying energy dissipation.\\
\textbf{(i)} Voids lead to a higher initial penetration depth $\delta$ in the DPL model and introduce wave-like variations as $k$ increases.\\
 \textbf{(j)} The inclusion of voids intensifies fluctuations in wave behavior, demonstrating a pronounced effect on penetration depth. Their presence significantly alters thermoelastic wave dynamics by accelerating initial propagation and extending penetration, while simultaneously affecting attenuation and specific loss in intricate ways. These results highlight the necessity of incorporating voids into thermoelastic models to ensure precise predictions of wave behavior and energy dissipation across different materials.\\
\textbf{(k)}The effects of diffusion on the propagation speed (V) decreases with increasing wave number k across all models, with the DPL model without diffusion exhibiting the highest initial value, highlighting the role of diffusion in slowing down wave propagation.\\
\textbf{(l)} Attenuation coefficient (Q) follows an oscillatory trend, with the TPL model without diffusion showing the highest initial value, suggesting that diffusion significantly influences energy dissipation within the system.\\
\textbf{(m)} Specific loss (SL) remains relatively constant over a broad range of k in the TPL and DPL models before gradually decreasing, whereas the LS model exhibits periodic variations, indicating a more dynamic interaction with diffusion effects.\\
\textbf{(n)} Penetration depth ($\delta$) also exhibits oscillatory behavior, with the DPL model with diffusion attaining the highest value, underscoring the impact of diffusion on wave penetration in thermoelastic media.\\
 \textbf{(o)} These observations collectively highlight the complex interplay between diffusion and thermoelastic parameters, demonstrating that diffusion generally enhances energy dissipation and wave penetration while reducing propagation speed. The distinct behaviors across the TPL, DPL, and LS models suggest that different thermoelastic theories respond uniquely to diffusion effects, which is crucial for applications involving wave propagation in nonlocal porous orthotropic materials.

\section*{Appendix A}

$R_{1}=\frac{R_{7}}{R_{6}}$,
$R_{2}=\frac{R_{8}}{R_{6}}$,
$R_{3}=\frac{R_{9}}{R_{6}}$,
$R_{4}=\frac{R_{10}}{R_{6}}$,
$R_{5}=\frac{R_{11}}{R_{6}}$,

$R_{6}=-a_{12} a_{26} a_{34} a_{45} a_{54}  +
    a_{12} a_{23} a_{34} a_{45} a_{5(10)} $

$R_{7}=-a_{16} a_{23} a_{34} a_{45} a_{52} +
    a_{13} a_{26} a_{34} a_{45} a_{52}  -
    a_{12} a_{26} a_{34} a_{45} a_{53}  -
    a_{12} a_{26} a_{34} a_{44} a_{54}  -
    a_{12} a_{26} a_{33} a_{45} a_{54}  +
    a_{16} a_{21} a_{34} a_{45} a_{54}  -
    a_{11} a_{26} a_{34} a_{45} a_{54}  +
    a_{12} a_{24} a_{36} a_{45} a_{54}  +
    a_{12} a_{25} a_{34} a_{46} a_{54}  +
    a_{12} a_{26} a_{32} a_{45} a_{56}  -
    a_{12} a_{23} a_{36} a_{45} a_{56}  +
    a_{12} a_{26} a_{34} a_{42} a_{58}  -
    a_{12} a_{23} a_{34} a_{46} a_{58}+
    a_{12} a_{23} a_{34} a_{45} a_{59}-
    a_{12} a_{25} a_{34} a_{42} a_{5(10)}+
    a_{12} a_{23} a_{34} a_{44} a_{5(10)}-\\
    a_{12} a_{24} a_{32} a_{45} a_{5(10)}+
    a_{12} a_{23} a_{33} a_{45} a_{5(10)}-
    a_{13} a_{21} a_{34} a_{45} a_{5(10)}  +
    a_{12} a_{22} a_{34} a_{45} a_{5(10)}  +
    a_{11} a_{23} a_{34} a_{45} a_{5(10)} $

 $R_{8}=-a_{16} a_{23} a_{34} a_{45} a_{51}  +
    a_{13} a_{26} a_{34} a_{45} a_{51}  +
    a_{16} a_{25} a_{34} a_{42} a_{52}  -
    a_{15} a_{26} a_{34} a_{42} a_{52}  -
    a_{16} a_{23} a_{34} a_{44} a_{52}  +
    a_{13} a_{26} a_{34} a_{44} a_{52}  +
    a_{16} a_{24} a_{32} a_{45} a_{52}  -
    a_{14} a_{26} a_{32} a_{45} a_{52}  -
    a_{16} a_{23} a_{33} a_{45} a_{52}  +
    a_{13} a_{26} a_{33} a_{45} a_{52}  -
    a_{16} a_{22} a_{34} a_{45} a_{52} +
    a_{14} a_{23} a_{36} a_{45} a_{52}  -
    a_{13} a_{24} a_{36} a_{45} a_{52}  +
    a_{15} a_{23} a_{34} a_{46} a_{52}  -
    a_{13} a_{25} a_{34} a_{46} a_{52}  -
    a_{12} a_{26} a_{34} a_{44} a_{53}  -
    a_{12} a_{26} a_{33} a_{45} a_{53}  +
    a_{16} a_{21} a_{34} a_{45} a_{53}  -
    a_{11} a_{26} a_{34} a_{45} a_{53}  +
    a_{12} a_{24} a_{36} a_{45} a_{53}  +
    a_{12} a_{25} a_{34} a_{46} a_{53}  -
    a_{16} a_{25} a_{34} a_{41} a_{54}  +
    a_{15} a_{26} a_{34} a_{41} a_{54}  +
    a_{12} a_{26} a_{35} a_{43} a_{54}  -
    a_{12} a_{25} a_{36} a_{43} a_{54}  -
    a_{12} a_{26} a_{33} a_{44} a_{54} +
    a_{16} a_{21} a_{34} a_{44} a_{54}  -
    a_{11} a_{26} a_{34} a_{44} a_{54}  +
    a_{12} a_{24} a_{36} a_{44} a_{54}  -
    a_{16} a_{24} a_{31} a_{45} a_{54}  +
    a_{14} a_{26} a_{31} a_{45} a_{54}  +
    a_{16} a_{21} a_{33} a_{45} a_{54}  -
    a_{11} a_{26} a_{33} a_{45} a_{54}  -
    a_{14} a_{21} a_{36} a_{45} a_{54}  +
    a_{11} a_{24} a_{36} a_{45} a_{54}  +
    a_{12} a_{25} a_{33} a_{46} a_{54}  -
    a_{15} a_{21} a_{34} a_{46} a_{54}  +
    a_{11} a_{25} a_{34} a_{46} a_{54}  -
    a_{12} a_{24} a_{35} a_{46} a_{54}  +
    a_{12} a_{26} a_{32} a_{45} a_{55}  -
    a_{12} a_{23} a_{36} a_{45} a_{55}  -
    a_{12} a_{26} a_{35} a_{42} a_{56}  +
 a_{12} a_{25} a_{36} a_{42} a_{56}  +
    a_{12} a_{26} a_{32} a_{44} a_{56}  -
    a_{12} a_{23} a_{36} a_{44} a_{56}  +
    a_{16} a_{23} a_{31} a_{45} a_{56}  -
    a_{13} a_{26} a_{31} a_{45} a_{56}  -
    a_{16} a_{21} a_{32} a_{45} a_{56}  +
    a_{11} a_{26} a_{32} a_{45} a_{56}  +
    a_{13} a_{21} a_{36} a_{45} a_{56}  -
    a_{12} a_{22} a_{36} a_{45} a_{56}  -
    a_{11} a_{23} a_{36} a_{45} a_{56}  -
    a_{12} a_{25} a_{32} a_{46} a_{56}  +
    a_{12} a_{23} a_{35} a_{46} a_{56}  +
    a_{12} a_{26} a_{34} a_{42} a_{57}  -
    a_{12} a_{23} a_{34} a_{46} a_{57}  +
    a_{16} a_{23} a_{34} a_{41} a_{58}  -
    a_{13} a_{26} a_{34} a_{41} a_{58}  +
 a_{12} a_{26} a_{33} a_{42} a_{58}  -
    a_{16} a_{21} a_{34} a_{42} a_{58}  +
    a_{11} a_{26} a_{34} a_{42} a_{58}  -
    a_{12} a_{24} a_{36} a_{42} a_{58}  -
    a_{12} a_{26} a_{32} a_{43} a_{58}  +
    a_{12} a_{23} a_{36} a_{43} a_{58}  +
    a_{12} a_{24} a_{32} a_{46} a_{58}  -
    a_{12} a_{23} a_{33} a_{46} a_{58}  +
    a_{13} a_{21} a_{34} a_{46} a_{58}  -
    a_{12} a_{22} a_{34} a_{46} a_{58}  -
    a_{11} a_{23} a_{34} a_{46} a_{58}  -
    a_{12} a_{25} a_{34} a_{42} a_{59}  +
    a_{12} a_{23} a_{34} a_{44} a_{59}  -
    a_{12} a_{24} a_{32} a_{45} a_{59} +
 a_{12} a_{23} a_{33} a_{45} a_{59}  -
    a_{13} a_{21} a_{34} a_{45} a_{59}  +
    a_{12} a_{22} a_{34} a_{45} a_{59}  +
    a_{11} a_{23} a_{34} a_{45} a_{59}  -
    a_{15} a_{23} a_{34} a_{41} a_{5(10)}  +
    a_{13} a_{25} a_{34} a_{41} a_{5(10)}  -
    a_{12} a_{25} a_{33} a_{42} a_{5(10)}  +
    a_{15} a_{21} a_{34} a_{42} a_{5(10)}  -
    a_{11} a_{25} a_{34} a_{42} a_{5(10)}  +
    a_{12} a_{24} a_{35} a_{42} a_{5(10)}  +
    a_{12} a_{25} a_{32} a_{43} a_{5(10)}  -
    a_{12} a_{23} a_{35} a_{43} a_{5(10)}  -
    a_{12} a_{24} a_{32} a_{44} a_{5(10)}  +
 a_{12} a_{23} a_{33} a_{44} a_{5(10)}  -
    a_{13} a_{21} a_{34} a_{44} a_{5(10)}  +
    a_{12} a_{22} a_{34} a_{44} a_{5(10)}  +
    a_{11} a_{23} a_{34} a_{44} a_{5(10)}  -
    a_{14} a_{23} a_{31} a_{45} a_{5(10)}  +
    a_{13} a_{24} a_{31} a_{45} a_{5(10)}  +
    a_{14} a_{21} a_{32} a_{45} a_{5(10)}  -
    a_{11} a_{24} a_{32} a_{45} a_{5(10)}  -
    a_{13} a_{21} a_{33} a_{45} a_{5(10)}  +
    a_{12} a_{22} a_{33} a_{45} a_{5(10)}  +
    a_{11} a_{23} a_{33} a_{45} a_{5(10)}  +
    a_{11} a_{22} a_{34} a_{45} a_{5(10)} $

$R_{9}=a_{16} a_{25}a_{34} a_{42}a_{51}  -
 a_{15} a_{26} a_{34} a_{42}a_{51}  -
 a_{16} a_{23} a_{34} a_{44}a_{51}  +
 a_{13} a_{26} a_{34}a_{44}a_{51}  +
 a_{16} a_{24} a_{32} a_{45} a_{51}  -
 a_{14} a_{26} a_{32} a_{45}a_{51}  -
 a_{16} a_{23} a_{33} a_{45}a_{51}  +
 a_{13} a_{26} a_{33} a_{45} a_{51}  -
 a_{16} a_{22} a_{34} a_{45}a_{51}  +
 a_{14} a_{23} a_{36} a_{45}a_{51}  -
 a_{13} a_{24} a_{36} a_{45} a_{51}  +
 a_{15} a_{23} a_{34} a_{46}a_{51}  -
 a_{13} a_{25} a_{34} a_{46}a_{51}  +
a_{16} a_{25} a_{33} a_{42} a_{52}  -
 a_{15} a_{26} a_{33} a_{42}a_{52}  -
 a_{16} a_{24} a_{35} a_{42}a_{52}  +
 a_{14} a_{26} a_{35} a_{42}a_{52}  +
 a_{15} a_{24} a_{36} a_{42}a_{52}  -
 a_{14} a_{25} a_{36} a_{42}a_{52}  -
 a_{16} a_{25} a_{32} a_{43}a_{52}  +
 a_{15} a_{ 26} a_{32} a_{43}a_{52}  +
 a_{16} a_{23} a_{35} a_{43}a_{52} -
 a_{13} a_{26} a_{35} a_{43} a_{52}  -
 a_{15} a_{23} a_{36} a_{43}a_{52}  +
 a_{13} a_{25} a_{36} a_{43} a_{52}  +
a_{16} a_{24} a_{32} a_{44} a_{52}  -
a_{14} a_{26} a_{32} a_{44} a_{52}  -
a_{16} a_{23} a_{33} a_{44} a_{52}  +
a_{13} a_{26} a_{33} a_{44} a_{52}  -
a_{16} a_{22} a_{34} a_{44} a_{52}  +
a_{14} a_{23} a_{36} a_{44} a_{52}  -
a_{13} a_{24} a_{36} a_{44} a_{52}  -
a_{16} a_{22} a_{33} a_{45} a_{52}  +
a_{14} a_{22} a_{36} a_{45} a_{52}  -
a_{15} a_{24} a_{32} a_{46} a_{52}  +
 a_{14} a_{25} a_{32} a_{46} a_{52}  +
a_{15} a_{23} a_{33} a_{46} a_{52}  -
a_{13} a_{25} a_{33} a_{46} a_{52}  +
a_{15} a_{22} a_{34} a_{46} a_{52}  -
a_{14} a_{23} a_{35} a_{46} a_{52}  +
a_{13} a_{24} a_{35} a_{46} a_{52}  -
a_{16} a_{25} a_{34} a_{41} a_{53}  +
a_{15} a_{26} a_{34} a_{41} a_{53}  +
a_{12} a_{26} a_{35} a_{43} a_{53}  -
a_{12} a_{25} a_{36} a_{43} a_{53}  -
a_{12} a_{26} a_{33} a_{44} a_{53}  +
a_{16} a_{21} a_{34} a_{44} a_{53}  -
a_{11} a_{26} a_{34} a_{44} a_{53}  +
a_{12} a_{24} a_{36} a_{44} a_{53}  -
 a_{16} a_{24} a_{31} a_{45} a_{53}  +
 a_{14} a_{26} a_{31} a_{45} a_{53}  +
a_{16} a_{21} a_{33} a_{45} a_{53}  -
a_{11} a_{26} a_{33} a_{45} a_{53}  -
a_{14} a_{21} a_{36} a_{45} a_{53}  +
a_{11} a_{24} a_{36} a_{45} a_{53}  +
a_{12} a_{25} a_{33} a_{46} a_{53}  -
a_{15} a_{21} a_{34} a_{46} a_{53}  +
a_{11} a_{25} a_{34} a_{46} a_{53}  -
a_{12} a_{24} a_{35} a_{46} a_{53}  -
a_{16} a_{25} a_{33} a_{41} a_{54}  +
a_{15} a_{26} a_{33} a_{41} a_{54}  +
a_{16} a_{24} a_{35} a_{41} a_{54}  -
a_{14} a_{26} a_{35} a_{41} a_{54}  -
a_{15} a_{24} a_{36} a_{41} a_{54}  +
a_{14} a_{25} a_{36} a_{41} a_{54}  +
a_{16} a_{25} a_{31} a_{43} a_{54}  -
a_{15} a_{26} a_{31} a_{43} a_{54}  -
a_{16} a_{21} a_{35} a_{43} a_{54}  +
a_{11} a_{26} a_{35} a_{43} a_{54}  +
a_{15} a_{21} a_{36} a_{43} a_{54}  -
a_{11} a_{25} a_{36} a_{43} a_{54}  -
a_{12} a_{26} a_{35} a_{42} a_{55}  +
a_{12} a_{25} a_{36} a_{42} a_{55}  +
a_{12} a_{26} a_{32} a_{44} a_{55}  -
a_{12} a_{23} a_{36} a_{44} a_{55}  +
a_{16} a_{23} a_{31} a_{45} a_{55}  -
a_{13} a_{26} a_{31} a_{45} a_{55}  -
a_{16} a_{21} a_{32} a_{45} a_{55}  +
a_{11} a_{26} a_{32} a_{45} a_{55}  +
a_{13} a_{21} a_{36} a_{45} a_{55}  -
a_{12} a_{22} a_{36} a_{45} a_{55}  -
a_{11} a_{23} a_{36} a_{45} a_{55}  -
a_{12} a_{25} a_{32} a_{46} a_{55}  +
a_{12} a_{23} a_{35} a_{46} a_{55}  +
 a_{16} a_{25} a_{32} a_{41}a_{56}  -
 a_{15} a_{26} a_{32} a_{41}a_{56}  -
 a_{16} a_{ 23} a_{35} a_{41} a_{56}  +
a_{13} a_{26} a_{35} a_{41}a_{56}  +
 a_{15} a_{23} a_{36} a_{41} a_{56}  -
a_{13} a_{25} a_{36} a_{41} a_{56} -
a_{16} a_{25} a_{31} a_{42} a_{56}  +
a_{15} a_{26} a_{31} a_{42} a_{56} +
a_{16} a_{21} a_{35} a_{42} a_{56}  -
a_{11} a_{26} a_{35} a_{42} a_{56}  -
a_{15} a_{21} a_{36} a_{42} a_{56}  +
a_{11} a_{25} a_{36} a_{42} a_{56}  +
a_{16} a_{23} a_{31} a_{44} a_{56}  -
a_{13} a_{26} a_{31} a_{44} a_{56}  -
a_{16} a_{21} a_{32} a_{44} a_{56} +
a_{11} a_{26} a_{32} a_{44} a_{56} +
a_{13} a_{21} a_{36} a_{44} a_{56}  -
a_{12} a_{22} a_{36} a_{44} a_{56}  -
a_{11} a_{23} a_{36} a_{44} a_{56}  +
a_{16} a_{22} a_{31} a_{45} a_{56}  -
a_{11} a_{22} a_{36} a_{45} a_{56}  -
a_{15} a_{23} a_{31} a_{46} a_{56}  +
a_{13} a_{25} a_{31} a_{46} a_{56}  +
a_{15} a_{21} a_{32} a_{46} a_{56}  -
a_{11} a_{25} a_{32} a_{46} a_{56}  -
a_{13} a_{21} a_{35} a_{46} a_{56}  +
a_{12} a_{22} a_{35} a_{46} a_{56}  +
a_{11} a_{23} a_{35} a_{46} a_{56}  -
a_{16} a_{23} a_{34} a_{41} a_{57}  -
a_{13} a_{26} a_{34} a_{41} a_{57} +
a_{12} a_{26} a_{33} a_{42} a_{57}  -
a_{16} a_{21} a_{34} a_{42} a_{57}  +
a_{11} a_{26} a_{34} a_{42} a_{57}  -
a_{12} a_{24} a_{36} a_{42} a_{57}  -
a_{12} a_{26} a_{32} a_{43} a_{57}  +
a_{12} a_{23} a_{36} a_{43} a_{57}  +
a_{12} a_{24} a_{32} a_{46} a_{57}  -
a_{12} a_{23} a_{33} a_{46} a_{57}  +
a_{13} a_{21} a_{34} a_{46} a_{57}  -
a_{12} a_{22} a_{34} a_{46} a_{57}  -
a_{11} a_{23} a_{34} a_{46} a_{57}  -
 a_{16} a_{ 24} a_{32} a_{41}a_{58}  +
 a_{14} a_{26} a_{32} a_{41} a_{58}  +
 a_{16} a_{23} a_{33} a_{41}a_{58}  -
 a_{13}a_{26}a_{33} a_{41} a_{58}  +
 a_{16} a_{22} a_{34} a_{41}a_{58}  -
 a_{14} a_{23} a_{36} a_{41} a_{58}  +
 a_{13} a_{24} a_{36} a_{41}a_{58}  +
 a_{16} a_{24} a_{31} a_{ 42}a_{58}  -
 a_{14} a_{26} a_{31} a_{42} a_{58}  -
a_{16} a_{21} a_{33} a_{42} a_{58}  +
a_{11} a_{26} a_{33} a_{42} a_{58}  +
a_{14} a_{21} a_{36} a_{42} a_{58}  -
a_{11} a_{24} a_{36} a_{42} a_{58}  -
a_{16} a_{23} a_{31} a_{43} a_{58}  +
a_{13} a_{26} a_{31} a_{43} a_{58}  +
a_{16} a_{21} a_{32} a_{43} a_{58}  -
a_{11} a_{26} a_{32} a_{43} a_{58}  -
a_{13} a_{21} a_{36} a_{43} a_{58}  +
a_{12} a_{22} a_{36} a_{43} a_{58}  +
a_{11} a_{23} a_{36} a_{43} a_{58}  +
a_{14} a_{23} a_{31} a_{46} a_{58}  -
a_{13} a_{24} a_{31} a_{46} a_{58}  -
a_{14} a_{21} a_{32} a_{46} a_{58}  +
a_{11} a_{24} a_{32} a_{46} a_{58}  +
a_{13} a_{21} a_{33} a_{46} a_{58}  -
a_{12} a_{22} a_{33} a_{46} a_{58}  -
a_{11} a_{23} a_{33} a_{46} a_{58}  -
a_{11} a_{22} a_{34} a_{46} a_{58}  -
a_{15} a_{23} a_{34} a_{41} a_{59}  +
a_{13} a_{25} a_{34} a_{41} a_{59}  -
a_{12} a_{25} a_{33} a_{42} a_{59}  +
a_{15} a_{21} a_{34} a_{42} a_{59}  -
a_{11} a_{25} a_{34} a_{42} a_{59}  +
a_{12} a_{24} a_{35} a_{42} a_{59}  +
a_{12} a_{25} a_{32} a_{43} a_{59}  -
a_{12} a_{23} a_{35} a_{43} a_{59}  +
a_{12} a_{24} a_{32} a_{44} a_{59} -
a_{12} a_{23} a_{33} a_{44} a_{59}  +
a_{13} a_{21} a_{34} a_{44} a_{59}  +
a_{12} a_{22} a_{34} a_{44} a_{59}  +
a_{11} a_{23} a_{34} a_{44} a_{59}  -
a_{14} a_{23} a_{31} a_{45} a_{59}  +
a_{13} a_{24} a_{31} a_{45} a_{59}  +
a_{14} a_{21} a_{32} a_{45} a_{59}  -
a_{11} a_{24} a_{32} a_{45} a_{59}  -
a_{13} a_{21} a_{33} a_{45} a_{59}  +
a_{12} a_{22} a_{33} a_{45} a_{59}  +
a_{11} a_{23} a_{33} a_{45} a_{59}  +
a_{11} a_{22} a_{34} a_{45} a_{59}  +
a_{14} a_{22} a_{31} a_{45} a_{5(10)}  -
a_{11} a_{22} a_{33} a_{45} a_{5(10)} $

$R_{10}=a_{16}a_{25}a_{33}a_{42}a_{51} -
 a_{15}a_{26}a_{33} a_{42} a_{51} -
 a_{16} a_{24}a_{35}a_{42}a_{51}  +
 a_{14}a_{26} a_{35}a_{42}a_{51} +
 a_{15} a_{24} a_{36}a_{42}a_{51} -
 a_{14}a_{25} a_{36} a_{42}a_{51}  -
a_{16} a_{25}a_{32} a_{43}a_{51}  +
 a_{15} a_{26} a_{32}a_{43}a_{51}  +
 a_{16} a_{23} a_{35} a_{43}a_{51}  -
 a_{13}a_{26} a_{35} a_{43} a_{51}  -
 a_{15}a_{23} a_{36} a_{43}a_{51}  +
 a_{13} a_{25} a_{36} a_{43}a_{51}  +
 a_{16} a_{24} a_{32} a_{44}a_{51}  -
a_{14} a_{26} a_{32} a_{44}a_{51}  -
 a_{16} a_{23}a_{33}a_{44}a_{51} +
 a_{13} a_{26}a_{33} a_{44} a_{51}  -
 a_{16} a_{22} a_{34} a_{44}a_{51} +
 a_{14} a_{23} a_{36} a_{44}a_{51}  -
 a_{13} a_{24} a_{36} a_{44}a_{51}  -
 a_{16} a_{22} a_{33} a_{45}a_{51}  +
 a_{14} a_{22}a_{36} a_{45}a_{51}  -
 a_{15} a_{24} a_{32} a_{46}a_{51}  +
 a_{14} a_{25} a_{32} a_{46}a_{51}  +
a_{15} a_{23} a_{33} a_{46}a_{51}  -
a_{13}a_{25} a_{33} a_{46}a_{51}  +
 a_{15} a_{22} a_{34} a_{46}a_{51} -
 a_{14}a_{23} a_{35} a_{46}a_{51} +
 a_{13} a_{24} a_{35} a_{46} a_{51}  +
a_{16} a_{22} a_{35} a_{43}a_{52}  -
 a_{15} a_{22} a_{36}a_{43}a_{52}  -
a_{16}a_{22} a_{33} a_{44}a_{52}  +
 a_{14} a_{22} a_{36} a_{44}a_{52}  +
 a_{15} a_{22} a_{33} a_{46}a_{52}  -
 a_{14} a_{22} a_{35} a_{46}a_{52}  -
 a_{16} a_{25} a_{33} a_{41}a_{53}  +
 a_{15} a_{26} a_{33} a_{41}a_{53}  +
 a_{16} a_{24} a_{35} a_{41}a_{53}  -
 a_{14} a_{26} a_{35} a_{41}a_{53}  -
 a_{15} a_{24} a_{36} a_{41}a_{53}  +
 a_{14} a_{25} a_{36} a_{41}a_{53}  +
 a_{16} a_{25} a_{31} a_{43}a_{53}  -
 a_{15} a_{26} a_{31}a_{43}a_{53}  -
 a_{16} a_{21} a_{35} a_{43}a_{53}  +
 a_{11} a_{26} a_{35} a_{43}a_{53}  +
 a_{15} a_{21} a_{36} a_{43} a_{53} -
 a_{11} a_{25} a_{36} a_{43}a_{53}  -
 a_{16} a_{24} a_{31} a_{44}a_{53}  +
 a_{14} a_{26} a_{31} a_{44}a_{53}  +
 a_{16} a_{21} a_{33} a_{44}a_{53}  -
 a_{11} a_{26} a_{33} a_{44}a_{53}  -
 a_{14} a_{21} a_{36} a_{44}a_{53}  +
 a_{11} a_{24} a_{36} a_{44}a_{53}  +
a_{15} a_{24} a_{31} a_{46}a_{53}  -
 a_{14} a_{25} a_{31} a_{46}a_{53}  -
a_{15} a_{21} a_{33} a_{46}a_{53}  +
a_{11} a_{25} a_{33} a_{46}a_{53}  +
 a_{14} a_{21} a_{35} a_{46} a_{53}  -
a_{11} a_{24} a_{35} a_{46}a_{53}  +
 a_{16} a_{25} a_{32} a_{41}a_{55}  -
 a_{15} a_{26} a_{32} a_{41}a_{55}  -
 a_{16} a_{23} a_{35} a_{41}a_{55}  +
 a_{13} a_{26} a_{35} a_{41}a_{55}  +
 a_{15} a_{23} a_{36} a_{41}a_{55}  -
 a_{13} a_{25} a_{36} a_{41}a_{55}  -
 a_{16} a_{25} a_{31} a_{42}a_{55}  +
 a_{15} a_{26} a_{31} a_{42}a_{55}  +
 a_{16} a_{21} a_{35} a_{42}a_{55}  -
 a_{11} a_{26} a_{35} a_{42}a_{55}  -
 a_{15} a_{21} a_{36} a_{42}a_{55}  +
 a_{11} a_{25} a_{36} a_{42}a_{55}  +
 a_{16} a_{23} a_{31} a_{44}a_{55}  -
 a_{13} a_{26} a_{31} a_{44}a_{55}  -
 a_{16} a_{21} a_{32} a_{44}a_{55} +
 a_{11} a_{26} a_{32} a_{44}a_{55}  +
 a_{13} a_{21} a_{36} a_{44}a_{55}  -
 a_{12} a_{22} a_{36} a_{44}a_{55}  -
 a_{11}a_{23} a_{36} a_{44}a_{55}  +
 a_{16} a_{22} a_{31} a_{45}a_{55}  -
 a_{11} a_{22} a_{36} a_{45}a_{55}  -
a_{15}a_{23} a_{31} a_{46}a_{55}  +
 a_{13} a_{25} a_{31} a_{46}a_{55}  +
 a_{15} a_{21} a_{32} a_{46}a_{55}  -
 a_{11} a_{25} a_{32} a_{46}a_{55}  -
 a_{13} a_{21} a_{35} a_{46}a_{55}  +
 a_{12} a_{22} a_{35} a_{46}a_{55}  +
a_{11} a_{23} a_{35} a_{46}a_{55}  -
a_{16} a_{22} a_{35} a_{41}a_{56} +
 a_{15} a_{22} a_{36} a_{41}a_{56}  +
 a_{16} a_{22} a_{31} a_{44}a_{56}  -
a_{11} a_{22} a_{36} a_{44}a_{56}  -
 a_{15} a_{22} a_{31} a_{46}a_{56}  +
 a_{11} a_{22} a_{35} a_{46}a_{56}  -
 a_{16} a_{24} a_{32} a_{41}a_{57}  +
 a_{14} a_{26} a_{32} a_{41}a_{57}  +
 a_{16} a_{23} a_{33} a_{41}a_{57}  -
 a_{13} a_{26} a_{33} a_{41}a_{57}  +
 a_{16} a_{22} a_{34} a_{41}a_{57}  -
 a_{14} a_{23} a_{36} a_{41}a_{57}  +
 a_{13} a_{24} a_{36} a_{41}a_{57}  +
 a_{16} a_{24} a_{31} a_{42}a_{57}  -
 a_{14} a_{26} a_{31} a_{42}a_{57}  -
 a_{16} a_{21} a_{33} a_{42}a_{57}  +
 a_{11} a_{26} a_{33} a_{42}a_{57}  +
 a_{14} a_{21} a_{36} a_{42}a_{57}  -
 a_{11} a_{24} a_{36} a_{42}a_{57}  -
 a_{16} a_{23} a_{31} a_{43}a_{57}  +
 a_{13} a_{26} a_{31} a_{43}a_{57}  +
 a_{16} a_{21} a_{32}a_{43}a_{57}  -
 a_{11} a_{26} a_{32} a_{43}a_{57}  -
 a_{13} a_{21} a_{36} a_{43} a_{57}  +
 a_{12} a_{22} a_{36} a_{43}a_{57}  +
 a_{11} a_{23} a_{36} a_{43}a_{57}  +
 a_{14} a_{23} a_{31} a_{46}a_{57}  -
 a_{13} a_{24} a_{31} a_{46}a_{57}  -
 a_{14} a_{21} a_{32} a_{46}a_{57}  +
 a_{11} a_{24} a_{32} a_{46}a_{57}  +
 a_{13} a_{21} a_{33} a_{46}a_{57}  -
 a_{12} a_{22} a_{33} a_{46}a_{57}  -
 a_{11} a_{23} a_{33} a_{46}a_{57}  -
 a_{11} a_{22}a_{34} a_{46}a_{57}  +
 a_{16} a_{22} a_{33} a_{41}a_{58}  -
 a_{14} a_{22} a_{36} a_{41}a_{58}  -
 a_{16} a_{22} a_{31} a_{43}a_{58}  +
 a_{11} a_{22} a_{36} a_{43}a_{58}  +
 a_{14} a_{22} a_{31} a_{46}a_{58}  -
 a_{11}a_{22} a_{33} a_{46}a_{58}  +
 a_{15} a_{24} a_{32} a_{41}a_{59}  -
 a_{14} a_{25} a_{32} a_{41}a_{59}  -
 a_{15} a_{23} a_{33} a_{41}a_{59}  +
 a_{13} a_{25} a_{33} a_{41}a_{59}  -
 a_{15} a_{22} a_{34} a_{41}a_{59}  +
 a_{14}a_{23}a_{35} a_{41}a_{59}  -
 a_{13} a_{24} a_{35} a_{41}a_{59}  -
 a_{15} a_{24} a_{31} a_{42}a_{59}  +
 a_{14} a_{25} a_{31} a_{42}a_{59}  +
 a_{15} a_{21}a_{33} a_{42}a_{59}  -
 a_{11} a_{25} a_{33} a_{42}a_{59}  -
 a_{14}a_{21} a_{35} a_{42}a_{59}  +
 a_{11} a_{24} a_{35} a_{42}a_{59}  +
 a_{15} a_{23} a_{31} a_{43}a_{59}  -
 a_{13} a_{25} a_{31} a_{43}a_{59}  -
 a_{15} a_{21} a_{32} a_{43}a_{59}  +
 a_{11} a_{25} a_{32} a_{43}a_{59}  +
 a_{13} a_{21} a_{35} a_{43}a_{59}  -
 a_{12} a_{22} a_{35} a_{43}a_{59}  -
 a_{11} a_{23} a_{35}a_{43} a_{59}  -
 a_{14} a_{23} a_{31} a_{ 44}a_{59}  +
 a_{13} a_{24} a_{31} a_{44}a_{59}  +
 a_{14} a_{21} a_{32} a_{44}a_{59}  -
 a_{11} a_{24} a_{32} a_{44}a_{59}  -
 a_{13} a_{21} a_{33} a_{44}a_{59}  +
 a_{12} a_{22} a_{33} a_{44}a_{59}  +
 a_{11} a_{23} a_{33} a_{44}a_{59}  +
 a_{11} a_{22} a_{34} a_{44}a_{59} -
 a_{14} a_{22} a_{31} a_{45}a_{59}  +
 a_{11} a_{22} a_{33}a_{45}a_{59}  -
 a_{15} a_{22} a_{33} a_{41}a_{5(10)}  +
 a_{14} a_{22} a_{35} a_{41} a_{5(10)}  +
 a_{15}a_{22}a_{31} a_{43}a_{5(10)} $

$ R_{11}=a_{16} a_{22} a_{35} a_{43}a_{51} -
 a_{15} a_{22} a_{36} a_{43}a_{51} -
 a_{16} a_{22} a_{33} a_{44}a_{51} +
 a_{14} a_{22} a_{36} a_{44}a_{51} +
 a_{15} a_{22} a_{33} a_{46}a_{51} -
 a_{14} a_{22} a_{35} a_{46}a_{51} -
 a_{16} a_{22} a_{35} a_{41}a_{55} +
 a_{15} a_{22} a_{36} a_{41}a_{55} +
 a_{16} a_{22} a_{31} a_{44}a_{55} -
 a_{11} a_{22} a_{36} a_{44}a_{55} -
 a_{15} a_{22} a_{31} a_{46}a_{55} +
 a_{11} a_{22} a_{35} a_{46}a_{55} +
 a_{16} a_{22} a_{33} a_{41}a_{57} -
 a_{14} a_{22} a_{36} a_{41}a_{57} -
a_{16} a_{22} a_{31} a_{43}a_{57} +
 a_{11}a_{22}a_{36} a_{43}a_{57} +
 a_{14} a_{22} a_{31} a_{46}a_{57} -
 a_{11} a_{22} a_{33} a_{46}a_{57} -
 a_{15} a_{22} a_{33} a_{41}a_{59} +
 a_{14} a_{22} a_{35} a_{41}a_{59} +
 a_{15} a_{22} a_{31} a_{43}a_{59} -
 a_{11} a_{22} a_{35} a_{43}a_{59} -
 a_{14} a_{22} a_{31} a_{44}a_{59} +
 a_{11}a_{22} a_{33} a_{44} a_{59}$\\
\\
   $ a_{11}=\rho_{1}{k^2}c^2+\varepsilon^2k^4c^2-k^2c_{11}$,
 $a_{12}=c_{55}-\varepsilon^2k^2c^2$,
 $a_{13}=ik(c_{13}+c_{55})$,
 $a_{14}=ikB_{1}$,
 $a_{15}=-ik\beta_{1}$,\\
 $a_{16}=-ikb_{1}$,
 $a_{21}=ik(c_{13}+c_{55})$,$a_{22}=\rho{k^2}c^2+\varepsilon^2k^4c^2-k^2c_{55}$,$a_{23}=c_{33}-\varepsilon^2k^2c^2$, $a_{24}=B_{3}$
 $a_{25}=-\beta_{3}$,$a_{26}=-b_{3}$,
 $a_{31}=-ikB_{1}$,$a_{32}=-B_{3}$,$a_{33}=\rho_{1} \chi \varepsilon^2k^4c^2-k^{2} A_{1}-\xi$,$a_{34}=A_{3}-\rho_{1} \chi \varepsilon^2k^2c^2$,
 $a_{35}=\nu_{1}$,$a_{36}=\nu_{2}$,
 $a_{41}=\beta_{1}T_{0}ck^{2}(-1+ikc\tau_{q})$,$a_{42}=ikcT_{0}\beta_{3}(1-ikc\tau_{q})$,$a_{43}=ikac\nu_{1}T_{0}$,\\
 $a_{44}=ik\rho_{1} C_{v}(1-ikc\tau_{q})-(1-ikc\tau_{T})K_{1}k^2$,$a_{45}=(1-ikc\tau_{T})K_{3}$,$a_{46}=ikacT_{0}(1-ikc\tau_{q})$
\\
 $a_{51}=k^4cd_{1}b_{1}(1-ikc\tau_{2})+ik^3b_{1}d^{\ast}_{1}(1-ikc\tau_{3})$,
$a_{52}=-[k^2cd_{3}b_{1}(1-ikc\tau_{2})+ikb_{1}d^{\ast}_{3}(1-ikc\tau_{3})]$,\\
$a_{53}=ik^3cd_{1}b_{3}(1-ikc\tau_{2})-k^2b_{3}d^{\ast}_{1}(1-ikc\tau_{3})$,
$a_{54}=ikcd_{3}b_{3}(1-ikc\tau_{2})-b_{3}d^{\ast}_{3}(1-ikc\tau_{3})$,\\
$a_{55}=-[ik^3c\nu_{2}d_{1}(1-ikc\tau_{2})-k^2\nu_{2}d^{\ast}_{1}(1-ikc\tau_{3})]$,
$a_{56}=[ikc\nu_{2}d_{3}(1-ikc\tau_{2})-\nu_{2}d^{\ast}_{3}(1-ikc\tau_{3})]$,\\
$a_{57}=-[ik^3c\nu_{3}d_{1}(1-ikc\tau_{2})-k^2\nu_{3}d^{\ast}_{1}(1-ikc\tau_{3})]$,
$a_{58}=[ikc\nu_{3}d_{3}(1-ikc\tau_{2})-\nu_{3}d^{\ast}_{3}(1-ikc\tau_{3})]$,\\
$a_{59}=[ik^3cbd_{1}(1-ikc\tau_{2})-k^2bd^{\ast}_{1}(1-ikc\tau_{3})-k^2c^2+ik^3c^3\tau_{2}+ik^5c^3\tau_{2}\varepsilon^2+\frac{k^4c^4\tau^2_{1}}{2}
+\frac{k^6c^4\varepsilon^2\tau^2_{1}}{2}]$,\\
$a_{5(10)}=-[ikcbd_{3}(1-ikc\tau_{2})-bd^{\ast}_{3}(1-ikc\tau_{3})+k^2c^2\varepsilon^2-ik^3c^3\varepsilon^2\tau_{2}-\frac{k^6c^4\varepsilon^2\tau^2_{1}}{2}]$\\

 \section*{Appendix B}
 $R'_{1}=\frac{R'_{7}}{R'_{6}}$,
$R'_{2}=\frac{R'_{8}}{R'_{6}}$,
$R'_{3}=\frac{R'_{9}}{R'_{6}}$,
$R'_{4}=\frac{R'_{10}}{R'_{6}}$,
$R'_{5}=\frac{R'_{11}}{R'_{6}}$,

$R'_{6}=-a'_{12} a'_{26} a'_{34} a'_{45} a'_{54}  +
    a'_{12} a'_{23} a'_{34} a'_{45} a'_{5(10)} $

$R'_{7}=-a'_{16} a'_{23} a'_{34} a'_{45} a'_{52} +
    a'_{13} a'_{26} a'_{34} a'_{45} a'_{52}  -
    a'_{12} a'_{26} a'_{34} a'_{45} a'_{53}  -
    a'_{12} a'_{26} a'_{34} a'_{44} a'_{54}  -
    a'_{12} a'_{26} a'_{33} a'_{45} a'_{54}  +
    a'_{16} a'_{21} a'_{34} a'_{45} a'_{54}  -
    a'_{11} a'_{26} a'_{34} a'_{45} a'_{54}  +
    a'_{12} a'_{24} a'_{36} a'_{45} a'_{54}  +
    a'_{12} a'_{25} a'_{34} a'_{46} a'_{54}  +
    a'_{12} a'_{26} a'_{32} a'_{45} a'_{56}  -
    a'_{12} a'_{23} a'_{36} a'_{45} a'_{56}  +
    a'_{12} a'_{26} a'_{34} a'_{42} a'_{58}  -
    a'_{12} a'_{23} a'_{34} a'_{46} a'_{58}+
    a'_{12} a'_{23} a'_{34} a'_{45} a'_{59}-
    a'_{12} a'_{25} a'_{34} a'_{42} a'_{5(10)}+
    a'_{12} a'_{23} a'_{34} a'_{44} a'_{5(10)}-\\
    a'_{12} a'_{24} a'_{32} a'_{45} a'_{5(10)}+
    a'_{12} a'_{23} a'_{33} a'_{45} a'_{5(10)}-
    a'_{13} a'_{21} a'_{34} a'_{45} a'_{5(10)}  +
    a'_{12} a'_{22} a'_{34} a'_{45} a'_{5(10)}  +
    a'_{11} a'_{23} a'_{34} a'_{45} a'_{5(10)} $

 $R'_{8}=-a'_{16} a'_{23} a'_{34} a'_{45} a'_{51}  +
    a'_{13} a'_{26} a'_{34} a'_{45} a'_{51}  +
    a'_{16} a'_{25} a'_{34} a'_{42} a'_{52}  -
    a'_{15} a'_{26} a'_{34} a'_{42} a'_{52}  -
    a'_{16} a'_{23} a'_{34} a'_{44} a'_{52}  +
    a'_{13} a'_{26} a'_{34} a'_{44} a'_{52}  +
    a'_{16} a'_{24} a'_{32} a'_{45} a'_{52}  -
    a'_{14} a'_{26} a'_{32} a'_{45} a'_{52}  -
    a'_{16} a'_{23} a'_{33} a'_{45} a'_{52}  +
    a'_{13} a'_{26} a'_{33} a'_{45} a'_{52}  -
    a'_{16} a'_{22} a'_{34} a'_{45} a'_{52} +
    a'_{14} a'_{23} a'_{36} a'_{45} a'_{52}  -
    a'_{13} a'_{24} a'_{36} a'_{45} a'_{52}  +
    a'_{15} a'_{23} a'_{34} a'_{46} a'_{52}  -
    a'_{13} a'_{25} a'_{34} a'_{46} a'_{52}  -
    a'_{12} a'_{26} a'_{34} a'_{44} a'_{53}  -
    a'_{12} a'_{26} a'_{33} a'_{45} a'_{53}  +
    a'_{16} a'_{21} a'_{34} a'_{45} a'_{53}  -
    a'_{11} a'_{26} a'_{34} a'_{45} a'_{53}  +
    a'_{12} a'_{24} a'_{36} a'_{45} a'_{53}  +
    a'_{12} a'_{25} a'_{34} a'_{46} a'_{53}  -
    a'_{16} a'_{25} a'_{34} a'_{41} a'_{54}  +
    a'_{15} a'_{26} a'_{34} a'_{41} a'_{54}  +
    a'_{12} a'_{26} a'_{35} a'_{43} a'_{54}  -
    a'_{12} a'_{25} a'_{36} a'_{43} a'_{54}  -
    a'_{12} a'_{26} a'_{33} a'_{44} a'_{54} +
    a'_{16} a'_{21} a'_{34} a'_{44} a'_{54}  -
    a'_{11} a'_{26} a'_{34} a'_{44} a'_{54}  +
    a'_{12} a'_{24} a'_{36} a'_{44} a'_{54}  -
    a'_{16} a'_{24} a'_{31} a'_{45} a'_{54}  +
    a'_{14} a'_{26} a'_{31} a'_{45} a'_{54}  +
    a'_{16} a'_{21} a'_{33} a'_{45} a'_{54}  -
    a'_{11} a'_{26} a'_{33} a'_{45} a'_{54}  -
    a'_{14} a'_{21} a'_{36} a'_{45} a'_{54}  +
    a'_{11} a'_{24} a'_{36} a'_{45} a'_{54}  +
    a'_{12} a'_{25} a'_{33} a'_{46} a'_{54}  -
    a'_{15} a'_{21} a'_{34} a'_{46} a'_{54}  +
    a'_{11} a'_{25} a'_{34} a'_{46} a'_{54}  -
    a'_{12} a'_{24} a'_{35} a'_{46} a'_{54}  +
    a'_{12} a'_{26} a'_{32} a'_{45} a'_{55}  -
    a'_{12} a'_{23} a'_{36} a'_{45} a'_{55}  -
    a'_{12} a'_{26} a'_{35} a'_{42} a'_{56}  +
    a'_{12} a'_{25} a'_{36} a'_{42} a'_{56}  +
    a'_{12} a'_{26} a'_{32} a'_{44} a'_{56}  -
    a'_{12} a'_{23} a'_{36} a'_{44} a'_{56}  +
    a'_{16} a'_{23} a'_{31} a'_{45} a'_{56}  -
    a'_{13} a'_{26} a'_{31} a'_{45} a'_{56}  -
    a'_{16} a'_{21} a'_{32} a'_{45} a'_{56}  +
    a'_{11} a'_{26} a'_{32} a'_{45} a'_{56}  +
    a'_{13} a'_{21} a'_{36} a'_{45} a'_{56}  -
    a'_{12} a'_{22} a'_{36} a'_{45} a'_{56}  -
    a'_{11} a'_{23} a'_{36} a'_{45} a'_{56}  -
    a'_{12} a'_{25} a'_{32} a'_{46} a'_{56}  +
    a'_{12} a'_{23} a'_{35} a'_{46} a'_{56}  +
    a'_{12} a'_{26} a'_{34} a'_{42} a'_{57}  -
    a'_{12} a'_{23} a'_{34} a'_{46} a'_{57}  +
    a'_{16} a'_{23} a'_{34} a'_{41} a'_{58}  -
    a'_{13} a'_{26} a'_{34} a'_{41} a'_{58}  +
    a'_{12} a'_{26} a'_{33} a'_{42} a'_{58}  -
    a'_{16} a'_{21} a'_{34} a'_{42} a'_{58}  +
    a'_{11} a'_{26} a'_{34} a'_{42} a'_{58}  -
    a'_{12} a'_{24} a'_{36} a'_{42} a'_{58}  -
    a'_{12} a'_{26} a'_{32} a'_{43} a'_{58}  +
    a'_{12} a'_{23} a'_{36} a'_{43} a'_{58}  +
    a'_{12} a'_{24} a'_{32} a'_{46} a'_{58}  -
    a'_{12} a'_{23} a'_{33} a'_{46} a'_{58}  +
    a'_{13} a'_{21} a'_{34} a'_{46} a'_{58}  -
    a'_{12} a'_{22} a'_{34} av_{46} a'_{58}  -
    a'_{11} a'_{23} a'_{34} a'_{46} a'_{58}  -
    a'_{12} a'_{25} a'_{34} a'_{42} a'_{59}  +
    a'_{12} a'_{23} a'_{34} a'_{44} a'_{59}  -
    a'_{12} a'_{24} a'_{32} a'_{45} a'_{59} +
    a'_{12} a'_{23} a'_{33} a'_{45} a'_{59}  -
    a'_{13} a'_{21} a'_{34} a'_{45} a'_{59}  +
    a'_{12} a'_{22} a'_{34} a'_{45} a'_{59}  +
    a'_{11} a'_{23} a'_{34} a'_{45} a'_{59}  -
    a'_{15} a'_{23} a'_{34} a'_{41} a'_{5(10)}  +
    a'_{13} a'_{25} a'_{34} a'_{41} a'_{5(10)}  -
    a'_{12} a'_{25} a'_{33} a'_{42} a'_{5(10)}  +
    a'_{15} a'_{21} a'_{34} a'_{42} a'_{5(10)}  -
    a'_{11} av_{25} a'_{34} a'_{42} a'_{5(10)}  +
    a'_{12} a'_{24} a'_{35} a'_{42} a'_{5(10)}  +
    a'_{12} a'_{25} a'_{32} a'_{43} a'_{5(10)}  -
    a'_{12} a'_{23} a'_{35} a'_{43} a'_{5(10)}  -
    a'_{12} a'_{24} a'_{32} a'_{44} a'_{5(10)}  +
    a'_{12} a'_{23} a'_{33} a'_{44} a'_{5(10)}  -
    a'_{13} a'_{21} a'_{34} a'_{44} a'_{5(10)}  +
    a'_{12} a'_{22} a'_{34} a'_{44} a'_{5(10)}  +
    a'_{11} a'_{23} a'_{34} a'_{44} a'_{5(10)}  -
    a'_{14} a'_{23} a'_{31} a'_{45} a'_{5(10)}  +
    a'_{13} a'_{24} a'_{31} a'_{45} a'_{5(10)}  +
    a'_{14} a'_{21} a'_{32} a'_{45} a'_{5(10)}  -
    a'_{11} a'_{24} a'_{32} a'_{45} a'_{5(10)}  -
    a'_{13} a'_{21} a'_{33} a'_{45} a'_{5(10)}  +
    a'_{12} a'_{22} a'_{33} a'_{45} a'_{5(10)}  +
    a'_{11} a'_{23} a'_{33} a'_{45} a'_{5(10)}  +
    a'_{11} a'_{22} a'_{34} a'_{45} a'_{5(10)} $

$R_{9}=a'_{16} a'_{25} a'_{34} a'_{42}a'_{51}  -
 a'_{15} a'_{26} a'_{34} a'_{42} a'_{51}  -
 a'_{16} a'_{23} a'_{34} a'_{44} a'_{51}  +
 a'_{13} a'_{26} a'_{34} a'_{44} a'_{51}  +
 a'_{16} a'_{24} a'_{32} a'_{45} a'_{51}  -
 a'_{14} a'_{26} a'_{32} a'_{45}a'_{51}  -
 a'_{16} a'_{23} a'_{33} a'_{45}a'_{51}  +
 a'_{13} a'_{26} a'_{33} a'_{45} a'_{51}  -
 a'_{16} a'_{22} a'_{34} a'_{45}a'_{51}  +
 a'_{14} a'_{23} a'_{36} a'_{45}a'_{51}  -
 a'_{13} a'_{24} a'_{36} a_{45} a'_{51}  +
 a'_{15} a'_{23} a'_{34} a'_{46}a'_{51}  -
 a'_{13} a'_{25} a'_{34} a'_{46}a'_{51}  +
 a'_{16} a'_{25} a'_{33} a'_{42} a'_{52}  -
 a'_{15} a'_{26} a'_{33} a'_{42}a'_{52}  -
 a'_{16} a'_{24} a'_{35} a'_{42}a'_{52}  +
 a'_{14} a'_{26} a'_{35} a'_{42}a'_{52}  +
 a'_{15} a'_{24} a'_{36} a'_{42}a'_{52}  -
 a'_{14} a'_{25} a'_{36} a'_{42}a'_{52}  -
 a'_{16} a'_{25} a'_{32} a'_{43}a'_{52}  +
 a'_{15} a'_{26} a'_{32} a'_{43}a'_{52}  +
 a'_{16} a'_{23} a'_{35} a'_{43}a'_{52} -
 a'_{13} a'_{26} a'_{35} a'_{43} a'_{52}  -
 a'_{15} a'_{23} a'_{36} a'_{43}a'_{52}  +
 a'_{13} a'_{25} a'_{36} a'_{43} a'_{52}  +
 a'_{16} a'_{24} a'_{32} a'_{44} a'_{52}  -
 a'_{14} a'_{26} a'_{32} a'_{44} a'_{52}  -
 a'_{16} a'_{23} a'_{33} a'_{44} a'_{52}  +
 a'_{13} a'_{26} a'_{33} a'_{44} a'_{52}  -
 a'_{16} a'_{22} a'_{34} a'_{44} a'_{52}  +
 a'_{14} a'_{23} a'_{36} a'_{44} a'_{52}  -
 a'_{13} a'_{24} a'_{36} a'_{44} a'_{52}  -
 a'_{16} a'_{22} a'_{33} a'_{45} a'_{52}  +
 a'_{14} a'_{22} a'_{36} a'_{45} a'_{52}  -
 a'_{15} a'_{24} a'_{32} a'_{46} a'_{52}  +
 a'_{14} a'_{25} a'_{32} a'_{46} a'_{52}  +
 a'_{15} a'_{23} a'_{33} a'_{46} a'_{52}  -
 a'_{13} a'_{25} a'_{33} a'_{46} a'_{52}  +
 a'_{15} a'_{22} a'_{34} a'_{46} a'_{52}  -
 a'_{14} a'_{23} a'_{35} a'_{46} a'_{52}  +
 a'_{13} a'_{24} a'_{35} a'_{46} a'_{52}  -
 a'_{16} a'_{25} a'_{34} a'_{41} a'_{53}  +
 a'_{15} a'_{26} a'_{34} a'_{41} a'_{53}  +
 a'_{12} a'_{26} a'_{35} a'_{43} a'_{53}  -
 a'_{12} a'_{25} a'_{36} a'_{43} a'_{53}  -
 a'_{12} a'_{26} a'_{33} a'_{44} a'_{53}  +
 a'_{16} a'_{21} a'_{34} a'_{44} a'_{53}  -
 a'_{11} a'_{26} a'_{34} a'_{44} a'_{53}  +
 a'_{12} a'_{24} a'_{36} a'_{44} a'_{53}  -
 a'_{16} a'_{24} a'_{31} a'_{45} a'_{53}  +
 a'_{14} a'_{26} a'_{31} a'_{45} a'_{53}  +
 a'_{16} a'_{21} a'_{33} a'_{45} a'_{53}  -
 a'_{11} a'_{26} a'_{33} a'_{45} a'_{53}  -
 a'_{14} a'_{21} a'_{36} a'_{45} a'_{53}  +
 a'_{11} a'_{24} a'_{36} a'_{45} a'_{53}  +
 a'_{12} a'_{25} a'_{33} a'_{46} a'_{53}  -
 a'_{15} a'_{21} a'_{34} a'_{46} a'_{53}  +
 a'_{11} a'_{25} a'_{34} a'_{46} a'_{53}  -
 a'_{12} a'_{24} a'_{35} a'_{46} a'_{53}  -
 a'_{16} a'_{25} a'_{33} a'_{41} a'_{54}  +
 a'_{15} a'_{26} a'_{33} a'_{41} a'_{54}  +
 a'_{16} a'_{24} a'_{35} a'_{41} a'_{54}  -
 a'_{14} a'_{26} a'_{35} a'_{41} a'_{54}  -
 a'_{15} a'_{24} a'_{36} a'_{41} a'_{54}  +
 a'_{14} a'_{25} a'_{36} a'_{41} a'_{54}  +
 a'_{16} a'_{25} a'_{31} a'_{43} a'_{54}  -
 a'_{15} a'_{26} a'_{31} a'_{43} a'_{54}  -
 a_{16}  a'_{21} a'_{35} a'_{43} a'_{54}  +
 a'_{11} a'_{26} a'_{35} a'_{43} a'_{54}  +
 a'_{15} a'_{21} a'_{36} a'_{43} a'_{54}  -
 a'_{11} a'_{25} a'_{36} a'_{43} a'_{54}  -
 a'_{12} a'_{26} a'_{35} a'_{42} a'_{55}  +
 a'_{12} a'_{25} a'_{36} a'_{42} a'_{55}  +
 a'_{12} a'_{26} a'_{32} a'_{44} a'_{55}  -
 a'_{12} a'_{23} a'_{36} a'_{44} a'_{55}  +
 a'_{16} a'_{23} a'_{31} a'_{45} a'_{55}  -
 a'_{13} a'_{26} a'_{31} a'_{45} a'_{55}  -
 a'_{16} a'_{21} a'_{32} a'_{45} a'_{55}  +
 a'_{11} a'_{26} a'_{32} a'_{45} a'_{55}  +
 a'_{13} a'_{21} a'_{36} a'_{45} a'_{55}  -
 a'_{12} a'_{22} a'_{36} a'_{45} a'_{55}  -
 a'_{11} a'_{23} a'_{36} a'_{45} a'_{55}  -
 a'_{12} a'_{25} a'_{32} a'_{46} a'_{55}  +
 a'_{12} a'_{23} a'_{35} a'_{46} a'_{55}  +
 a'_{16} a'_{25} a'_{32} a'_{41} a'_{56}  -
 a'_{15} a'_{26} a'_{32} a'_{41} a'_{56}  -
 a'_{16} a'_{ 23} a'_{35} a'_{41} a'_{56}  +
 a'_{13} a'_{26} a'_{35} a'_{41} a'_{56}  +
 a'_{15} a'_{23} a'_{36} a'_{41} a'_{56}  -
 a'_{13} a'_{25} a'_{36} a'_{41} a'_{56} -
 a'_{16} a'_{25} a'_{31} a'_{42} a'_{56}  +
 a'_{15} a'_{26} a'_{31} a'_{42} a'_{56} +
 a'_{16} a'_{21} a'_{35} a'_{42} a'_{56}  -
 a'_{11} a'_{26} a'_{35} a'_{42} a'_{56}  -
 a'_{15} a'_{21} a'_{36} a'_{42} a'_{56}  +
 a'_{11} a'_{25} a'_{36} a'_{42} a'_{56}  +
 a'_{16} a'_{23} a'_{31} a'_{44} a'_{56}  -
 a'_{13} a'_{26} a'_{31} a'_{44} a'_{56}  -
 a'_{16} a'_{21} a'_{32} a'_{44} a'_{56} +
 a'_{11} a'_{26} a'_{32} a'_{44} a'_{56} +
 a'_{13} a'_{21} a'_{36} a'_{44} a'_{56}  -
 a'_{12} a'_{22} a'_{36} a'_{44} a'_{56}  -
 a'_{11} a'_{23} a'_{36} a'_{44} a'_{56}  +
 a'_{16} a'_{22} a'_{31} a'_{45} a'_{56}  -
 a'_{11} a'_{22} a'_{36} a'_{45} a'_{56}  -
 a'_{15} a'_{23} a'_{31} a'_{46} a'_{56}  +
 a'_{13} a'_{25} a'_{31} a'_{46} a'_{56}  +
 a'_{15} a'_{21} a'_{32} a'_{46} a'_{56}  -
 a'_{11} a'_{25} a'_{32} a'_{46} a'_{56}  -
 a'_{13} a'_{21} a'_{35} a'_{46} a'_{56}  +
 a'_{12} a'_{22} a'_{35} a'_{46} a'_{56}  +
 a'_{11} a'_{23} a'_{35} a'_{46} a'_{56}  -
 a'_{16} a'_{23} a'_{34} a'_{41} a'_{57}  -
 a'_{13} a'_{26} a'_{34} a'_{41} a'_{57} +
 a'_{12} a'_{26} a'_{33} a'_{42} a'_{57}  -
 a'_{16} a'_{21} a'_{34} a'_{42} a'_{57}  +
 a'_{11} a'_{26} a'_{34} a'_{42} a'_{57}  -
 a'_{12} a'_{24} a'_{36} a'_{42} a'_{57}  -
 a'_{12} a'_{26} a'_{32} a'_{43} a'_{57}  +
 a'_{12} a'_{23} a'_{36} a'_{43} a'_{57}  +
 a'_{12} a'_{24} a'_{32} a'_{46} a'_{57}  -
 a'_{12} a'_{23} a'_{33} a'_{46} a'_{57}  +
 a'_{13} a'_{21} a'_{34} a'_{46} a'_{57}  -
 a'_{12} a'_{22} a'_{34} a'_{46} a'_{57}  -
 a'_{11} a'_{23} a'_{34} a'_{46} a'_{57}  -
 a'_{16} a'_{24} a'_{32} a'_{41} a'_{58}  +
 a'_{14} a'_{26} a'_{32} a'_{41} a'_{58}  +
 a'_{16} a'_{23} a'_{33} a'_{41} a'_{58}  -
 a'_{13} a'_{26} a'_{33} a'_{41} a'_{58}  +
 a'_{16} a'_{22} a'_{34} a'_{41} a'_{58}  -
 a'_{14} a'_{23} a'_{36} a'_{41} a'_{58}  +
 a'_{13} a'_{24} a'_{36} a'_{41} a'_{58}  +
 a'_{16} a'_{24} a'_{31} a'_{42} a'_{58}  -
 a'_{14} a'_{26} a'_{31} a'_{42} a'_{58}  -
 a'_{16} a'_{21} a'_{33} a'_{42} a'_{58}  +
 a'_{11} a'_{26} a'_{33} a'_{42} a'_{58}  +
 a'_{14} a'_{21} a'_{36} a'_{42} a'_{58}  -
 a'_{11} a'_{24} a'_{36} a'_{42} a'_{58}  -
 a'_{16} a'_{23} a'_{31} a'_{43} a'_{58}  +
 a'_{13} a'_{26} a'_{31} a'_{43} a'_{58}  +
 a'_{16} a'_{21} a'_{32} a'_{43} a'_{58}  -
 a'_{11} a'_{26} a'_{32} a'_{43} a'_{58}  -
 a'_{13} a'_{21} a'_{36} a'_{43} a'_{58}  +
 a'_{12} a'_{22} a'_{36} a'_{43} a'_{58}  +
 a'_{11} a'_{23} a'_{36} a'_{43} a'_{58}  +
 a'_{14} a'_{23} a'_{31} a'_{46} a'_{58}  -
 a'_{13} a'_{24} a'_{31} a'_{46} a'_{58}  -
 a'_{14} a'_{21} a'_{32} a'_{46} a'_{58}  +
 a'_{11} a'_{24} a'_{32} a'_{46} a'_{58}  +
 a'_{13} a'_{21} a'_{33} a'_{46} a'_{58}  -
 a'_{12} a'_{22} a'_{33} a'_{46} a'_{58}  -
 a'_{11} a'_{23} a'_{33} a'_{46} a'_{58}  -
 a'_{11} a'_{22} a'_{34} a'_{46} a'_{58}  -
 a'_{15} a_{23}  a'_{34} a'_{41} a'_{59}  +
 a'_{13} a'_{25} a'_{34} a'_{41} a'_{59}  -
 a'_{12} a'_{25} a'_{33} a'_{42} a'_{59}  +
 a'_{15} a'_{21} a'_{34} a'_{42} a'_{59}  -
 a'_{11} a'_{25} a'_{34} a'_{42} a'_{59}  +
 a'_{12} a'_{24} a'_{35} a'_{42} a'_{59}  +
 a'_{12} a'_{25} a'_{32} a'_{43} a'_{59}  -
 a'_{12} a'_{23} a'_{35} a'_{43} a'_{59}  +
 a'_{12} a'_{24} a'_{32} a'_{44} a'_{59} -
 a'_{12} a'_{23} a'_{33} a'_{44} a'_{59}  +
 a'_{13} a'_{21} a'_{34} a'_{44} a'_{59}  +
 a'_{12} a'_{22} a'_{34} a'_{44} a'_{59}  +
 a'_{11} a'_{23} a'_{34} a'_{44} a'_{59}  -
 a'_{14} a'_{23} a'_{31} a'_{45} a'_{59}  +
 a'_{13} a'_{24} a'_{31} a'_{45} a'_{59}  +
 a'_{14} a'_{21} a'_{32} a'_{45} a'_{59}  -
 a'_{11} a'_{24} a'_{32} a'_{45} a'_{59}  -
 a'_{13} a'_{21} a'_{33} a'_{45} a'_{59}  +
 a'_{12} a'_{22} a'_{33} a'_{45} a'_{59}  +
 a'_{11} a'_{23} a'_{33} a'_{45} a'_{59}  +
 a'_{11} a'_{22} a'_{34} a'_{45} a'_{59}  +
 a'_{14} a'_{22} a'_{31} a'_{45} a'_{5(10)}  -
 a'_{11} a'_{22} a'_{33} a'_{45} a'_{5(10)} $

$R'_{10}=a'_{16}a'_{25}a'_{33}a'_{42}a'_{51} -
 a'_{15}a'_{26}a'_{33} a'_{42} a'_{51} -
 a'_{16} a'_{24}a'_{35}a'_{42}a'_{51}  +
 a'_{14}a'_{26} a'_{35}a'_{42}a'_{51} +
 a'_{15} a'_{24} a'_{36}a'_{42}a'_{51} -
 a'_{14}a'_{25} a'_{36} a'_{42}a'_{51}  -
a'_{16} a'_{25}a'_{32} a'_{43}a'_{51}  +
 a'_{15} a'_{26} a'_{32}a'_{43}a'_{51}  +
 a'_{16} a'_{23} a'_{35} a'_{43}a'_{51}  -
 a'_{13}a'_{26} a'_{35} a'_{43} a'_{51}  -
 a'_{15}a'_{23} a'_{36} a'_{43}a'_{51}  +
 a'_{13} a'_{25} a'_{36} a'_{43}a'_{51}  +
 a'_{16} a_{24} a'_{32} a'_{44}a'_{51}  -
a'_{14} a_{26} a'_{32} a'_{44}a'_{51}  -
 a'_{16} a_{23}a'_{33}a'_{44}a'_{51} +
 a'_{13} a'_{26}a'_{33} a'_{44} a'_{51}  -
 a'_{16} a'_{22} a'_{34} a'_{44}a'_{51} +
 a'_{14} a'_{23} a'_{36} a'_{44}a'_{51}  -
 a'_{13} a'_{24} a'_{36} a'_{44}a'_{51}  -
 a'_{16} a'_{22} a'_{33} a'_{45}a'_{51}  +
 a'_{14} a'_{22}a'_{36} a'_{45}a'_{51}  -
 a'_{15} a'_{24} a'_{32} a'_{46}a'_{51}  +
 a'_{14} a'_{25} a'_{32} a'_{46}a'_{51}  +
a'_{15} a'_{23} a_{33} a'_{46}a'_{51}  -
a'_{13}a'_{25} a_{33} a'_{46}a'_{51}  +
 a'_{15} a'_{22} a'_{34} a'_{46}a'_{51} -
 a'_{14}a'_{23} a'_{35} a_{46}a_{51} +
 a'_{13} a'_{24} a'_{35} a'_{46} a'_{51}  +
a'_{16} a'_{22} a'_{35} a'_{43}a'_{52}  -
 a'_{15} a'_{22} a'_{36}a'_{43}a'_{52}  -
a'_{16}a'_{22} a'_{33} a'_{44}a'_{52}  +
 a'_{14} a'_{22} a'_{36} a'_{44}a'_{52}  +
 a'_{15} a'_{22} a'_{33} a'_{46}a'_{52}  -
 a'_{14} a'_{22} a'_{35} a'_{46}a'_{52}  -
 a'_{16} a'_{25} a'_{33} a'_{41}a'_{53}  +
 a'_{15} a'_{26} a'_{33} a'_{41}a'_{53}  +
 a'_{16} a'_{24} a'_{35} a'_{41}a'_{53}  -
 a'_{14} a'_{26} a'_{35} a'_{41}a'_{53}  -
 a'_{15} a'_{24} a'_{36} a'_{41}a'_{53}  +
 a'_{14} a'_{25} a'_{36} a'_{41}a'_{53}  +
 a'_{16} a'_{25} a'_{31} a'_{43}a'_{53}  -
 a'_{15} a'_{26} a'_{31}a'_{43}a'_{53}  -
 a'_{16} a'_{21} a'_{35} a'_{43}a'_{53}  +
 a'_{11} a'_{26} a'_{35} a'_{43}a'_{53}  +
 a'_{15} a'_{21} a'_{36} a'_{43} a'_{53} -
 a'_{11} a'_{25} a'_{36} a'_{43}a'_{53}  -
 a'_{16} a'_{24} a'_{31} a'_{44}a'_{53}  +
 a'_{14} a'_{26} a'_{31} a'_{44}a'_{53}  +
 a'_{16} a'_{21} a'_{33} a'_{44}a'_{53}  -
 a'_{11} a'_{26} a'_{33} a'_{44}a'_{53}  -
 a'_{14} a'_{21} a'_{36} a'_{44}a'_{53}  +
 a'_{11} a'_{24} a'_{36} a'_{44}a'_{53}  +
 a'_{15} a'_{24} a'_{31} a'_{46}a'_{53}  -
 a'_{14} a'_{25} a'_{31} a'_{46}a'_{53}  -
 a'_{15} a'_{21} a'_{33} a'_{46}a'_{53}  +
 a'_{11} a'_{25} a'_{33} a'_{46}a'_{53}  +
 a'_{14} a'_{21} a'_{35} a'_{46} a'_{53}  -
 a'_{11} a'_{24} a'_{35} a'_{46}a'_{53}  +
 a'_{16} a'_{25} a'_{32} a'_{41}a'_{55}  -
 a'_{15} a_{26}  a'_{32}  a'_{41}a'_{55}  -
 a'_{16} a'_{23} a'_{35} a'_{41}a'_{55}  +
 a'_{13} a'_{26} a'_{35} a'_{41}a'_{55}  +
 a'_{15} a'_{23} a'_{36} a'_{41}a'_{55}  -
 a'_{13} a'_{25} a'_{36} a'_{41}a'_{55}  -
 a'_{16} a'_{25} a'_{31} a'_{42}a'_{55}  +
 a'_{15} a'_{26} a'_{31} a'_{42}a'_{55}  +
 a'_{16} a'_{21} a'_{35} a'_{42}a'_{55}  -
 a'_{11} a'_{26} a'_{35} a'_{42}a'_{55}  -
 a'_{15} a'_{21} a'_{36} a'_{42}a'_{55}  +
 a'_{11} a'_{25} a'_{36} a'_{42}a'_{55}  +
 a'_{16} a'_{23} a'_{31} a'_{44}a'_{55}  -
 a'_{13} a'_{26} a'_{31} a'_{44}a'_{55}  -
 a'_{16} a'_{21} a'_{32} a'_{44}a'_{55} +
 a'_{11} a'_{26} a'_{32} a'_{44}a'_{55}  +
 a'_{13} a'_{21} a'_{36} a'_{44}a'_{55}  -
 a'_{12} a'_{22} a'_{36} a_{44}a'_{55}  -
 a'_{11}a'_{23} a'_{36} a'_{44}a'_{55}  +
 a'_{16} a'_{22} a'_{31} a'_{45}a'_{55}  -
 a'_{11} a'_{22} a'_{36} a'_{45}a'_{55}  -
 a'_{15}a'_{23} a'_{31} a'_{46}a'_{55}  +
 a'_{13} a'_{25} a'_{31} a'_{46}a'_{55}  +
 a'_{15} a'_{21} a'_{32} a'_{46}a'_{55}  -
 a'_{11} a'_{25} a'_{32} a'_{46}a'_{55}  -
 a'_{13} a'_{21} a'_{35} a'_{46}a'_{55}  +
 a'_{12} a'_{22} a'_{35} a'_{46}a'_{55}  +
 a'_{11} a'_{23} a'_{35} a'_{46}a'_{55}  -
 a'_{16} a'_{22} a'_{35} a'_{41}a'_{56} +
 a'_{15} a'_{22} a'_{36} a'_{41}a'_{56}  +
 a'_{16} a'_{22} a'_{31} a'_{44}a'_{56}  -
 a'_{11} a'_{22}  a'_{36} a'_{44}a'_{56}  -
 a'_{15} a'_{22} a'_{31} a'_{46}a'_{56}  +
 a'_{11} a'_{22} a'_{35} a'_{46}a'_{56}  -
 a'_{16} a'_{24} a'_{32} a'_{41}a'_{57}  +
 a'_{14} a'_{26} a'_{32} a'_{41}a'_{57}  +
 a'_{16} a'_{23} a'_{33} a'_{41}a'_{57}  -
 a'_{13} a'_{26} a'_{33} a'_{41}a'_{57}  +
 a'_{16} a'_{22} a'_{34} a'_{41}a'_{57}  -
 a'_{14} a'_{23} a'_{36} a'_{41}a'_{57}  +
 a'_{13} a'_{24} a'_{36} a'_{41}a'_{57}  +
 a'_{16} a'_{24} a'_{31} a'_{42}a'_{57}  -
 a'_{14} a'_{26} a'_{31} a'_{42}a'_{57}  -
 a'_{16} a'_{21} a'_{33} a'_{42}a'_{57}  +
 a'_{11} a'_{26} a'_{33} a'_{42}a'_{57}  +
 a'_{14} a'_{21} a'_{36} a'_{42}a'_{57}  -
 a'_{11} a'_{24} a'_{36} a'_{42}a'_{57}  -
 a'_{16} a'_{23} a'_{31} a'_{43}a'_{57}  +
 a'_{13} a'_{26} a'_{31} a'_{43}a'_{57}  +
 a'_{16} a'_{21} a'_{32}a'_{43}a'_{57}  -
 a'_{11} a'_{26} a'_{32} a'_{43}a'_{57}  -
 a'_{13} a'_{21} a'_{36} a'_{43} a'_{57}  +
 a'_{12} a'_{22} a'_{36} a'_{43}a'_{57}  +
 a'_{11} a'_{23} a'_{36} a'_{43}a'_{57}  +
 a'_{14} a'_{23} a'_{31} a'_{46}a'_{57}  -
 a'_{13} a'_{24} a'_{31} a'_{46}a_{57}  -
 a'_{14} a'_{21} a'_{32} a'_{46}a'_{57}  +
 a'_{11} a'_{24} a'_{32} a'_{46}a'_{57}  +
 a'_{13} a'_{21} a'_{33} a'_{46}a'_{57}  -
 a'_{12} a'_{22} a'_{33} a'_{46}a'_{57}  -
 a'_{11} a'_{23} a'_{33} a'_{46}a'_{57}  -
 a'_{11} a'_{22}a'_{34} a'_{46}a'_{57}  +
 a'_{16} a'_{22} a'_{33} a'_{41}a'_{58}  -
 a'_{14} a'_{22} a'_{36} a'_{41}a'_{58}  -
 a'_{16} a'_{22} a'_{31} a'_{43}a'_{58}  +
 a'_{11} a'_{22} a'_{36} a'_{43}a'_{58}  +
 a'_{14} a'_{22} a'_{31} a'_{46}a'_{58}  -
 a'_{11} a'_{22} a'_{33} a'_{46}a'_{58}  +
 a'_{15} a'_{24} a'_{32} a'_{41}a'_{59}  -
 a'_{14} a'_{25} a'_{32} a'_{41}a'_{59}  -
 a'_{15} a'_{23} a'_{33} a'_{41}a'_{59}  +
 a'_{13} a'_{25} a'_{33} a'_{41}a'_{59}  -
 a'_{15} a'_{22} a'_{34} a'_{41}a'_{59}  +
 a'_{14} a'_{23} a'_{35} a'_{41}a'_{59}  -
 a'_{13} a'_{24} a'_{35} a'_{41}a'_{59}  -
 a'_{15} a'_{24} a'_{31} a'_{42}a'_{59}  +
 a'_{14} a'_{25} a'_{31} a'_{42}a'_{59}  +
 a'_{15} a'_{21} a'_{33} a'_{42}a'_{59}  -
 a'_{11} a'_{25} a'_{33} a'_{42}a'_{59}  -
 a'_{14} a'_{21} a'_{35} a'_{42}a'_{59}  +
 a'_{11} a'_{24} a'_{35} a'_{42}a'_{59}  +
 a'_{15} a'_{23} a'_{31} a'_{43}a'_{59}  -
 a'_{13} a'_{25} a'_{31} a'_{43}a'_{59}  -
 a'_{15} a'_{21} a'_{32} a'_{43}a'_{59}  +
 a'_{11} a'_{25} a'_{32} a'_{43}a'_{59}  +
 a'_{13} a'_{21} a'_{35} a'_{43}a'_{59}  -
 a'_{12} a'_{22} a'_{35} a'_{43}a'_{59}  -
 a'_{11} a'_{23} a'_{35} a'_{43} a'_{59}  -
 a'_{14} a'_{23} a'_{31} a'_{ 44}a'_{59}  +
 a'_{13} a'_{24} a'_{31} a'_{44}a'_{59}  +
 a'_{14} a'_{21} a'_{32} a'_{44}a'_{59}  -
 a'_{11} a'_{24} a'_{32} a'_{44}a'_{59}  -
 a'_{13} a'_{21} a'_{33} a'_{44}a'_{59}  +
 a'_{12} a'_{22} a'_{33} a'_{44}a'_{59}  +
 a'_{11} a'_{23} a'_{33} a'_{44}a'_{59}  +
 a'_{11} a'_{22} a'_{34} a'_{44}a'_{59} -
 a'_{14} a'_{22} a'_{31} a'_{45}a'_{59}  +
 a'_{11} a'_{22} a'_{33} a'_{45}a'_{59}  -
 a'_{15} a'_{22} a'_{33} a'_{41}a'_{5(10)}  +
 a'_{14} a'_{22} a'_{35} a'_{41} a'_{5(10)}  +
 a'_{15}a'_{22}a'_{31} a'_{43}a'_{5(10)} $

$ R_{11}=a'_{16} a'_{22} a'_{35} a'_{43}a'_{51} -
 a'_{15} a'_{22} a'_{36} a'_{43}a'_{51} -
 a'_{16} a'_{22} a'_{33} a'_{44}a_{51} +
 a'_{14} a'_{22} a'_{36} a'_{44}a'_{51} +
 a'_{15} a'_{22} a'_{33} a'_{46}a'_{51} -
 a'_{14} a'_{22} a'_{35} a'_{46}a'_{51} -
 a'_{16} a'_{22} a'_{35} a'_{41}a'_{55} +
 a'_{15} a'_{22} a'_{36} a'_{41}a'_{55} +
 a'_{16} a'_{22} a'_{31} a'_{44}a'_{55} -
 a'_{11} a'_{22} a'_{36} a'_{44}a'_{55} -
 a'_{15} a'_{22} a'_{31} a'_{46}a'_{55} +
 a'_{11} a'_{22} a'_{35} a'_{46}a'_{55} +
 a'_{16} a'_{22} a'_{33} a'_{41}a'_{57} -
 a'_{14} a'_{22} a'_{36} a'_{41}a'_{57} -
 a'_{16} a'_{22} a'_{31} a'_{43}a'_{57} +
 a'_{11}a'_{22}a'_{36} a'_{43}a'_{57} +
 a'_{14} a'_{22} a'_{31} a'_{46}a'_{57} -
 a'_{11} a'_{22} a'_{33} a'_{46}a'_{57} -
 a'_{15} a'_{22} a'_{33} a'_{41}a'_{59} +
 a'_{14} a'_{22} a'_{35} a'_{41}a'_{59} +
 a'_{15} a'_{22} a'_{31} a'_{43}a'_{59} -
 a'_{11} a'_{22} a'_{35} a'_{43}a'_{59} -
 a'_{14} a'_{22} a'_{31} a'_{44}a'_{59} +
 a'_{11} a'_{22} a'_{33} a'_{44}a'_{59}$\\
\\
   $ a'_{11}=\rho_{2}{k^2}c^2+\varepsilon^2k^4c^2-k^2c'_{11}$,
 $a'_{12}=c_{55}-\varepsilon^2k^2c^2$,
 $a'_{13}=ik(c'_{13}+c'_{55})$,
 $a'_{14}=ikB'_{1}$,
 $a'_{15}=-ik\beta'_{1}$,\\
 $a'_{16}=-ikb'_{1}$,
 $a'_{21}=ik(c'_{13}+c'_{55})$,$a_{22}=\rho_{2}{k^2}c^2+\varepsilon^2k^4c^2-k^2c'_{55}$,$a'_{23}=c'_{33}-\varepsilon^2k^2c^2$, $a'_{24}=B'_{3}$
 $a'_{25}=-\beta'_{3}$,$a'_{26}=-b'_{3}$
 $a'_{31}=-ikB'_{1}$,$a'_{32}=-B'_{3}$,$a'_{33}=\rho_{2} \chi' \varepsilon^2k^4c^2-k^{2} A'_{1}-\xi$,$a'_{34}=A'_{3}-\rho_{2} \chi \varepsilon^2k^2c^2$
 $a'_{35}=\nu'_{1}$,$a'_{36}=\nu'_{2}$
 $a'_{41}=\beta'_{1}T_{0}ck^{2}(-1+ikc\tau_{q})$,$a'_{42}=ikcT_{0}\beta'_{3}(1-ikc\tau_{q})$,$a'_{43}=ikac\nu'_{1}T_{0}$,\\
 $a'_{44}=ik\rho_{2} C'_{v}(1-ikc\tau_{q})-(1-ikc\tau_{T})K'_{1}k^2$,$a'_{45}=(1-ikc\tau_{T})K'_{3}$,$a'_{46}=ikacT_{0}(1-ikc\tau_{q})$
\\
 $a'_{51}=k^4cd'_{1}b'_{1}(1-ikc\tau_{2})+ik^3b'_{1}d'^{\ast}_{1}(1-ikc\tau_{3})$,
$a'_{52}=-[k^2cd'_{3}b'_{1}(1-ikc\tau_{2})+ikb'_{1}d'^{\ast}_{3}(1-ikc\tau_{3})]$,\\
$a'_{53}=ik^3cd'_{1}b'_{3}(1-ikc\tau_{2})-k^2b'_{3}d'^{\ast}_{1}(1-ikc\tau_{3})$,
$a'_{54}=ikcd'_{3}b'_{3}(1-ikc\tau_{2})-b'_{3}d'^{\ast}_{3}(1-ikc\tau_{3})$,\\
$a'_{55}=-[ik^3c\nu'_{2}d'_{1}(1-ikc\tau_{2})-k^2\nu'_{2}d'^{\ast}_{1}(1-ikc\tau_{3})]$,
$a'_{56}=[ikc\nu'_{2}d'_{3}(1-ikc\tau_{2})-\nu'_{2}d^{\ast}_{3}(1-ikc\tau_{3})]$,\\
$a'_{57}=-[ik^3c\nu'_{3}d'_{1}(1-ikc\tau_{2})-k^2\nu'_{3}d^{\ast}_{1}(1-ikc\tau_{3})]$,
$a'_{58}=[ikc\nu'_{3}d'_{3}(1-ikc\tau_{2})-\nu'_{3}d^{\ast}_{3}(1-ikc\tau_{3})]$,\\
$a'_{59}=[ik^3cb'd'_{1}(1-ikc\tau_{2})-k^2b'd'^{\ast}_{1}(1-ikc\tau_{3})-k^2c^2+ik^3c^3\tau_{2}+ik^5c^3\tau_{2}\varepsilon^2+\frac{k^4c^4\tau^2_{1}}{2}
+\frac{k^6c^4\varepsilon^2\tau^2_{1}}{2}]$,\\
$a'_{5(10)}=-[ikcb'd'_{3}(1-ikc\tau_{2})-b'd'^{\ast}_{3}(1-ikc\tau_{3})+k^2c^2\varepsilon^2-ik^3c^3\varepsilon^2\tau_{2}-\frac{k^6c^4\varepsilon^2\tau^2_{1}}{2}]$\\

 $f_{n}=Q_{1}+\frac{Q_{2}(Q_{1}-P_{1})}{P_{2}-Q_{2}}$,\\
 $j_{n}=\frac{Q_{1}-P_{1}}{P_{2}-Q_{2}}$,\\
 $k_{n}=-\frac{(\Omega_{5}\Omega_{6}-\Omega_{1}\Omega_{10})}{\Omega_{5}\Omega_{9}-\Omega_{4}\Omega_{10}}
 -\frac{(\Omega_{5}\Omega_{7}-\Omega_{2}\Omega_{10})}{\Omega_{5}\Omega_{9}-\Omega_{4}\Omega_{10}}\left(Q_{1}+\frac{Q_{2}(Q_{1}-P_{1})}{P_{2}-Q_{2}}\right)
 -\frac{(\Omega_{5}\Omega_{8}-\Omega_{3}\Omega_{10})}{\Omega_{5}\Omega_{9}-\Omega_{4}\Omega_{10}}\frac{Q_{1}-P_{1}}{P_{2}-Q_{2}}$,\\
 $l_{n}=-\frac{\Omega_{1}}{\Omega_{5}}-\frac{\Omega_{2}}{\Omega_{5}}\times f_{n}-\frac{\Omega_{3}}{\Omega_{5}}\times j_{n}-\frac{\Omega_{4}}{\Omega_{5}}\times k_{n}$,\\
 $P_{1}=-\frac{(\Omega_{5}\Omega_{11}-\Omega_{1}\Omega_{15})(\Omega_{5}\Omega_{9}-\Omega_{4}\Omega_{10})-
 (\Omega_{5}\Omega_{14}-\Omega_{4}\Omega_{15})(\Omega_{5}\Omega_{6}-\Omega_{1}\Omega_{10})}
 {(\Omega_{5}\Omega_{12}-\Omega_{2}\Omega_{15})(\Omega_{5}\Omega_{9}-\Omega_{4}\Omega_{10})-
 (\Omega_{5}\Omega_{7}-\Omega_{2}\Omega_{10})(\Omega_{5}\Omega_{14}-\Omega_{4}\Omega_{15})}$
 \\
 $P_{2}=-\frac{(\Omega_{5}\Omega_{13}-\Omega_{3}\Omega_{15})(\Omega_{5}\Omega_{9}-\Omega_{4}\Omega_{10})-
 (\Omega_{5}\Omega_{8}-\Omega_{3}\Omega_{10})(\Omega_{5}\Omega_{14}-\Omega_{14}\Omega_{15})}
 {(\Omega_{5}\Omega_{12}-\Omega_{2}\Omega_{15})(\Omega_{5}\Omega_{9}-\Omega_{4}\Omega_{10})-
 (\Omega_{5}\Omega_{7}-\Omega_{2}\Omega_{10})(\Omega_{5}\Omega_{14}-\Omega_{4}\Omega_{15})}$\\
 $Q_{1}=-\frac{(\Omega_{5}\Omega_{11}-\Omega_{1}\Omega_{15})(\Omega_{5}\Omega_{9}-\Omega_{4}\Omega_{10})-
 (\Omega_{5}\Omega_{14}-\Omega_{4}\Omega_{15})(\Omega_{5}\Omega_{6}-\Omega_{1}\Omega_{10})}
 {(\Omega_{5}\Omega_{12}-\Omega_{2}\Omega_{15})(\Omega_{5}\Omega_{9}-\Omega_{4}\Omega_{10})-
 (\Omega_{5}\Omega_{7}-\Omega_{2}\Omega_{10})(\Omega_{5}\Omega_{14}-\Omega_{4}\Omega_{15})}$\\
 $Q_{2}=-\frac{(\Omega_{5}\Omega_{11}-\Omega_{1}\Omega_{15})(\Omega_{5}\Omega_{9}-\Omega_{4}\Omega_{10})-
 (\Omega_{5}\Omega_{14}-\Omega_{4}\Omega_{15})(\Omega_{5}\Omega_{6}-\Omega_{1}\Omega_{10})}
 {(\Omega_{5}\Omega_{12}-\Omega_{2}\Omega_{15})(\Omega_{5}\Omega_{9}-\Omega_{4}\Omega_{10})-
 (\Omega_{5}\Omega_{7}-\Omega_{2}\Omega_{10})(\Omega_{5}\Omega_{14}-\Omega_{4}\Omega_{15})}$
\\
$\Omega_{1}=\rho_{2} k^2 c^2+\varepsilon^2 k^4 c^2-k^2c'_{11}+\lambda^2_{n}(c'_{55}-\varepsilon^2 k^2 c^2),\Omega_{2}=-ik\lambda_{n}(c'_{13}+c'_{55}),\Omega_{3}=ikB'_{1},\\
 \Omega_{4}=-ik\beta'_{1},\Omega_{5}=-ikb'_{1}$,
$\Omega_{6}=-ik\lambda_{n}(c'_{13}+c'_{55}),\Omega_{7}=\rho_{2} k^2c^2,
\Omega_{8}=-\lambda_{n}B'_{3}, \Omega_{9}=\lambda_{n}\beta'_{3},\Omega_{10}=\lambda_{n}b'_{3}$\\
$\Omega_{11}=ikB'_{1},\Omega_{12}=-\lambda_{n}B'_{3}f_{n},\Omega_{13}=-[(\rho_{2} \chi k^2 c^2 +\rho \chi \varepsilon^2 k^4 c^2-k^2A_{1}-\xi')+\lambda^2_{n}(A_{3}-\rho_{2} \chi' \varepsilon^2 k^2 c^2)],\\ \Omega_{14}=-\nu'_{1},\Omega_{15}=\nu'_{2}$
$\Omega_{16}=T_{0}\beta_{1}ck^2(-1+ikc\tau_{q}),\Omega_{17}=-ikcT_{0}\beta'_{3}\lambda_{n}(1-ikc\tau_{q}),\Omega_{18}=ikac\nu'_{3}T_{0},\\ \Omega_{19}=ikc\rho C'_{v}(1-ikc\tau_{q})-K'_{1}k^2 (1-ikc\tau_{T})+K'_{3}\lambda^2_{n}(1-ikc\tau_{T}),\Omega_{20}=ikacT_{0}(1-ikc\tau_{q}).$\\
    \section*{Appendix C}
$\beth_{11}=cos\lambda_{1}H+\daleth_{1}sin\lambda_{1}H-\frac{g^{(1)}_{1}}{g^{(1)}_{5}}(cos\lambda_{5}H+\daleth_{5}sin\lambda_{5}H)$,
$\beth_{12}=cos\lambda_{2}H+\daleth_{2}sin\lambda_{2}H-\frac{g^{(1)}_{2}}{g^{(1)}_{5}}(cos\lambda_{5}H+\daleth_{5}sin\lambda_{5}H)$,\\
$\beth_{13}=cos\lambda_{3}H+\daleth_{3}sin\lambda_{3}H-\frac{g^{(1)}_{3}}{g^{(1)}_{5}}(cos\lambda_{5}H+\daleth_{5}sin\lambda_{5}H)$,
$\beth_{14}=cos\lambda_{4}H+\daleth_{4}sin\lambda_{4}H-\frac{g^{(1)}_{4}}{g^{(1)}_{5}}(cos\lambda_{5}H+\daleth_{5}sin\lambda_{5}H)$,\\
$\beth_{21}=c^{(1)}_{1}cos\lambda_{1}H+\daleth_{1}d^{(1)}_{1}sin\lambda_{1}H-\frac{g^{(1)}_{1}}{g^{(1)}_{5}}(c^{(1)}_{5}cos\lambda_{5}H+d^{(1)}_{5}\daleth_{5}sin\lambda_{5}H)$,\\
$\beth_{22}=c^{(1)}_{2}cos\lambda_{2}H+\daleth_{2}d^{(1)}_{2}sin\lambda_{1}H-\frac{g^{(1)}_{2}}{g^{(1)}_{5}}(c^{(1)}_{5}cos\lambda_{5}H+d^{(1)}_{5}\daleth_{5}sin\lambda_{5}H)$,\\
$\beth_{23}=c^{(1)}_{3}cos\lambda_{3}H+\daleth_{3}d^{(1)}_{3}sin\lambda_{1}H-\frac{g^{(1)}_{3}}{g^{(1)}_{5}}(c^{(1)}_{5}cos\lambda_{5}H+d^{(1)}_{5}\daleth_{5}sin\lambda_{5}H)$,\\
$\beth_{24}=c^{(1)}_{4}cos\lambda_{4}H+\daleth_{4}d^{(1)}_{4}sin\lambda_{1}H-\frac{g^{(1)}_{4}}{g^{(1)}_{5}}(c^{(1)}_{5}cos\lambda_{5}H+d^{(1)}_{5}\daleth_{5}sin\lambda_{5}H)$,\\
$\beth_{31}=[(\lambda_{1}\daleth_{1}+ikg^{(1)}_{1})cos\lambda_{1}H+(-\lambda_{1}+ik\daleth_{1}b^{(1)}_{1})sin\lambda_{1}H]c_{55}-\frac{g^{(1)}_{1}}{g^{(1)}_{5}}
[(\lambda_{5}\daleth_{5}+ikc^{(1)}_{5})cos\lambda_{5}H+(-\lambda_{5}+ik\daleth_{5}b^{(1)}_{5})sin\lambda_{5}H]c_{55}$,\\
$\beth_{32}=[(\lambda_{2}\daleth_{2}+ikg^{(1)}_{2})cos\lambda_{2}H+(-\lambda_{2}+ik\daleth_{2}b^{(1)}_{2})sin\lambda_{2}H]c_{55}-\frac{g^{(1)}_{2}}{g^{(1)}_{5}}
[(\lambda_{5}\daleth_{5}+ikc^{(1)}_{5})cos\lambda_{5}H+(-\lambda_{5}+ik\daleth_{5}b^{(1)}_{5})sin\lambda_{5}H]c_{55}$,\\
$\beth_{33}=[(\lambda_{3}\daleth_{3}+ikg^{(1)}_{3})cos\lambda_{3}H+(-\lambda_{3}+ik\daleth_{3}b^{(1)}_{3})sin\lambda_{3}H]c_{55}-\frac{g^{(1)}_{3}}{g^{(1)}_{5}}
[(\lambda_{5}\daleth_{5}+ikc^{(1)}_{5})cos\lambda_{5}H+(-\lambda_{5}+ik\daleth_{5}b^{(1)}_{5})sin\lambda_{5}H]c_{55}$,\\
$\beth_{34}=[(\lambda_{4}\daleth_{4}+ikg^{(1)}_{4})cos\lambda_{4}H+(-\lambda_{4}+ik\daleth_{4}b^{(1)}_{4})sin\lambda_{4}H]c_{55}-\frac{g^{(1)}_{4}}{g^{(1)}_{5}}
[(\lambda_{5}\daleth_{5}+ikc^{(1)}_{5})cos\lambda_{5}H+(-\lambda_{5}+ik\daleth_{5}b^{(1)}_{5})sin\lambda_{5}H]c_{55}$,\\
$\beth_{35}=c'_{55}(ikf_{1}-\varrho_{1})$,
$\beth_{36}=c'_{55}(ikf_{2}-\varrho_{2})$,
$\beth_{37}=c'_{55}(ikf_{3}-\varrho_{3})$,
$\beth_{38}=c'_{55}(ikf_{4}-\varrho_{4})$,
$\beth_{39}=c'_{55}(ikf_{5}-\varrho_{5})$,
$\beth_{41}=(ikc_{11}+\lambda_{1}c_{33}\daleth_{1}d^{(1)}_{1}+B_{3}e^{(1)}_{1}-\beta_{3}g^{(1)}_{1}-b_{3}i^{(1)}_{1})cos\lambda_{1}H
+(ikc_{11}\daleth_{1}-\lambda_{1}c_{33}c^{(1)}_{1}+B_{3}\daleth_{1}f^{(1)}_{1}-\beta_{3}\daleth_{1}h^{(1)}_{1}-b_{3}\daleth_{1}j^{(1)}_{1})sin\lambda_{1}H
-\frac{g^{(1)}_{1}}{g^{(1)}_{5}}[(ikc_{11}+\lambda_{5}c_{33}\daleth_{5}d^{(1)}_{5}+B_{3}e^{(1)}_{5}-\beta_{3}g^{(1)}_{5}-b_{3}i^{(1)}_{5})cos\lambda_{5}H
+(ikc_{11}\daleth_{5}-\lambda_{5}c_{33}c^{(1)}_{5}+B_{3}\daleth_{5}f^{(1)}_{5}-\beta_{3}\daleth_{5}h^{(1)}_{5}-b_{3}\daleth_{1}j^{(1)}_{5})sin\lambda_{5}H]$,\\
$\daleth_{42}=(ikc_{11}+\lambda_{2}c_{33}\daleth_{2}d^{(1)}_{2}+B_{3}e^{(1)}_{2}-\beta_{3}g^{(1)}_{2}-b_{3}i^{(1)}_{2})cos\lambda_{2}H
+(ikc_{11}\daleth_{2}-\lambda_{2}c_{33}c^{(1)}_{2}+B_{3}\daleth_{2}f^{(1)}_{2}-\beta_{3}\daleth_{2}h^{(1)}_{2}-b_{3}\daleth_{2}j^{(1)}_{2})sin\lambda_{2}H
-\frac{g^{(1)}_{2}}{g^{(1)}_{5}}[(ikc_{11}+\lambda_{5}c_{33}\daleth_{5}d^{(1)}_{5}+B_{3}e^{(1)}_{5}-\beta_{3}g^{(1)}_{5}-b_{3}i^{(1)}_{5})cos\lambda_{5}H
+(ikc_{11}\daleth_{5}-\lambda_{5}c_{33}c^{(1)}_{5}+B_{3}\daleth_{5}f^{(1)}_{5}-\beta_{3}\daleth_{5}h^{(1)}_{5}-b_{3}\daleth_{1}j^{(1)}_{5})sin\lambda_{5}H]$,\\
$\beth_{43}=(ikc_{11}+\lambda_{3}c_{33}\daleth_{3}d^{(1)}_{3}+B_{3}e^{(1)}_{3}-\beta_{3}g^{(1)}_{3}-b_{3}i^{(1)}_{3})cos\lambda_{3}H
+(ikc_{11}\daleth_{3}-\lambda_{3}c_{33}c^{(1)}_{3}+B_{3}\daleth_{3}f^{(1)}_{3}-\beta_{3}\daleth_{3}h^{(1)}_{3}-b_{3}\daleth_{3}j^{(1)}_{3})sin\lambda_{3}H
-\frac{g^{(1)}_{3}}{g^{(1)}_{5}}[(ikc_{11}+\lambda_{5}c_{33}\daleth_{5}d^{(1)}_{5}+B_{3}e^{(1)}_{5}-\beta_{3}g^{(1)}_{5}-b_{3}i^{(1)}_{5})cos\lambda_{5}H
+(ikc_{11}\daleth_{5}-\lambda_{5}c_{33}c^{(1)}_{5}+B_{3}\daleth_{5}f^{(1)}_{5}-\beta_{3}\daleth_{5}h^{(1)}_{5}-b_{3}\daleth_{1}j^{(1)}_{5})sin\lambda_{5}H]$,\\
$\beth_{44}=(ikc_{11}+\lambda_{4}c_{33}\daleth_{4}d^{(1)}_{4}+B_{3}e^{(1)}_{4}-\beta_{3}g^{(1)}_{4}-b_{3}i^{(1)}_{4})cos\lambda_{4}H
+(ikc_{11}\daleth_{4}-\lambda_{4}c_{33}c^{(1)}_{4}+B_{3}\daleth_{1}f^{(1)}_{4}-\beta_{3}\daleth_{1}h^{(1)}_{4}-b_{3}\daleth_{4}j^{(1)}_{4})sin\lambda_{4}H
-\frac{g^{(1)}_{4}}{g^{(1)}_{5}}[(ikc_{11}+\lambda_{5}c_{33}\daleth_{5}d^{(1)}_{5}+B_{3}e^{(1)}_{5}-\beta_{3}g^{(1)}_{5}-b_{3}i^{(1)}_{5})cos\lambda_{5}H
+(ikc_{11}\daleth_{5}-\lambda_{5}c_{33}c^{(1)}_{5}+B_{3}\daleth_{5}f^{(1)}_{5}-\beta_{3}\daleth_{5}h^{(1)}_{5}-b_{3}\daleth_{1}j^{(1)}_{5})sin\lambda_{5}H]$,\\
$\beth_{45}=(ikc'_{13}-c'_{33}\varrho_{1}f_{1}+B'_{3}j_{1}-\beta'_{3}k_{1}-b'_{3}l_{1})$,
$\beth_{46}=(ikc'_{13}-c'_{33}\varrho_{2}f_{2}+B'_{3}j_{2}-\beta'_{3}k_{2}-b'_{3}l_{2})$,\\
$\beth_{47}=(ikc'_{13}-c'_{33}\varrho_{3}f_{3}+B'_{3}j_{3}-\beta'_{3}k_{3}-b'_{3}l_{3})$,
$\beth_{48}=(ikc'_{13}-c'_{33}\varrho_{4}f_{4}+B'_{3}j_{4}-\beta'_{3}k_{4}-b'_{3}l_{4})$,\\
$\beth_{49}=(ikc'_{13}-c'_{33}\varrho_{5}f_{5}+B'_{3}j_{5}-\beta'_{3}k_{5}-b'_{3}l_{5})$,\\
$\beth_{51}=A_{3}(\lambda_{1}\daleth_{1}f^{(1)}_{1}cos\lambda_{1}H-\lambda_{1}e^{(1)}_{1}sin\lambda_{1}H)
-\frac{g^{(1)}_{1}}{g^{(1)}_{5}}[A_{3}(\lambda_{5}\daleth_{5}f^{(1)}_{5}cos\lambda_{5}H-\lambda_{1}e^{(1)}_{5}sin\lambda_{5}H)]$,\\
$\beth_{52}=A_{3}(\lambda_{2}\daleth_{2}f^{(1)}_{2}cos\lambda_{2}H-\lambda_{2}e^{(1)}_{2}sin\lambda_{2}H)
-\frac{g^{(1)}_{2}}{g^{(1)}_{5}}[A_{3}(\lambda_{5}\daleth_{5}f^{(1)}_{5}cos\lambda_{5}H-\lambda_{1}e^{(1)}_{5}sin\lambda_{5}H)]$,\\
$\beth_{53}=A_{3}(\lambda_{3}\daleth_{3}f^{(1)}_{3}cos\lambda_{3}H-\lambda_{3}e^{(1)}_{3}sin\lambda_{3}H)
-\frac{g^{(1)}_{3}}{g^{(1)}_{5}}[A_{3}(\lambda_{5}\daleth_{5}f^{(1)}_{5}cos\lambda_{5}H-\lambda_{1}e^{(1)}_{5}sin\lambda_{5}H)]$,\\
$\beth_{54}=A_{3}(\lambda_{4}\daleth_{4}f^{(1)}_{4}cos\lambda_{4}H-\lambda_{4}e^{(1)}_{4}sin\lambda_{4}H)
-\frac{g^{(1)}_{4}}{g^{(1)}_{5}}[A_{3}(\lambda_{5}\daleth_{5}f^{(1)}_{5}cos\lambda_{5}H-\lambda_{1}e^{(1)}_{5}sin\lambda_{5}H)]$,\\
$\beth_{55}=-\varrho_{1}j_{1}A'_{3}$,
$\beth_{56}=-\varrho_{2}j_{2}A'_{3}$,
$\beth_{57}=-\varrho_{3}j_{3}A'_{3}$,
$\beth_{58}=-\varrho_{4}j_{4}A'_{3}$,
$\beth_{59}=-\varrho_{5}j_{5}A'_{3}$,\\
$\beth_{61}=e^{(1)}_{1}cos\lambda_{1}H+\daleth_{1}f^{(1)}_{1}sin\lambda_{1}H-\frac{g^{(1)}_{1}}{g^{(1)}_{5}}(e^{(1)}_{5}cos\lambda_{5}H+f^{(1)}_{5}\daleth_{5}sin\lambda_{5}H)$,\\
$\beth_{62}=e^{(1)}_{2}cos\lambda_{2}H+\daleth_{2}f^{(1)}_{2}sin\lambda_{2}H-\frac{g^{(1)}_{2}}{g^{(1)}_{5}}(e^{(1)}_{5}cos\lambda_{5}H+f^{(1)}_{5}\daleth_{5}sin\lambda_{5}H)$,\\
$\beth_{63}=e^{(1)}_{3}cos\lambda_{3}H+\daleth_{3}f^{(1)}_{3}sin\lambda_{3}H-\frac{g^{(1)}_{3}}{g^{(1)}_{5}}(e^{(1)}_{5}cos\lambda_{5}H+f^{(1)}_{5}\daleth_{5}sin\lambda_{5}H)$,\\
$\beth_{64}=e^{(1)}_{4}cos\lambda_{4}H+\daleth_{4}f^{(1)}_{4}sin\lambda_{4}H-\frac{g^{(1)}_{4}}{g^{(1)}_{5}}(e^{(1)}_{5}cos\lambda_{5}H+f^{(1)}_{5}\daleth_{5}sin\lambda_{5}H)$,\\
$\beth_{71}=g^{(1)}_{1}cos\lambda_{1}H+\daleth_{1}h^{(1)}_{1}sin\lambda_{1}H-\frac{g^{(1)}_{1}}{g^{(1)}_{5}}(g^{(1)}_{5}cos\lambda_{5}H+h^{(1)}_{5}\daleth_{5}sin\lambda_{5}H)$,\\
$\beth_{72}=g^{(1)}_{2}cos\lambda_{2}H+\daleth_{2}h^{(1)}_{2}sin\lambda_{2}H-\frac{g^{(1)}_{2}}{g^{(1)}_{5}}(g^{(1)}_{5}cos\lambda_{5}H+h^{(1)}_{5}\daleth_{5}sin\lambda_{5}H)$,\\
$\beth_{73}=g^{(1)}_{3}cos\lambda_{3}H+\daleth_{3}h^{(1)}_{3}sin\lambda_{3}H-\frac{g^{(1)}_{3}}{g^{(1)}_{5}}(g^{(1)}_{5}cos\lambda_{5}H+h^{(1)}_{5}\daleth_{5}sin\lambda_{5}H)$,\\
$\beth_{74}=g^{(1)}_{4}cos\lambda_{4}H+\daleth_{4}h^{(1)}_{4}sin\lambda_{4}H-\frac{g^{(1)}_{4}}{g^{(1)}_{5}}(g^{(1)}_{5}cos\lambda_{5}H+h^{(1)}_{5}\daleth_{5}sin\lambda_{5}H)$,\\
$\beth_{81}=(ikb_{1}-b_{3}\lambda_{1}e^{(1)}_{1}-\nu_{2}\lambda_{1}g^{(1)}_{1}-\nu_{3}\lambda_{1}i^{(1)}_{1}
-b_{3}\lambda^2_{1}d^{(1)}_{1}\daleth^2_{1})-\frac{g^{(1)}_{1}}{g^{(1)}_{5}}(ikb_{1}-b_{3}\lambda_{5}e^{(1)}_{5}-\nu_{2}\lambda_{5}g^{(1)}_{5}-\nu_{3}\lambda_{5}i^{(1)}_{5}
-b_{3}\lambda^2_{5}d^{(1)}_{5}\daleth^2_{5})$,\\
$\beth_{82}=(ikb_{1}-b_{3}\lambda_{2}e^{(1)}_{2}-\nu_{2}\lambda_{2}g^{(1)}_{2}-\nu_{3}\lambda_{2}i^{(1)}_{2}
-b_{3}\lambda^2_{2}d^{(1)}_{2}\daleth^2_{2})-\frac{g^{(1)}_{2}}{g^{(1)}_{5}}(ikb_{1}-b_{3}\lambda_{5}e^{(1)}_{5}-\nu_{2}\lambda_{5}g^{(1)}_{5}-\nu_{3}\lambda_{5}i^{(1)}_{5}
-b_{3}\lambda^2_{5}d^{(1)}_{5}\daleth^2_{5})$,\\
$\beth_{83}=(ikb_{1}-b_{3}\lambda_{3}e^{(1)}_{3}-\nu_{2}\lambda_{3}g^{(1)}_{3}-\nu_{3}\lambda_{1}i^{(1)}_{3}
-b_{3}\lambda^2_{3}d^{(1)}_{3}\daleth^2_{3})-\frac{g^{(1)}_{3}}{g^{(1)}_{5}}(ikb_{1}-b_{3}\lambda_{5}e^{(1)}_{5}-\nu_{2}\lambda_{5}g^{(1)}_{5}-\nu_{3}\lambda_{5}i^{(1)}_{5}
-b_{3}\lambda^2_{5}d^{(1)}_{5}\daleth^2_{5})$,\\
$\beth_{84}=(ikb_{1}-b_{3}\lambda_{4}e^{(1)}_{4}-\nu_{2}\lambda_{4}g^{(1)}_{4}-\nu_{3}\lambda_{4}i^{(1)}_{4}
-b_{3}\lambda^2_{4}d^{(1)}_{4}\daleth^2_{4})-\frac{g^{(1)}_{4}}{g^{(1)}_{5}}(ikb_{1}-b_{3}\lambda_{5}e^{(1)}_{5}-\nu_{2}\lambda_{5}g^{(1)}_{5}-\nu_{3}\lambda_{5}i^{(1)}_{5}
-b_{3}\lambda^2_{5}d^{(1)}_{5}\daleth^2_{5})$,\\
$\beth_{85}=(ikb'_{1}\varrho_{1}-b'_{3}f_{1}\varrho^{2}_{1}+\nu'_{2}\varrho_{1}j_{1}+\nu'_{3}\varrho_{1}k_{1}-b'\varrho_{1}l_{1})$,\\
$\beth_{86}=(ikb'_{1}\varrho_{2}-b'_{3}f_{2}\varrho^{2}_{2}+\nu'_{2}\varrho_{2}j_{2}+\nu'_{3}\varrho_{2}k_{2}-b'\varrho_{2}l_{2})$,\\
$\beth_{87}=(ikb'_{1}\varrho_{3}-b'_{3}f_{3}\varrho^{2}_{3}+\nu'_{2}\varrho_{3}j_{3}+\nu'_{3}\varrho_{3}k_{3}-b'\varrho_{3}l_{3})$,\\
$\beth_{88}=(ikb'_{1}\varrho_{4}-b'_{3}f_{4}\varrho^{2}_{4}+\nu'_{2}\varrho_{4}j_{4}+\nu'_{3}\varrho_{4}k_{4}-b'\varrho_{4}l_{4})$,\\
$\beth_{89}=(ikb'_{1}\varrho_{5}-b'_{3}f_{5}\varrho^{2}_{5}+\nu'_{2}\varrho_{5}j_{5}+\nu'_{3}\varrho_{5}k_{5}-b'\varrho_{5}l_{5})$,\\
$\beth_{91}=i^{(1)}_{1}cos\lambda_{1}H+\daleth_{1}j^{(1)}_{1}sin\lambda_{1}H-\frac{g^{(1)}_{1}}{i^{(1)}_{5}}(g^{(1)}_{5}cos\lambda_{5}H+j^{(1)}_{5}\daleth_{5}sin\lambda_{5}H)$,\\
$\beth_{92}=i^{(1)}_{2}cos\lambda_{2}H+\daleth_{2}j^{(1)}_{2}sin\lambda_{1}H-\frac{g^{(1)}_{2}}{i^{(1)}_{5}}(g^{(1)}_{5}cos\lambda_{5}H+j^{(1)}_{5}\daleth_{5}sin\lambda_{5}H)$,\\
$\beth_{93}=i^{(1)}_{3}cos\lambda_{3}H+\daleth_{3}j^{(1)}_{3}sin\lambda_{1}H-\frac{g^{(1)}_{3}}{i^{(1)}_{5}}(g^{(1)}_{5}cos\lambda_{5}H+j^{(1)}_{5}\daleth_{5}sin\lambda_{5}H)$,\\
$\beth_{94}=i^{(1)}_{4}cos\lambda_{4}H+\daleth_{4}j^{(1)}_{4}sin\lambda_{1}H-\frac{g^{(1)}_{4}}{i^{(1)}_{5}}(g^{(1)}_{5}cos\lambda_{5}H+j^{(1)}_{5}\daleth_{5}sin\lambda_{5}H)$,\\

$X_{1}=a_{11}a_{26}\lambda_{n}-a_{16}a_{21}\lambda_{n}-a_{12}a_{26}\lambda^3_{n}$,
$X_{2}=a_{16}a_{22}-a_{16}a_{23}\lambda^2_{n}+a_{13}a_{26}\lambda^2_{n}$,
$X_{3}=a_{14}a_{26}\lambda_{n}-a_{16}a_{24}\lambda_{n}$,\\
$X_{4}=a_{14}a_{26}\lambda_{n}-a_{16}a_{25}\lambda_{n}$,
$X_{5}=-a_{11}a_{26}\lambda_{n}+a_{16}a_{21}\lambda_{n}+a_{12}a_{26}\lambda^3_{n}$,
$X_{6}=a_{16}a_{22}-a_{16}a_{23}\lambda^2_{n}+a_{13}a_{26}\lambda^3_{n}$,\\
$X_{7}=a_{16}a_{24}\lambda^2_{n}-a_{14}a_{26}\lambda_{n}$,
$X_{8}=a_{16}a_{25}\lambda_{n}-a_{15}a_{26}\lambda_{n}$,
$X_{9}=a_{16}a_{31}-a_{11}a_{36}+a_{12}a_{36}\lambda^2_{n}$,
$X_{10}=a_{16}a_{32}\lambda_{n}-a_{13}a_{36}\lambda_{n}$,
$X_{11}=a_{16}a_{33}-a_{16}a_{34}\lambda^2_{n}$,
$X_{12}=a_{16}a_{35}-a_{15}a_{36}$,
$X_{13}=a_{16}a_{31}-a_{11}a_{36}-a_{12}a_{36}\lambda^2_{n}$,\\
$X_{14}=a_{13}a_{36}-a_{14}a_{16}\lambda_{n}$,
$X_{15}=a_{16}n_{15}-a_{14}a_{36}-a_{16}a_{34}\lambda^2_{n}$,
$X_{16}=a_{16}a_{35}-a_{15}a_{36}$,
$X_{17}=a_{16}a_{41}-a_{11}a_{46}+a_{12}a_{46}\lambda^2_{n}$,
$X_{18}=a_{16}a_{42}\lambda_{n}-a_{13}a_{46}\lambda_{n}$,
$X_{19}=a_{16}a_{43}-a_{14}a_{46}$,
$X_{20}=a_{16}a_{44}-a_{15}a_{46}-a_{16}a_{45}\lambda^2_{n}$,\\
$X_{21}=a_{16}n_{41}-n_{1}a_{46}+a_{12}a_{46}\lambda^2_{n}$,
$X_{22}=-a_{16}a_{42}\lambda_{n}+a_{13}a_{46}\lambda_{n}$,
$X_{23}=a_{16}a_{43}-a_{14}a_{46}$,
$X_{24}=a_{16}a_{44}-a_{15}a_{46}-a_{16}a_{45}\lambda^2_{n}$,\\

\section*{Appendix D}

 $M_{1}=\left|\begin{array}{ccccccccc}
 \beth_{22} & \beth_{23} & \beth_{24} & -f_{1} & -f_{2} & -f_{3} & -f_{4} & -f_{5}\\
 \beth_{32} & \beth_{33} & \beth_{34} & -\beth_{35} & -\beth_{36} & -\beth_{37} & -\beth_{38} & \beth_{39}\\
 \beth_{42} & \beth_{43} & \beth_{44} & -\beth_{45} & -\beth_{46} & -\beth_{47} & -\beth_{48} & \beth_{49}\\
 \beth_{52} & \beth_{53} & \beth_{54} & -\beth_{55} & -\beth_{56} & -\beth_{57} & -\beth_{58} & \beth_{59}\\
 \beth_{62} & \beth_{63} & \beth_{64} & -j_{1} & -j_{2} & -j_{3} & -j_{4} & -j_{5}\\
 \beth_{72} & \beth_{73} & \beth_{74} & -k_{1} & -k_{2} & -k_{3} & -k_{4} & -k_{5}\\
 \beth_{82} & \beth_{83} & \beth_{84} & -\beth_{85} & -\beth_{86} & -\beth_{87} & -\beth_{88} & \beth_{89}\\
 \beth_{92} & \beth_{93} & \beth_{94} & -l_{1} & -l_{2} & -l_{3} & -l_{4} & -l_{5}
          \end{array}\right|$,\\

          $M_{2}=\left|\begin{array}{ccccccccc}
 \beth_{21} & \beth_{23} & \beth_{24} & -f_{1} & -f_{2} & -f_{3} & -f_{4} & -f_{5}\\
 \beth_{31} & \beth_{33} & \beth_{34} & -\beth_{35} & -\beth_{36} & -\beth_{37} & -\beth_{38} & \beth_{39}\\
 \beth_{41} & \beth_{43} & \beth_{44} & -\beth_{45} & -\beth_{46} & -\beth_{47} & -\beth_{48} & \beth_{49}\\
 \beth_{51} & \beth_{53} & \beth_{54} & -\beth_{55} & -\beth_{56} & -\beth_{57} & -\beth_{58} & \beth_{59}\\
 \beth_{61} & \beth_{63} & \beth_{64} & -j_{1} & -j_{2} & -j_{3} & -j_{4} & -j_{5}\\
 \beth_{71} & \beth_{73} & \beth_{74} & -k_{1} & -k_{2} & -k_{3} & -k_{4} & -k_{5}\\
 \beth_{81} & \beth_{83} & \beth_{84} & -\beth_{85} & -\beth_{86} & -\beth_{87} & -\beth_{88} & \beth_{89}\\
 \beth_{91} & \beth_{93} & \beth_{94} & -l_{1} & -l_{2} & -l_{3} & -l_{4} & -l_{5}
          \end{array}\right|$,\\

           $M_{3}=\left|\begin{array}{ccccccccc}
 \beth_{21} & \beth_{22} & \beth_{24} & -f_{1} & -f_{2} & -f_{3} & -f_{4} & -f_{5}\\
 \beth_{31} & \beth_{32} & \beth_{34} & -\beth_{35} & -\beth_{36} & -\beth_{37} & -\beth_{38} & \beth_{39}\\
 \beth_{41} & \beth_{42} & \beth_{44} & -\beth_{45} & -\beth_{46} & -\beth_{47} & -\beth_{48} & \beth_{49}\\
 \beth_{51} & \beth_{52} & \beth_{54} & -\beth_{55} & -\beth_{56} & -\beth_{57} & -\beth_{58} & \beth_{59}\\
 \beth_{61} & \beth_{62} & \beth_{64} & -j_{1} & -j_{2} & -j_{3} & -j_{4} & -j_{5}\\
 \beth_{71} & \beth_{72} & \beth_{74} & -k_{1} & -k_{2} & -k_{3} & -k_{4} & -k_{5}\\
 \beth_{81} & \beth_{82} & \beth_{84} & -\beth_{85} & -\beth_{86} & -\beth_{87} & -\beth_{88} & \beth_{89}\\
 \beth_{91} & \beth_{92} & \beth_{94} & -l_{1} & -l_{2} & -l_{3} & -l_{4} & -l_{5}
          \end{array}\right|$,\\
          $M_{4}=\left|\begin{array}{ccccccccc}
 \beth_{21} & \beth_{22} & \beth_{23} & -f_{1} & -f_{2} & -f_{3} & -f_{4} & -f_{5}\\
 \beth_{31} & \beth_{32} & \beth_{33} & -\beth_{35} & -\beth_{36} & -\beth_{37} & -\beth_{38} & \beth_{39}\\
 \beth_{41} & \beth_{42} & \beth_{43} & -\beth_{45} & -\beth_{46} & -\beth_{47} & -\beth_{48} & \beth_{49}\\
 \beth_{51} & \beth_{52} & \beth_{53} & -\beth_{55} & -\beth_{56} & -\beth_{57} & -\beth_{58} & \beth_{59}\\
 \beth_{61} & \beth_{62} & \beth_{63} & -j_{1} & -j_{2} & -j_{3} & -j_{4} & -j_{5}\\
 \beth_{71} & \beth_{72} & \beth_{73} & -k_{1} & -k_{2} & -k_{3} & -k_{4} & -k_{5}\\
 \beth_{81} & \beth_{82} & \beth_{83} & -\beth_{85} & -\beth_{86} & -\beth_{87} & -\beth_{88} & \beth_{89}\\
 \beth_{91} & \beth_{92} & \beth_{93} & -l_{1} & -l_{2} & -l_{3} & -l_{4} & -l_{5}
          \end{array}\right|$,\\

          $M_{5}=\left|\begin{array}{ccccccccc}
 \beth_{21} & \beth_{22} & \beth_{23} & \beth_{24} & -f_{2} & -f_{3} & -f_{4} & -f_{5}\\
 \beth_{31} & \beth_{32} & \beth_{33} & \beth_{34} & -\beth_{36} & -\beth_{37} & -\beth_{38} & \beth_{39}\\
 \beth_{41} & \beth_{42} & \beth_{43} & \beth_{44} & -\beth_{46} & -\beth_{47} & -\beth_{48} & \beth_{49}\\
 \beth_{51} & \beth_{52} & \beth_{53} & \beth_{54} & -\beth_{56} & -\beth_{57} & -\beth_{58} & \beth_{59}\\
 \beth_{61} & \beth_{62} & \beth_{63} & \beth_{64} & -j_{2} & -j_{3} & -j_{4} & -j_{5}\\
 \beth_{71} & \beth_{72} & \beth_{73} & \beth_{74} & -k_{2} & -k_{3} & -k_{4} & -k_{5}\\
 \beth_{81} & \beth_{82} & \beth_{83} & \beth_{84} & -\beth_{86} & -\beth_{87} & -\beth_{88} & \beth_{89}\\
 \beth_{91} & \beth_{92} & \beth_{93} & \beth_{94} & -l_{2} & -l_{3} & -l_{4} & -l_{5}
          \end{array}\right|$,\\
          $M_{6}=\left|\begin{array}{ccccccccc}
 \beth_{21} & \beth_{22} & \beth_{23} & \beth_{24} & -f_{1} & -f_{3} & -f_{4} & -f_{5}\\
 \beth_{31} & \beth_{32} & \beth_{33} & \beth_{34} & -\beth_{35} & -\beth_{37} & -\beth_{38} & \beth_{39}\\
 \beth_{41} & \beth_{42} & \beth_{43} & \beth_{44} & -\beth_{45} & -\beth_{47} & -\beth_{48} & \beth_{49}\\
 \beth_{51} & \beth_{52} & \beth_{53} & \beth_{54} & -\beth_{55} & -\beth_{57} & -\beth_{58} & \beth_{59}\\
 \beth_{61} & \beth_{62} & \beth_{63} & \beth_{64} & -j_{1} & -j_{3} & -j_{4} & -j_{5}\\
 \beth_{71} & \beth_{72} & \beth_{73} & \beth_{74} & -k_{1} & -k_{3} & -k_{4} & -k_{5}\\
 \beth_{81} & \beth_{82} & \beth_{83} & \beth_{84} & -\beth_{85} & -\beth_{87} & -\beth_{88} & \beth_{89}\\
 \beth_{91} & \beth_{92} & \beth_{93} & \beth_{94} & -l_{1} & -l_{3} & -l_{4} & -l_{5}
          \end{array}\right|$,\\

          $M_{7}=\left|\begin{array}{ccccccccc}
 \beth_{21} & \beth_{22} & \beth_{23} & \beth_{24} & -f_{1} & -f_{2} & -f_{4} & -f_{5}\\
 \beth_{31} & \beth_{32} & \beth_{33} & \beth_{34} & -\beth_{35} & -\beth_{36} & -\beth_{38} & \beth_{39}\\
 \beth_{41} & \beth_{42} & \beth_{43} & \beth_{44} & -\beth_{45} & -\beth_{46} & -\beth_{48} & \beth_{49}\\
 \beth_{51} & \beth_{52} & \beth_{53} & \beth_{54} & -\beth_{55} & -\beth_{56} & -\beth_{58} & \beth_{59}\\
 \beth_{61} & \beth_{62} & \beth_{63} & \beth_{64} & -j_{1} & -j_{2} & -j_{4} & -j_{5}\\
 \beth_{71} & \beth_{72} & \beth_{73} & \beth_{74} & -k_{1} & -k_{2} & -k_{4} & -k_{5}\\
 \beth_{81} & \beth_{82} & \beth_{83} & \beth_{84} & -\beth_{85} & -\beth_{86} & -\beth_{88} & \beth_{89}\\
 \beth_{91} & \beth_{92} & \beth_{93} & \beth_{94} & -l_{1} & -l_{2} & -l_{4} & -l_{5}
          \end{array}\right|$,\\
          $M_{8}=\left|\begin{array}{ccccccccc}
 \beth_{21} & \beth_{22} & \beth_{23} & \beth_{24} & -f_{1} & -f_{2} & -f_{3} & -f_{5}\\
 \beth_{31} & \beth_{32} & \beth_{33} & \beth_{34} & -\beth_{35} & -\beth_{36} & -\beth_{37} & \beth_{39}\\
 \beth_{41} & \beth_{42} & \beth_{43} & \beth_{44} & -\beth_{45} & -\beth_{46} & -\beth_{47} & \beth_{49}\\
 \beth_{51} & \beth_{52} & \beth_{53} & \beth_{54} & -\beth_{55} & -\beth_{56} & -\beth_{57} & \beth_{59}\\
 \beth_{61} & \beth_{62} & \beth_{63} & \beth_{64} & -j_{1} & -j_{2} & -j_{3} & -j_{5}\\
 \beth_{71} & \beth_{72} & \beth_{73} & \beth_{74} & -k_{1} & -k_{2} & -k_{3} & -k_{5}\\
 \beth_{81} & \beth_{82} & \beth_{83} & \beth_{84} & -\beth_{85} & -\beth_{86} & -\beth_{87} & \beth_{89}\\
 \beth_{91} & \beth_{92} & \beth_{93} & \beth_{94} & -l_{1} & -l_{2} & -l_{3} & -l_{5}
          \end{array}\right|$,\\

           $M_{9}=\left|\begin{array}{ccccccccc}
 \beth_{21} & \beth_{22} & \beth_{23} & \beth_{24} & -f_{1} & -f_{2} & -f_{3} & -f_{4}\\
 \beth_{31} & \beth_{32} & \beth_{33} & \beth_{34} & -\beth_{35} & -\beth_{36} & -\beth_{37} & \beth_{38}\\
 \beth_{41} & \beth_{42} & \beth_{43} & \beth_{44} & -\beth_{45} & -\beth_{46} & -\beth_{47} & \beth_{48}\\
 \beth_{51} & \beth_{52} & \beth_{53} & \beth_{54} & -\beth_{55} & -\beth_{56} & -\beth_{57} & \beth_{58}\\
 \beth_{61} & \beth_{62} & \beth_{63} & \beth_{64} & -j_{1} & -j_{2} & -j_{3} & -j_{4}\\
 \beth_{71} & \beth_{72} & \beth_{73} & \beth_{74} & -k_{1} & -k_{2} & -k_{3} & -k_{4}\\
 \beth_{81} & \beth_{82} & \beth_{83} & \beth_{84} & -\beth_{85} & -\beth_{86} & -\beth_{87} & \beth_{88}\\
 \beth_{91} & \beth_{92} & \beth_{93} & \beth_{94} & -l_{1} & -l_{2} & -l_{3} & -l_{4}
          \end{array}\right|$,\\

    \section*{Declaration of competing interest}

 The authors state that they have no known financial interests or personal relationships that could have influenced the work presented in this paper.

    \section*{Funding}

    Research work of Abhishek Mallick is financially supported by UGC, New Delhi, India.
\section*{Data availability}
There is no associated data with the manuscript.
   
 \end{document}